\documentclass[fleqn,usenatbib,useAMS]{mnras}

\usepackage{graphicx}	
\usepackage{amsmath}	
\usepackage{amssymb}	
\usepackage{multicol}        
\usepackage{bm}		
\usepackage{pdflscape}	
\usepackage{longtable} 
\usepackage{blindtext}
\usepackage{multicol}
\usepackage{lscape}
\usepackage{caption}
\usepackage[skip=0.5ex]{subcaption}

\usepackage[T1]{fontenc}
\usepackage{ae,aecompl}

\usepackage{newtxtext,newtxmath}

\title[Optical properties of PS sources]{Optical properties of Peaked Spectrum radio sources}

\author[Nascimento et al.]{R. S. Nascimento$^{1}$\thanks{Contact e-mail: \href{mailto:rnascimento@lna.br}{rnascimento@lna.br}}, A. Rodr\'iguez-Ardila$^{1,2}$, L. Dahmer-Hahn$^{1,6}$, M. A. Fonseca-Faria $^{2}$
\newauthor R. Riffel$^{3}$,  M. Marinello$^{1}$, T. Beuchert$^{4}$, 
J. R. Callingham$^{5}$\\
$^{1}$ Laborat\'orio Nacional de Astrof\'isica, Rua dos Estados Unidos, 154, Itajub\'a/MG, 37504-364, Brazil\\
$^{2}$ Instituto Nacional de Pesquisas Espaciais, Av. dos Astronautas, 1758 - Jardim da Granja S\~ao Jos\'e dos Campos/SP - CEP 12227-010, Brazil\\
$^{3}$ Universidade Federal do Rio Grande do Sul, Porto Alegre, RS, Brazil\\
$^4$ European Southern Observatory, Garching, Germany \\
$^5$ Leiden Observatory, Leiden University, PO Box 9513, 2300 RA Leiden, The Netherlands\\
$^6$ Shanghai Astronomical Observatory, Chinese Academy of Sciences, 80 Nandan Road, Shanghai 200030, China}
\date{Accepted 2021 December 22. Received 2021 December 21; in original form 2021 July 07}

\pubyear{2015}

\begin{document}
\label{firstpage}
\pagerange{\pageref{firstpage}--\pageref{lastpage}}
\maketitle

\begin{abstract}
In this work, we study the optical properties of compact radio sources selected from the literature in order to determine the impact of the radio-jet in their circumnuclear environment. Our sample includes 58 Compact Steep Spectrum (CSS) and GigaHertz Peaked Spectrum (GPS) and 14 Megahertz-Peaked spectrum (MPS) radio sources located at $z\leq 1$. The radio luminosity ($L_R$) of the sample varies between Log\,L$_R\sim$ 23.2 and 27.7 W\,Hz$^{-1}$. We obtained optical spectra for all sources from SDSS-DR12 and performed a stellar population synthesis using the {\sc starlight} code. We derived stellar masses (M$_\star$), ages $\langle t_\star \rangle$, star formation rates (SFR), metallicities $\langle Z_\star \rangle$ and internal reddening A$_V$ for all young AGNs of our sample. A visual inspection of the SDSS images was made to assign a morphological class for each source. Our results indicate that the sample is dominated by intermediate to old stellar populations and there is no strong correlation between optical and radio properties of these sources. Also, we found that young AGNs can be hosted by elliptical, spiral and interacting galaxies, confirming recent findings. When comparing the optical properties of CSS/GPS and MPS sources, we do not find any significant difference. Finally, the Mid-Infrared WISE colours analysis suggest that compact radio sources defined as powerful AGNs are, in general, gas-rich systems. 
\end{abstract}

\begin{keywords}
galaxies, active galactic nuclei
\end{keywords}




\section{Introduction}

It is now recognized that active galactic nuclei (AGN) feedback plays a fundamental role in galaxy evolution \citep[][]{B06,B08,Si07,H15,H19,B20}. Quasar winds and jets from powerful AGN are believed to be important in shaping the galaxy luminosity function and in establishing the observed correlations between the properties of the bulge and the supermassive black hole \citep[SMBH][and references therein]{SR98,T02,Croton06,DM05,DM08,F12,KH13,Crain15,KP15}. The feedback effects are not restricted to galactic scales, though. On sub-kpc scales, they may have a significant impact upon the evolution of the host galaxy, particularly in the earliest stages of jet evolution via interactions between radio jets and the interstellar medium \citep[ISM,][]{Dal13,Morganti13,Cal15,Col18,Keim19,Sobo19,Orienti20}.

In this context, GigaHertz Peak Spectrum (GPS) and Compact Steep Spectrum (CSS) sources, considered as young AGNs (age of $\sim$10$^{2-3}$ years, \citealt{Mur03, PC03}), are exquisite laboratories to study the early stages of jet evolution and the interactions between the radio source and the ISM. Due to the compactness (lobe separation $<$100~pc) and low advance speeds of their radio-jet/lobe structures \citep{GP09}, they can provide information about the energy that is deposited into the ISM by the expanding radio source. They are, therefore, a unique class to search for winds during the onset of jet formation and to understand feedback process by winds and jets at the early stages of evolution of compact accreting objects.

In the early 1990s, when the first studies on the evolution of compact radio sources appeared, it was believed that CSS and GPS sources were part of an evolutionary sequence in which GPS evolved into CSS and, later, depending on the power of the AGN, become 
larger-sized extended sources \citep{fanti95+,odea98}.
Later studies of larger, more complete peaked-spectrum samples extending to lower flux densities have changed this paradigm. The recent review paper by \citet{OdeaSaikia}  suggests that many of the low-luminosity compact sources may not evolve into larger sources, being confined to the small dimensions by the dense interstellar medium of the host galaxies \citep{odea91+,Sobo19}. Alternatively, they may represent sources going through intermittent cycles of low-level activity \citep{kunert10+,Cal15}
 

Also, recent studies have revealed a class of radio-loud AGN with the same spectral shape as CSS and GPS sources, but with an observed turnover frequency below 1\,GHz. These objects, called Megahertz-Peaked spectrum sources \citep[][]{Fal04,Copp15,Copp16}, are believed to be a combination of nearby CSS and GPS sources and High-frequency peaked sources at high redshift ($z > 2$).

Although these sources are currently the focus of many authors to understand the physics involving the jet launching and feedback effects on the ISM \citep[][]{H07a,H09a,H11,SH13,Son,R16,J19,Sant20,W20}, few studies have concentrated on the analysis of the stellar population of their hosts. Early studies on the stellar population were limited to objects with a strong UV-excess emission and revealed the presence of young stellar populations (YSP) in a significant fraction of radio galaxy hosts. For example, \citet{Ta02} found evidence of YSP with ages in the range 0.1 - 2\,Gyr in 15-50 percent of their complete sample of 2-Jy radio galaxies at intermediate redshift. \citet{W04} pointed out the presence of YSP (0.05 - 2\,Gyr) in 25 per cent of a subsample of low-luminosity radio galaxies selected from a 2-Jy sample. \citet{R05} detected a higher fraction of YSP ($<$ 1\,Gyr) in a sample of 20 local radio galaxies.

Thereafter, \citet{H07} examined 12 powerful radio sources at low and intermediate redshift ($z < 0.7$) and found evidence that a combination of AGN-related continuum and an old (12.5\,Gyr) stellar population provides a good fit to the spectra of three of the 12 sources, including one of the two CSS sources in the sample. For the remaining sources they found strong evidence for YSP (0.02 - 1.5\,Gyr). \citet{L08} imaged a sample of nine GPS and CSS in the near-UV with Hubble Space Telescope Advanced Camera for Surveys to search for star formation regions in the host galaxies. They detected near-UV emission in seven of the sources, consistent with star formation occurring in a burst $\leq$ 10\,Myr ago. They also claimed that once the radio source ages are much smaller than this timescale ($10^3$ to $10^6$ yr), there might be a minimum delay of 10\,Myr between the onset of the starburst and the radio activity. \citet{W08} analysed two powerful radio galaxies at intermediate redshift with clear evidence of YSP ($<$ 5\,Gyr) in their optical spectra. They found that both galaxies have a massive intermediate-age (0.2 - 1.2\,Gyr) YSP, but just one of them is dominated by younger ($<$ 0.1\,Gyr) YSP component, suggesting that the sources represent different stages in the same evolutionary scenario. \citet{Son} investigated the relationship between the emission-line properties, black hole accretion and the radio properties of a sample of 34 low-redshift young radio galaxies but they did not access the presence of young stellar population in the host galaxies. 
\citet{R16} used MUSE-VLT data to analyse the spectroscopy and kinematics of the GPS radio source PKS\,B1934-63. They performed a stellar population synthesis using {\sc starlight} package \citep{C07} and found that the main galaxy and its companion are dominated by old stellar populations (mean ages $\sim$ 9 and 5\,Gyr) plus $\sim$ 0.1\% of very young ($\leq$ 40\,Myr) stars. \citet{L20} investigated the optical properties of a large sample of young AGN based on the spectra in SDSS-DR12 \citep{DR12}. They used the Penalized Pixel-Fitting \citep[pPXF,][]{Cap04} code to fit the continuum and measure the stellar velocity dispersion in 41 normal galaxies, but they did not discuss the age of stellar population in the host galaxies.

Note that most of the works described above have dealt with stellar population studies on a small number of sources by means of broad-band colours. With the increasing availability of powerful stellar population synthesis codes (i.e {\sc starlight}, pPXF) and optical spectroscopy for a sizeable number of objects from public surveys as the Sloan Digital Sky Survey \citep[SDSS,][]{Y00} the stellar content of CSS/GPS can now be studied in detail.

In this work, we carry out a detailed study of the stellar population in a sample of CSS/GPS galaxies limited by volume in order to (i) determine the general properties of the stellar population in these sources; (ii) investigate the impact of the radio-jet on the circumnuclear environment of these objects. In Section 2 we present the sample selection and data used. Section 3 describes the approach employed. In Section 4 we show the results from the fitting of the stellar continuum. Section 5 and 6 give a discussion about the morphology of the host galaxies and the Mid-Infrared (MIR) properties of the sources. Final remarks are given in Section 7. Throughout the paper we adopt a $\Lambda$CDM cosmology with $H_0=69.8{\rm km}\,s^{-1}\,{\rm Mpc}^{-1}$, $\Omega_{\rm M}$ = 0.3 and $\Omega_\lambda$ = 0.7.

\section{Peaked-spectrum Sample Selection}

Our sample of peaked-spectrum radio sources was obtained from different radio-selected catalogues publicly available in the literature \citep{S89,F1,F2,St,PT,K02,S1,S2,T09,H10,K10,Son,J16,J19}. Most of these catalogues were constructed using surveys that observed the sky around or above 1 GHz - which corresponds to the turnover frequency in the radio spectra. After removing duplicate objects we built a sample of 204 objects with redshift range restricted to $z < 1$. This choice was made in order to include narrow lines from [\ion{O}{ii}] to [\ion{Ar}{iii}] (3727~\AA\ - 7136~\AA~in the rest frame) in the observed wavelength range covered by SDSS. We looked for optical counterparts from the SDSS-DR12 \citep{DR12} for these objects. A maximum search radius of 8 arcseconds was adopted to match the sources in both FIRST and SDSS catalogs. Of the 204 sources, 84 have a SDSS spectroscopic counterpart and out of these, 75 have spectra with signal-to-noise ratio (SNR) higher than 10. This threshold is necessary in order to warrant a proper fit of the continuum shape and stellar absorption features \citep[see][for a discussion]{C04a,C05a}. Each spectrum was corrected for Galactic extinction using the maps of \citet{S98} and shifted to the rest-frame wavelength using the redshift information in the header of the SDSS spectra. After this last correction, we confirmed that all spectra were indeed in the rest-frame system by visually inspecting the position of the emission lines. 

\begin{figure}
\includegraphics[width=\columnwidth]{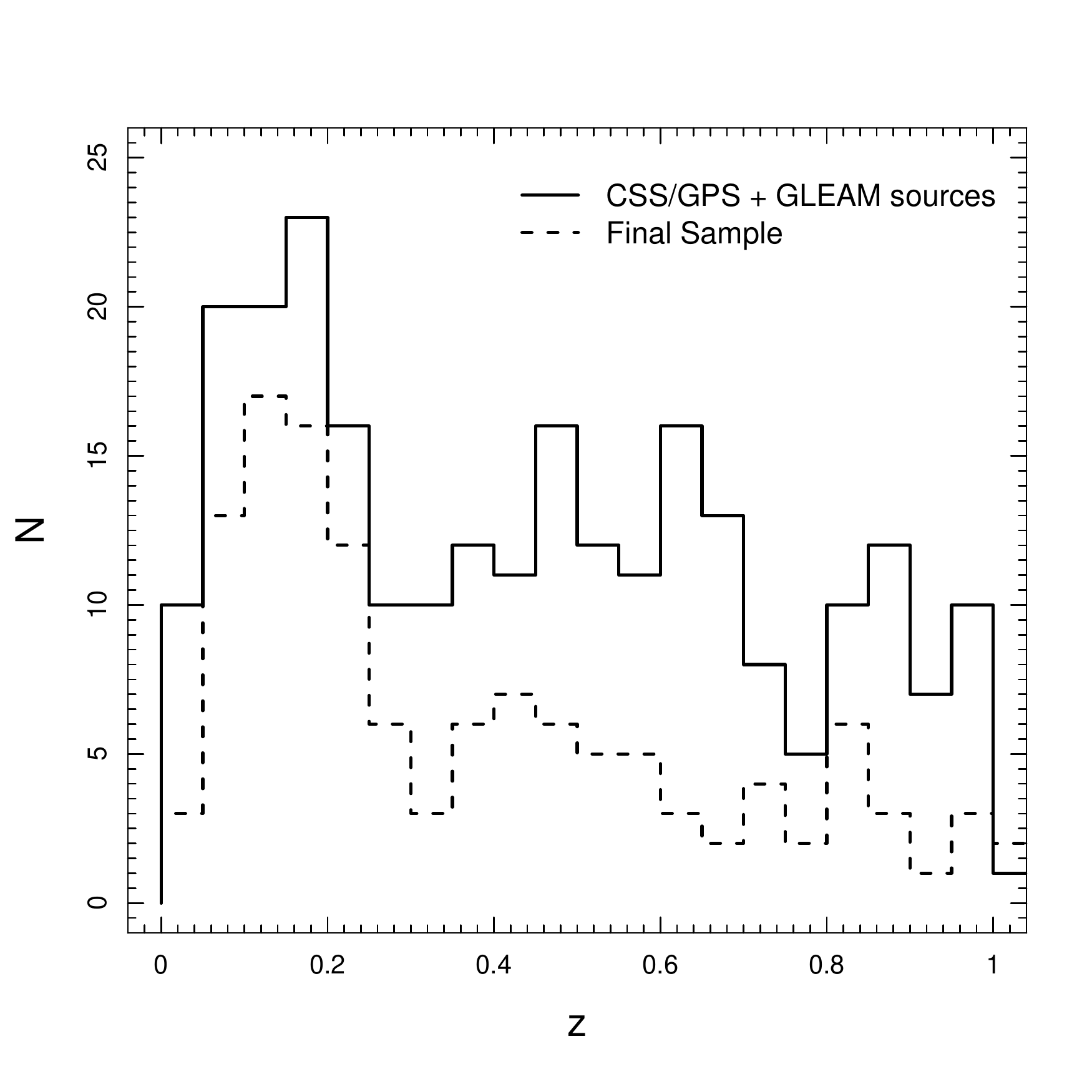}
\caption{Redshift distribution of our sample. The solid and dashed lines represent, respectively, the z distribution for the entire sample of CSS/GPS sources, for objects with a counterpart in the SDSS and those with a signal / noise ratio greater than 10.}\label{fig:zsample}
\end{figure}

We also included low-frequency sources in our sample in order to see whether they have similar properties to those of well-known CSS and GPS. For this purpose, we used a sample of MPS sources \citep[][]{Callingham17} identified from the GaLactic and Extragalactic All-sky Murchison Widefield Array Survey \citep[GLEAM,][]{Wayth15,HW17}. All 1483 objects in this sample have flux density measurements between 72\,MHz and 843\,MHz/1.4\,GHz. Of these, 154 are located at $z < 1$ and 23 have SDSS-DR12 optical spectra. We illustrate in Fig.~\ref{fig:zsample} the redshift distribution of the entire sample of CSS/GPS + GLEAM sources (solid line) and our final sample (dashed line) selected using the SNR criterion. Although the SNR cut removes most galaxies at high redshift ($z > 0.5$), it can be seen that our sub-sample well represents the distribution of objects within the initially available sample.

\begin{figure}
\includegraphics[width=\columnwidth]{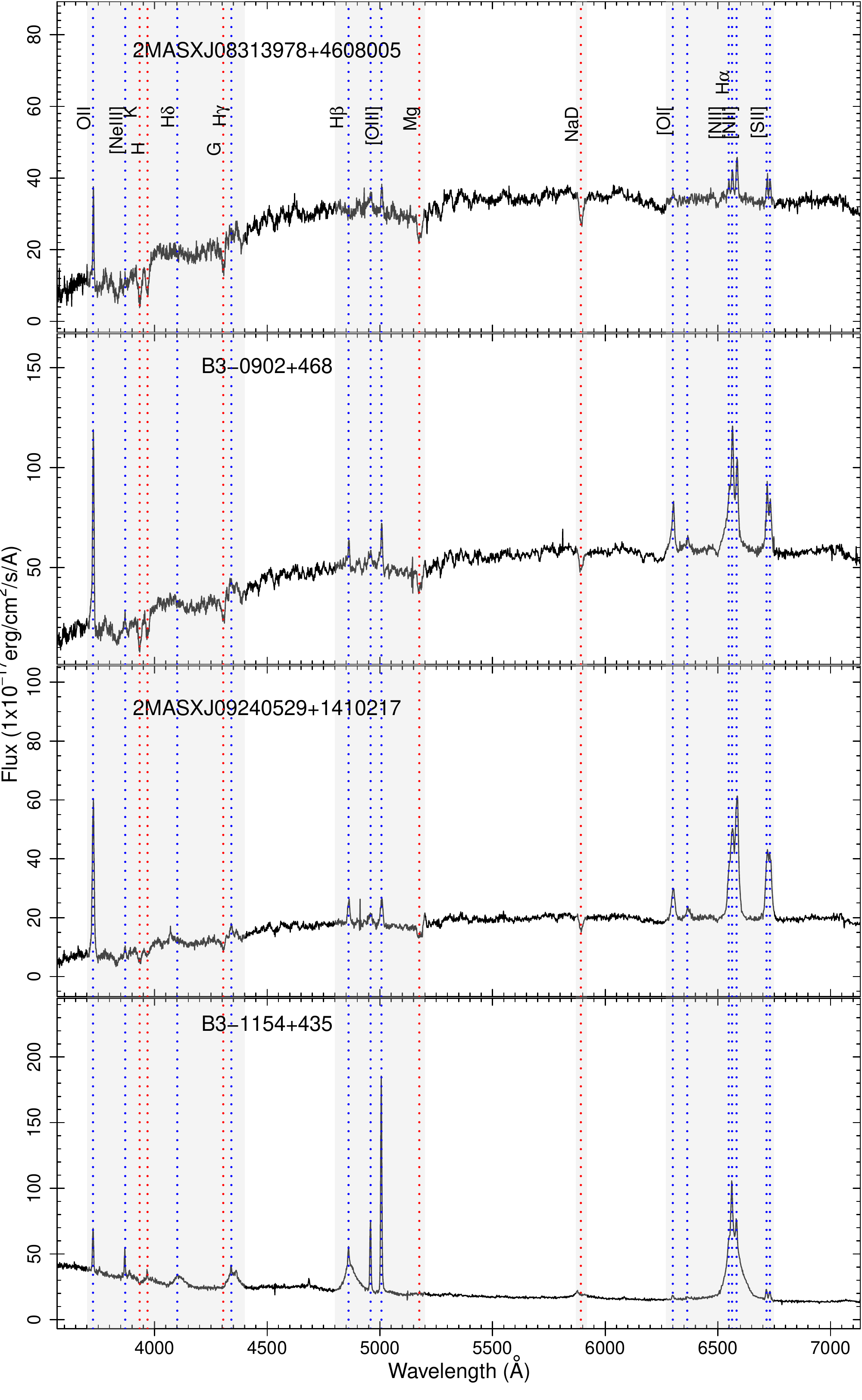}
\caption{An example of the CSS/GPS sample in the galaxy rest-frame. In the upper three spectra, the continuum emission is dominated by stellar absorption features while in the bottom one, the continuum emission is dominated by the featureless continuum emitted by the AGN. The main absorption (red) and emission (blue) lines are marked.}
\label{fig:sample}
\end{figure}

Fig.~\ref{fig:sample} shows four examples of peaked-spectrum sources spectra in the galaxy rest-frame organized in decreasing contribution of the stellar continuum (top to bottom). For instance, in the top panel we display an object dominated by the stellar continuum, with just a few emission lines visible. In the second and third rows, we show objects with decreasing stellar continuum contribution, whereas both the featureless continuum (FC) from the AGN, represented by a power-law continuum, and the emission lines become more prominent. In the bottom panel, we show an object with continuum emission dominated by a featureless continuum, with little signature of stellar absorption features. In contrast, the emission line spectrum is quite prominent, with strong emission lines, some of them displaying broad components, indicating the presence of an active galactic nucleus (AGN). We removed the later type of objects from our analysis, as the light continuum is dominated by the Featureless continuum produced by the AGN.

Table \ref{tab:radio} and \ref{tab_mps:radio} list, respectively, the main properties of the 84 CSS/GPS and the 21 MPS sources in our sample. The radio and MIR data were obtained from Faint Images of the Radio Sky at Twenty-cm survey \citep[FIRST,][]{B95} and Wide-field Infrared Survey Explorer \citep[WISE,][]{Wr10}, respectively. The peak and integrated flux densities, presented in column 5 and 6, were gathered directly from the FIRST catalogue and derived by fitting an elliptical Gaussian model to the source. On the other hand, some sources are outside the FIRST survey coverage area and, in these cases, the integrated flux was obtained from the NRAO VLA Sky Survey (NVSS). In this table, we also depicted the classification of the source according to the position in the WISE colour diagram. See Sect. 6 for a detailed description of the MIR data. We marked in the first column of Table \ref{tab:radio} the objects dominated by a featureless continuum. In these later sources, the continuum from the AGN is so strong that dilutes the stellar absorption features of the underlying stellar population, and thus were removed from our analysis. The gas properties of the whole sample will be analysed in a forthcoming paper.

For all said above, we remove from the subsequent analysis objects with $z > 1$, those dominated by a power-law continuum and those with poor SNR (SNR $<$ 10). Our final sample is composed of 58 CSS/GPS  and 14 MPS sources.

 \begin{table*}
  \caption{Properties of CSS/GPS sources in our sample. (1) Radio source; (2) and (3) Coordinates; (4) redshift; (5) and (6) Peak and Integrated Flux obtained from FIRST survey; (7) and (8) WISE colours ; (9) AGN classification defined from WISE colour diagram: W - weak AGN, S - strong AGN.}
  \label{tab:radio}
  \setlength{\tabcolsep}{0.06in}
  \begin{center}
  \begin{tabular}{l c c c r r c c c}
    \hline \hline
    Source & Ra    & DEC    & z & Peak Flux & Int. Flux & W1-W2 & W2-W3 & AGN class \\ 
       & J2000 & J2000  &   & (mJy/bm)    & (mJy)       &   (mag)    &   (mag)    & \\ 
    \hline
    4C+00.02 						& 00:22:25.42 & 00:14:56.06 & 0.30 & 2827.23 & 2919.39 &  0.47 & 2.44 &   W \\ 
	4C+00.03 					   	& 00:28:33.42 & 00:55:11.02 & 0.10 & 219.20 & 237.22   &  0.31 & 3.02 &   W \\ 
	B2-0026+34					   	& 00:29:14.20 & 34:56:32.00 & 0.52 & -- 	& 1902.90 & 0.74 & 2.49 &   S \\
	3C049$^\dagger$ 				& 01:41:09.10 & 13:53:28.00 & 0.62 & -- 	& 2740.20 & 1.02 & 3.09 &   S \\
	FBQSJ07411$^\dagger$ 	        & 07:41:10.70 & 31:12:00.31 & 0.63 & 2051.49 & 2071.25  & 1.09 & 2.57 &   S \\
	B3-0754+40 					    & 07:57:56.69 & 39:59:36.01 & 0.07 & 98.26 	 & 99.20    & 0.49 & 3.59 &   W \\ 
	4C+47.27$^\dagger$ 				& 08:04:13.91 & 47:04:43.07 & 0.51 & 847.18 & 869.59    & 0.98 & 2.88 &   S \\
	4C+40.20 						& 08:12:53.13 & 40:19:00.09 & 0.55 & 1058.83 & 1067.88  & 1.15 & 2.66 &   S \\
	4C+07.22 						& 08:13:23.76 & 07:34:05.76 & 0.11 & 434.89 & 462.93    & 0.37 & 2.02 &   W \\ 
	2MASXJ0831			            & 08:31:39.81 & 46:08:00.75 & 0.13 & 124.48 & 130.74    & 0.15 & 0.72 &   W \\ 
	2MASXJ0834 			            & 08:34:11.11 & 58:03:21.45 & 0.09 & 50.87 	& 53.04     & 0.07 & 1.13 &   W \\ 
	B3-0833+44 				    	& 08:36:37.83 & 44:01:09.43 & 0.06 & 132.42 & 139.30    & 0.06 & 2.48 &   W \\ 
	SDSSJ08382$^\dagger$ 		    & 08:38:25.00 & 37:10:37.00 & 0.40 & -- 	& 340.20        & 0.46 & 1.99 &   W \\ 
	SDSSJ08485$^\dagger$ 		    & 08:48:56.60 & 01:36:48.00 & 0.35 & -- 	& 87.60        & 0.31 & 2.87 &   W \\ 
	SDSSJ08531$^\dagger$ 		    & 08:53:14.20 & 02:14:54.00 & 0.46 & -- 	&  115.40        & 0.08 & 3.51 &   W \\ 
	3C213.1 						& 09:01:05.32 & 29:01:46.46 & 0.19 & 731.69 & 1492.63   & 0.50 & 2.47 &   W \\ 
	B3-0902+46 					    & 09:06:15.52 & 46:36:18.98 & 0.08 & 306.81 & 313.57    & 0.30 & 2.41 &   W \\ 
	3C216$^\dagger$ 				& 09:09:33.50 & 42:53:46.54 & 0.67 & 3444.13 & 4009.52  & 1.06 & 3.19 &   S \\
	SDSS091734$^\dagger$ 		    & 09:17:34.81 & 50:16:38.15 & 0.63 & 104.04 & 109.07    & 1.08 & 2.74 &   S \\
	2MASXJ0924 			            & 09:24:05.29 & 14:10:21.58 & 0.14 & 105.20 & 108.25    &  0.17 & 0.73 &   W \\ 
	SDSSJ09260$^\dagger$	        & 09:26:07.99 & 07:45:26.58 & 0.44 & 158.49 & 193.76    &  0.90 & 2.14 &   W \\ 
	2MASSJ0934			            & 09:34:30.74 & 03:05:45.49 & 0.23 & 279.90 & 292.09    &  0.32 & 1.75 &   W \\ 
	J0945+1737 						& 09:45:21.36 & 17:37:53.37 & 0.13 & 38.68 & 44.46      &  1.90 & 4.32 &   S \\
	FBQSJ09452 			            & 09:45:25.88 & 35:21:03.46 & 0.21 & 140.45 & 147.55    &  0.57 & 2.43 &   S \\
	J0958+1439 						& 09:58:16.89 & 14:39:24.05 & 0.11 & 10.28 & 10.40      &  1.31 & 4.19 &   S \\
	J1000+1242 						& 10:00:13.17 & 12:42:26.46 & 0.15 & 25.70 & 31.75      &  1.48 & 4.03 &   S \\
	3C237$^\dagger$ 				& 10:08:00.00 & 07:30:16.00 & 0.88 & -- & 6522.1            &  0.46 & 3.33 &   W \\ 
	4C+14.36 						& 10:09:55.51 & 14:01:54.33 & 0.21 & 995.39 & 1044.74   &  0.59 & 2.07 &   W \\ 
	J1010+1413 						& 10:10:22.95 & 14:13:01.11 & 0.20 & 8.09 & 8.82        &  1.85 & 4.42 &   S \\
	J1010+0612 						& 10:10:43.38 & 06:12:01.28 & 0.10 & 92.07 & 99.26      &  1.07 & 3.38 &   S \\
	4C+39.29 						& 10:17:14.18 & 39:01:22.79 & 0.21 & 1096.50 & 1416.72  &  0.36 & 3.51 &   W \\ 
	2MASXJ1026 			            & 10:26:18.27 & 45:42:29.45 & 0.15 & 101.48 & 105.18    &  0.28 & 1.25 &   W \\ 
	4C+39.32 						& 10:28:44.30 & 38:44:36.67 & 0.36 & 643.47 & 690.53    &  0.61 & 2.57 &   S \\
	SDSSJ10350 	            	    & 10:35:07.06 & 56:28:46.82 & 0.46 & 1738.85 & 1829.35  &  1.00 & 2.92 &   S \\
	UGC05771 						& 10:37:19.34 & 43:35:15.07 & 0.02 & 125.68 & 128.88    & -0.02 & 1.89 &   W \\ 
	4C+30.19 						& 10:40:29.96 & 29:57:57.98 & 0.09 & 364.02 & 388.37    &  0.14 & 2.43 &   W \\ 
	2MASSJ1056$^\dagger$ 			& 10:56:28.21 & 50:19:52.24 & 0.82 & 78.86 & 80.75      &  1.24 & 2.42 &   S \\
	J1100+0846 						& 11:00:12.40 & 08:46:16.16 & 0.10 & 58.54 & 61.26      &  1.49 & 3.75 &   S \\
	WISEJ111036$^\dagger$ 	        & 11:10:36.31 & 48:17:52.54 & 0.74 & 522.00 & 541.27    &  0.92 & 2.83 &   S \\
	4C+14.41 						& 11:20:27.81 & 14:20:54.99 & 0.36 & 2404.34 & 2438.53  &  0.41 & 3.60 &   W \\ 
	B3-1128+455 					& 11:31:38.92 & 45:14:51.00 & 0.40 & 1954.42 & 2027.97  &  0.46 & 3.55 &   W \\ 
	SDSSJ114311		                & 11:43:11.03 & 05:35:15.87 & 0.50 & 193.77 & 203.63    &  0.81 & 2.50 &   S \\
	4C+46.23 						& 11:43:39.86 & 46:21:23.57 & 0.12 & 292.80 & 452.69    &  0.15 & 0.87 &   W \\ 
	$[\text{HB}89]1153^\dagger$     & 11:56:18.74 & 31:28:04.75 & 0.42 & 2852.38 & 2960.92  &  1.03 & 3.22 &   S \\
	B3-1154+435$^\dagger$ 			& 11:57:27.60 & 43:18:06.77 & 0.23 & 246.92 & 255.92    &  0.94 & 2.89 &   S \\
	B2-1201+39$^\dagger$			& 12:04:06.86 & 39:12:18.21 & 0.44 & 439.00 & 482.69    &  0.53 & 2.61 &   W \\ 
	3C268.3$^\dagger$				& 12:06:24.70 & 64:13:37.00 & 0.37 & -- &  3719.4            &  1.08 & 2.25 &   S \\
	B3-1206+415 					& 12:09:02.80 & 41:15:59.41 & 0.10 & 145.25 & 146.75    &  0.10 & 1.07 &   W \\ 
	B3-1241+411 					& 12:44:19.99 & 40:51:37.22 & 0.25 & 345.89 & 367.63    &  0.66 & 2.68 &   S \\
	B3-1242+410$^\dagger$			& 12:44:49.20 & 40:48:06.35 & 0.81 & 1329.52 & 1369.07  &  0.80 & 2.72 &   S \\
	4C+49.25 						& 12:47:07.37 & 49:00:18.27 & 0.21 & 1037.73 & 1212.70  &  0.62 & 2.54 &   S \\
	SBS1250+568$^\dagger$ 			& 12:52:26.32 & 56:34:19.65 & 0.32 & 2258.50 & 2442.12  &  1.20 & 3.24 &   S \\
	B3-1308+451 					& 13:10:56.99 & 44:51:46.61 & 0.39 & 96.60 & 100.41     &  0.20 & 0.97 &   W \\ 
	J1316+1753 						& 13:16:42.91 & 17:53:32.35 & 0.15 & 10.66 & 11.42      &  1.13 & 3.86 &   S \\
	B3-1315+41 					    & 13:17:39.21 & 41:15:45.98 & 0.07 & 247.40 & 248.86    &  0.07 & 1.38 &   W \\ 
	2MASXJ1324		        	    & 13:24:19.70 & 04:19:07.21 & 0.26 & 127.92 & 155.15    &  0.27 & 1.12 &   W \\ 
	3C286$^\dagger$ 				& 13:31:08.28 & 30:30:32.95 & 0.85 & 14774.42 & 15025.79 & 1.02 & 2.54 &   S \\
	J1338+1503 						& 13:38:06.52 & 15:03:56.10 & 0.19 & 2.15 & 2.42        & 1.41 & 4.04 &   S \\
	4C+12.50 						& 13:47:33.38 & 12:17:24.09 & 0.12 & 4823.33 & 4859.88  & 1.31 & 3.93 &   S \\
	J1356+1026 						& 13:56:46.12 & 10:26:09.19 & 0.12 & 57.90 & 59.58      & 1.40 & 3.90 &   S \\
	4C+62.22 						& 14:00:28.69 & 62:10:38.41 & 0.43 & 4256.16 & 4375.44  & 1.21 & 2.59 &   S \\
	2MASXJ1400	                    & 14:00:51.62 & 52:16:06.61 & 0.12 & 169.91 & 174.49    & 0.23 & 2.42 &   W \\ 
    \hline
  \end{tabular}
  \end{center}
 \end{table*}
 \addtocounter{table}{-1}

\begin{table*}
\setlength{\tabcolsep}{0.06in} \caption{Continued\dots
\label{tab:radio}}
\begin{center}
\begin{tabular}{l c c c r r c c c}
    \hline \hline
    Source & Ra    & DEC    & z & Peak Flux & Int. Flux & W1-W2 & W2-W3 & AGN class \\ 
            & J2000 & J2000  &   & (mJy/bm)    & (mJy)       &   (mag)    &   (mag) & \\ 
    \hline \hline
	B3-1402+415$^\dagger$			& 14:04:16.30 & 41:17:49.00 & 0.36 & -- & 216.50            & 0.24 & 2.29 &   W \\ 
	MRK0668$^\dagger$ 				& 14:07:00.39 & 28:27:14.78 & 0.08 & 819.00 & 829.58    & 1.02 & 3.06 &   S \\
	2MASXiJ140			            & 14:09:42.46 & 36:04:16.00 & 0.15 & 140.77 & 143.24    & 0.73 & 2.83 &   S \\
	SDSSJ14132		                & 14:13:27.21 & 55:05:29.29 & 0.28 & 125.77 & 128.48    & 0.39 & 2.96 &   W \\ 
	B3-1412+461 					& 14:14:14.85 & 45:54:48.79 & 0.19 & 396.65 & 413.14    & 0.21 & 2.20 &   W \\ 
	SDSSJ14210 	            	    & 14:21:04.25 & 05:08:44.93 & 0.46 & 283.14 & 284.44    & 0.19 & 1.84 &   W \\ 
	J1430+1339 						& 14:30:29.97 & 13:39:12.36 & 0.09 & 13.49 & 26.41      & 1.17 & 3.36 &   S \\
	2MASXJ1435 			            & 14:35:21.68 & 50:51:22.81 & 0.10 & 138.16 & 140.96    & 0.15 & 2.16 &   W \\ 
	B3-1445+410 					& 14:47:12.73 & 40:47:45.65 & 0.20 & 122.71 & 396.94    & 0.53 & 2.14 &   W \\ 
	2MASXj1511		                & 15:11:41.25 & 05:18:09.45 & 0.08 & 75.20 & 77.46      & 1.23 & 2.90 &   S \\
	2MASXJ1530                  	& 15:30:16.24 & 37:58:31.16 & 0.15 & 98.53 & 99.82      & 0.50 & 3.27 &   W \\ 
	B2-1542+39$^\dagger$            & 15:43:49.50 & 38:56:01.00 & 0.55 & -- &  180.9            & 0.29 & 2.31 &   W \\ 
	PKS1543+00 	                    & 15:46:09.53 & 00:26:24.72 & 0.55 & 1844.40 & 1871.70  & 0.65 & 2.70 &   S \\
	SDSSJ15504$^\dagger$            & 15:50:43.90 & 45:36:24.00 & 0.50 & -- &  48.80 &  0.00 & -- &   W \\ 
	2MASSJ1559                  	& 15:59:27.66 & 53:30:54.66 & 0.18 & 170.43 & 182.35 & 0.38 & 2.86 &   W \\ 
	4C+52.37 		                & 16:02:46.39 & 52:43:58.66 & 0.11 & 557.47 & 575.70 & 0.16 & 2.86 &   W \\ 
	PKS1607+26 		                & 16:09:13.33 & 26:41:29.00 & 0.47 & 4735.33 & 4845.05 & 0.29 & 2.68 &   W \\
	4C+40.34 	                	& 16:11:48.55 & 40:40:20.90 & 0.15 & 553.55 & 553.08 & 0.37 & 2.21 &   W \\ 
	SDSSJ16431$^\dagger$            & 16:43:11.35 & 31:56:18.04 & 0.59 & 112.81 & 120.23 & 0.96 & 2.87 &   S \\
	3C346 		                    & 16:43:48.70 & 17:15:49.14 & 0.16 & 1685.26 & 3675.06 & 0.62 & 2.62 &   S \\
	2MASSJ1730$^\dagger$            & 17:30:52.67 & 60:25:16.73 & 0.73 & 71.00 & 72.29 & 1.09 & 2.37 &   S \\
	$[\text{HB}89]2247^\dagger$     & 22:50:25.36 & 14:19:51.88 & 0.23 & 1955.43 & 1981.15 & 0.94 & 2.80 &   S \\
    \hline
\end{tabular}
\begin{minipage}[t]{0.75\textwidth}{\raggedright The $\dagger$ symbol depicts the objects dominated by a non-stellar continuum. These objects were removed from the stellar population synthesis and will be analyzed in a forthcoming paper. The integrated flux of the sources B2-0026+34, 3C049, SDSSJ08382, SDSSJ08485, SDSSJ08531, 3C237, 3C268.3, B3-1402+415, B2-1542+39, and SDSSJ15504 were gathered from the NVSS catalogue. This is because they are outside the FIRST survey coverage area.\par}
\end{minipage}
\end{center}
\end{table*}

\begin{table*}
\centering
\caption{Same as Table \ref{tab:radio} but for the Megahertz-Peaked Spectrum sources in our sample.}
\label{tab_mps:radio}
\begin{tabular}{lcccrrccc}
  \hline\hline
  Source & Ra & DEC & z & Peak Flux & Int. Flux & W1-W2 & W2-W3 & AGN class \\ 
         & J2000 & J2000  &   & (mJy/bm)    & (mJy)       &   (mag)    &   (mag)    & \\ 
  \hline
    WISEAJ023516.81         & 02:35:16.822 & -1:00:51.50 & 0.25 &  174.45 &  176.35 &  0.27 & 2.47 & W \\ 
    $[$HB89$]$0829+046$^\dagger$  & 08:31:48.859 & 04:29:39.23 & 0.17 &  729.40 &  738.57 &  0.99 & 2.56 & S \\ 
    PKS0827+23$^\dagger$     & 08:30:21.694 & 23:23:25.72 & 0.57 & 1033.58 & 1076.92 &  0.70 & 2.23 & S \\ 
    OJ+287$^\dagger$          & 08:54:48.884 & 20:06:30.53 & 0.31 & 1135.80 & 1182.12 &  1.05 & 2.60 & S \\ 
    WISEAJ085010.43           & 08:50:10.419 & 07:47:58.40 & 0.18 &  217.82 &  223.70 &  0.78 & 3.00 & S \\ 
    WISEAJ085323.42           & 08:53:23.431 & 09:27:44.23 & 0.12 &  122.31 & 123.72  &  1.26 & 2.46 & S \\ 
    4C+17.44                  & 08:21:44.034 & 17:48:20.30 & 0.30 & 1875.82 & 1960.04 &  0.65 & 3.71 & W \\ 
    WISEAJ082926.57           & 08:29:26.606 & 01:07:06.66 & 0.39 &  129.16 &  131.56 &  0.28 & 2.85 & W \\ 
    WISEAJ083121.42           & 08:31:21.422 & 07:55:45.77 & 0.50 &  332.07 &  339.56 &  0.17 & 3.44 & W \\ 
    PKS0829+18                & 08:32:16.038 & 18:32:12.16 & 0.15 &  852.06 &  874.24 &  0.55 & 2.77 & W \\ 
    FBQSJ095649.8$^\dagger$   & 09:56:49.873 & 25:15:15.96 & 0.71 & 1041.77 & 1069.98 &  1.16 & 2.89 & S \\ 
    PKS0937+033               & 09:39:45.181 & 03:04:26.65 & 0.54 &  451.78 &  469.48 &  0.30 & 3.16 & W \\ 
    4C+05.47 $^\dagger$       & 11:00:11.516 & 04:44:01.18 & 0.89 &  562.22 &  575.23 &  1.59 & 3.98 & S \\ 
    WISEAJ115654.48           & 11:56:54.481 & 09:32:41.20 & 0.22 &  464.18 &  466.47 &  0.34 & 2.21 & W \\ 
    $[$HB89$]$1222+037$^\dagger$ & 12:24:52.427 & 03:30:50.35 & 0.95 & 1299.49 & 1348.77 &  1.17 & 3.04 & S \\ 
    $[$HB89$]$1252+119$^\dagger$ & 12:54:38.247 & 11:41:05.94 & 0.87 &  702.86 &  747.64 &  1.20 & 2.73 & S \\ 
    WISEAJ122933              & 12:29:33.489 & 16:01:57.34 & 0.41 &  390.64 &  398.14 &  0.50 & 2.65 & W \\ 
    WISEAJ144635.35$^\dagger$ & 14:46:35.357 & 17:21:07.35 & 0.10 &  539.38 &  552.32 &  0.99 & 3.10 & S\\
    NGC5506                   & 14:13:14.877 & -3:12:27.56 & 0.01 &  315.24 &  337.87 &  1.16 & 2.50 & S \\
    $[$HB89$]$1413+135        & 14:15:58.822 & 13:20:23.81 & 0.25 & 1141.91 & 1178.86 &  1.30 & 2.90 & S \\ 
    3C316                     & 15:16:56.588 & 18:30:21.77 & 0.58 & 1296.39 & 1363.18 &  1.27 & 4.02 & S \\ 
    3C315                     & 15:13:40.082 & 26:07:30.79 & 0.11 &  292.82 &  344.67 &  0.10 & 0.70 & W \\ 
   \hline
\end{tabular}
\end{table*}

\section{Stellar Population Synthesis}

In order to characterize the stellar properties of the radio-sources, we carry out a stellar population synthesis using the {\sc starlight} spectral synthesis code \citep{C04a,C05a,A07}. The procedures followed are described in \citet{R09} and \citet{D14}, including, for example, the methods used to handle extinction and emission lines.

Briefly, {\sc starlight} fits an observed spectrum ($O_\lambda$) with a combination, in different proportions, of simple stellar populations (SSPs). The code solves the following equation for a model spectrum M$_\lambda$ \citep[][]{C05}:

\begin{equation}
M_\lambda = M_{\lambda0}\left [\sum_{j=1}^{N\star} x_j\, b_{j,\lambda}\, r_\lambda \right] \otimes G(v_\star,\sigma_\star),
\end{equation}

\noindent where $M_{\lambda0}$ is the synthetic flux at the normalization wavelength; $x_j$ is the population vector; $b_{j,\lambda}\,r_\lambda$ is the reddened spectrum of the {\it j}th SSP normalized at the $\lambda_0$; $r_\lambda = 10^{-0.4(A_\lambda - A_{\lambda0})}$ is the reddening term; $\otimes$ denotes the convolution operator and $G(v_\star,\sigma_\star)$ is the Gaussian distribution used to model the line-of-sight stellar motion, centered at velocity $v_\star$ and with dispersion $\sigma_\star$. The final fit is carried out searching for the minimum between the observed and the model spectrum as follow:

\begin{equation}
    \chi^2 = \sum_k [(O_\lambda - M_\lambda) w_\lambda]^2,
\end{equation}

\noindent where emission lines and spurious features are masked out by using $w_\lambda = 0$. Examples of masked regions are indicated in the bottom panels of Fig \ref{fig:fit1}. Besides $\chi^2$, {\sc starlight} code uses the Allan Deviation (ADEV) parameter to measure the robustness of the fit. It provides the percent mean deviation $\vert O_\lambda - M_\lambda\vert/O_\lambda$ over all fitted pixels. The better  the spectral fitting is, $\chi^2$ approaches one and the ADEV approaches zero.. For a detailed description  of {\sc starlight} see its manual\footnote{http://astro.ufsc.br/starlight/}. For applications in the optical, see \citet[][]{C05,C10}. 

We fed {\sc starlight} with the GRANADA + MILES (GM) library as described by \citet{CF+14}. This base was constructed by combining simple stellar population spectra from \citet{miles}, which starts at an age of 63\,Myr, with synthetic stellar spectra from \citet{gonzalezdelgado+05} for younger ages. The Salpeter Initial Mass Function is adopted and the evolutionary tracks are based on \citet[][]{Gi00} isochrones. The final set of models employed consisted of 21 ages\footnote{0.001, 0.00562, 0.01, 0.0141, 0.02, 0.0316, 0.0562, 0.1, 0.2, 0.316, 0.398, 0.501, 0.631, 0.708, 0,794, 0.891, 1.0, 2.0, 5.01, 8.91 and 12.6\,Gyr} and 4 metallicities\footnote{Z = 0.2, 0.4, 1 and 1.5\,Z$\odot$}. In order to reproduce the featureless continuum emitted by the AGN, we also included a power-law with $F_\nu \propto \nu^{-1.5}$. The law by \citet{CCM} was used to model the dust extinction. 
\par
Also, since small variations are washed away by noise present in real data, we followed \citet[][]{R09} and defined the light fraction population vectors\footnote{We used the nomenclature population vector to specify bins of the stellar population with a determined range of ages.} as follows: $x_{\rm y}$ ($t\,\leq$ 50\,Myr), $x_{\rm i}$ (50\,Myr $< t \leq$ 2\,Gyr) and $x_{\rm o}$ ($t\,>$ 2\,Gyr) to represent the young, intermediate and old stellar population vectors, respectively. The mass components of the population vectors ($m_{y}$, $m_{i}$, $m_{o}$) were defined using the same bins previously described. The light- and mass-weighted population vectors for the CSS/GPS sample are presented in Columns 3-8 of Table \ref{tab:starlight}. For the MPS sample see Table \ref{tab_mps:radio}. We illustrate the contribution of the population vectors over the whole sample (CSS + MPS sources) in the upper panel of Figure~\ref{fig:hpop}. As we can see in the top-left panel of this figure, in most sources the $x_y$ and $x_i$ components contribute with less than $\sim$ 30 percent to the integrated flux. The old-age component, however, is well distributed with a maximum at $\sim$ 90 per cent. In terms of mass-weighted components (top-right hand of Fig.~\ref{fig:hpop}), the contribution of $m_y$ and $m_i$ is very small compared to $m_o$, which is higher than 70 per cent for most sources.

\subsection{Mean stellar age and metallicity}

\begin{figure*}
	\includegraphics[width=0.7\textwidth]{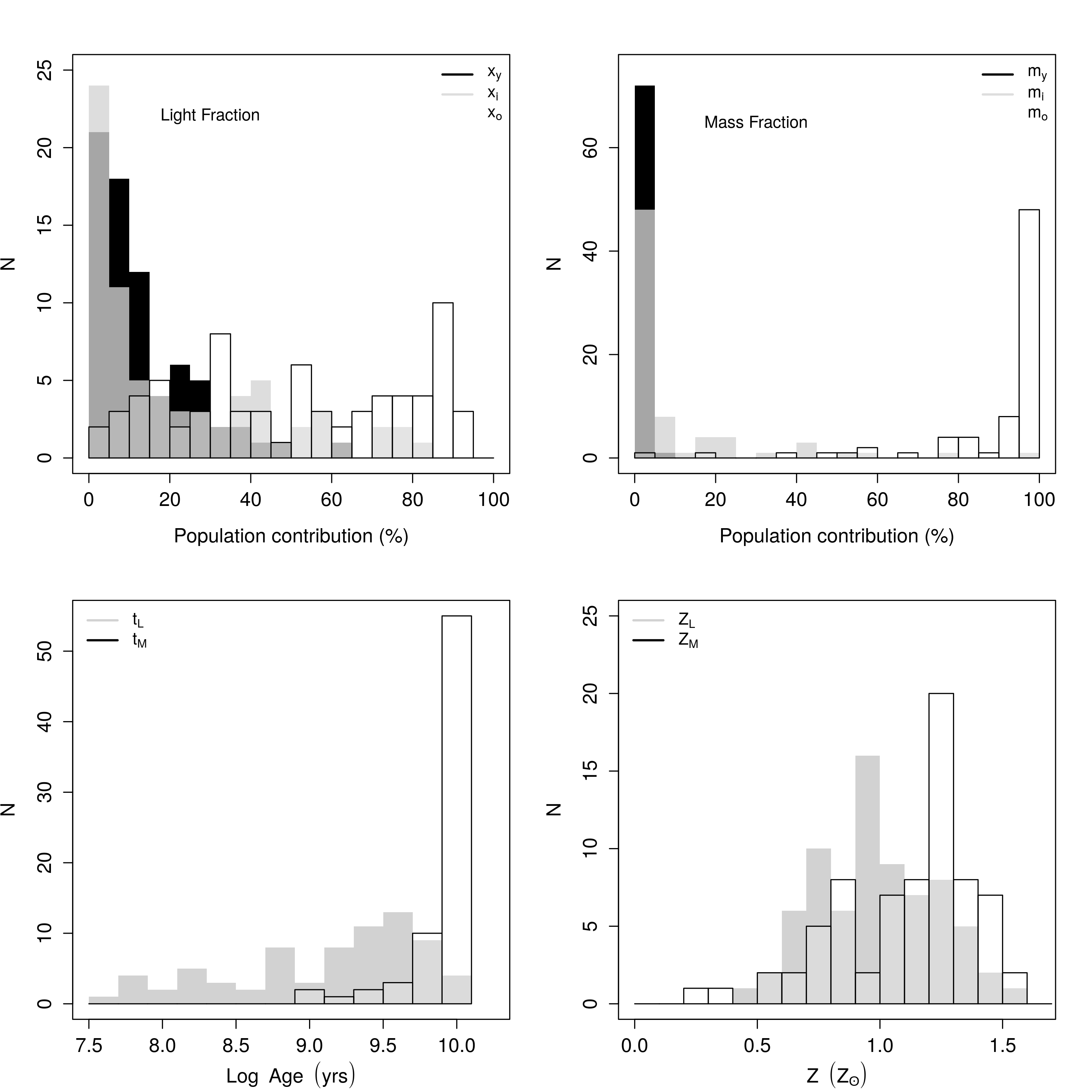}
	\caption{Up: Contribution of the population vector component in terms of flux (top-left panel) and mass (top-right panel). Bottom: Distribution of average ages (Right panel) and metallicities (left panel). Black, grey and white represent, respectively, the young, intermediate-age and old components.}
	\label{fig:hpop}
\end{figure*}

The mean stellar age $\langle t \rangle$ and metallicity $\langle Z \rangle$ are fundamental parameters to describe the mixture of stellar population in a galaxy. They can be defined in terms of light fraction:

\begin{equation}
\langle {\rm log}\, t_\star\rangle_L = \sum_{j=1}^{N\star} x_j \, {\rm log} t_j \quad \textrm{and} \quad  \langle {\rm Z}_\star\rangle_L = \sum_{j=1}^{N\star} x_j \, {\rm Z}_j
\label{eq:tzL}
\end{equation}

\noindent and mass fraction:

\begin{equation}
\langle {\rm log}\, t_\star\rangle_M = \sum_{j=1}^{N\star} m_j \, {\rm log} t_j \quad \textrm{and} \quad  \langle {\rm Z}_\star\rangle_M = \sum_{j=1}^{N\star} m_j \, {\rm Z}_M
\label{eq:tzM}
\end{equation}

\noindent The values found for the above parameters are listed in Columns 10\,-\,13 of Tables \ref{tab:starlight} and \ref{tab:starlight_ps}. The light-weighted mean age for our sources are spread among young to old age stellar population, while for the mass-weighted mean ages the old population dominates (see bottom-left panel of Fig.~\ref{fig:hpop} for clarity). Also, in the bottom-right panel of Figure~\ref{fig:hpop} we show the mass- and flux-weighted mean metallicities. According to the two-sample Kolmogorov-Smirnov test, the mean metallicity distributions are distinct ($\rho$-value = 0.01) with the light-weighted values dominated by solar metallicities and the mass-weighted ones indicating a super-solar metallicities. The discrepancy observed between age and metallicity when weighting by mass or by light, is related with the non-constant stellar M/L ratio. Whereas younger or less metallic populations are more luminous, older or more metallic populations require a much larger amount of mass to produce the same luminosity. 

\begin{table*}
  \caption{Stellar Population Synthesis Results for CSS/GPS sources}
  \label{tab:starlight}
  \setlength{\tabcolsep}{0.06in}
  \begin{center}
  \begin{tabular}{lrrrrrrcccccccccl}
    \hline \hline
        (1) & (2) & (3) & (4) & (5) & (6) & (7) & (8) & (9) & (10) & (11) & (12) & (13) & (14) & (15) & (16) & (17) \\
    Object  & FC  & {\it x}$_y$ & {\it x}$_i$ & {\it x}$_o$ & {\it m}$_y$ & {\it m}$_i$ & {\it m}$_o$ & A$_{\rm V}$ & $\langle \rm{log}\, t_\star \rangle_L$ & $\langle \rm{log}\,t_\star \rangle_M$ & $\langle\rm{Z_\star}\rangle_L$ & $\langle\rm{Z_\star}\rangle_M$ & M$_\star$ & SFR   & Log L$_R$  & Morph \\
    & (\%) & (\%)  &(\%)   &(\%)  &(\%) & (\%) & (\%) & (mag) &   (yr) &  (yr)      & (*) & (*) & (M$_\odot$) & (M$_\odot$/yr)  & (W/Hz) & \\ 
    \hline\hline
2MASSJ0934 & 3.35 & 10.93 & 12.92 & 72.81 & 0.03 & 2.41 & 97.56 & 0.28 & 9.48 & 9.98 & 0.02 & 0.02 & 11.59 & 2.16 & 25.41 & E \\ 
2MASSJ1559 & 7.53 & 5.05 & 44.00 & 43.42 & 0.02 & 19.07 & 80.91 & -0.22 & 9.38 & 9.80 & 0.02 & 0.02 & 11.03 & 0.36 & 25.03 & I/M \\ 
2MASXJ0831 & 0.29 & 5.13 & 3.32 & 91.26 & 0.01 & 0.59 & 99.40 & 0.05 & 9.81 & 10.02 & 0.03 & 0.03 & 11.81 & 1.34 & 24.64 & E \\ 
2MASXJ0834 & 0.00 & 13.38 & 0.28 & 86.33 & 0.02 & 0.01 & 99.97 & 0.41 & 9.60 & 10.08 & 0.02 & 0.02 & 12.02 & 4.33 & 23.98 & E \\ 
2MASXJ0924 & 15.50 & 4.62 & 8.34 & 71.55 & 0.01 & 0.89 & 99.10 & 0.20 & 9.72 & 10.01 & 0.03 & 0.03 & 11.00 & 0.20 & 24.59 & E \\ 
2MASXJ1026 & 0.09 & 12.79 & 0.29 & 86.83 & 0.02 & 0.01 & 99.98 & 0.19 & 9.58 & 10.06 & 0.03 & 0.03 & 11.70 & 1.81 & 24.66 & E \\ 
2MASXJ1324 & 6.53 & 8.66 & 0.49 & 84.32 & 0.02 & 0.11 & 99.88 & 0.34 & 9.63 & 9.99 & 0.02 & 0.02 & 12.13 & 4.76 & 25.25 & E \\ 
2MASXJ1400 & 2.94 & 10.74 & 24.73 & 61.58 & 0.04 & 1.31 & 98.65 & 0.55 & 9.28 & 9.97 & 0.02 & 0.02 & 11.70 & 3.84 & 24.68 & E \\ 
2MASXJ1435 & 3.67 & 13.88 & 14.36 & 68.09 & 0.04 & 0.80 & 99.16 & 0.67 & 9.35 & 10.02 & 0.02 & 0.02 & 11.72 & 5.07 & 24.45 & S \\ 
2MASXJ1530 & 6.70 & 21.01 & 36.10 & 36.20 & 0.12 & 4.73 & 95.16 & 0.93 & 8.83 & 9.96 & 0.02 & 0.03 & 11.50 & 8.15 & 24.64 & S \\ 
2MASXiJ140 & 5.52 & 5.68  & 2.94  & 85.86 & 0.01 & 0.94 & 99.05 & 0.68 & 9.74 & 9.99 & 0.02 & 0.02 & 11.67 & 0.90 & 24.78 & E \\ 
2MASXj1511 & 38.02 & 0.90 & 1.46 & 59.62 & 0.00 & 0.55 & 99.45 & 0.41 & 9.91 & 10.00 & 0.01 & 0.02 & 11.24 & 0.05 & 24.04 & S \\ 
3C213.1 & 2.92 & 6.63 & 59.43 & 31.01 & 0.04 & 24.29 & 75.67 & 0.32 & 9.11 & 9.76 & 0.02 & 0.02 & 11.17 & 1.45 & 26.00 & S \\ 
3C346 & 24.84 & 13.05 & 0.65 & 61.46 & 0.04 & 0.03 & 99.94 & 0.30 & 9.41 & 9.99 & 0.02 & 0.02 & 11.87 & 5.73 & 26.25 & I/M \\ 
4C+00.02 & 1.24 & 28.96 & 51.38 & 18.42 & 2.74 & 41.57 & 55.69 & 0.50 & 8.38 & 9.41 & 0.02 & 0.02 & 10.76 & 5.75 & 26.64 & P \\ 
4C+00.03 & 18.33 & 1.94 & 64.06 & 15.67 & 0.02 & 21.75 & 78.23 & 0.21 & 9.03 & 9.74 & 0.03 & 0.03 & 10.72 & 0.20 & 24.71 & S \\ 
4C+07.22 & 9.80 & 5.84 & 14.18 & 70.17 & 0.01 & 1.09 & 98.90 & 0.66 & 9.61 & 10.01 & 0.02 & 0.03 & 11.99 & 2.42 & 25.06 & S \\ 
4C+12.50 & 29.08 & 15.80 & 0.20 & 54.92 & 0.05 & 0.10 & 99.84 & 0.75 & 9.12 & 9.97 & 0.02 & 0.02 & 11.94 & 9.95 & 26.15 & I/M \\ 
4C+14.36 & 16.64 & 9.04 & 0.73 & 73.59 & 0.02 & 0.05 & 99.93 & 0.05 & 9.58 & 9.99 & 0.03 & 0.03 & 11.57 & 1.34 & 25.92 & I/M \\ 
4C+14.41 & 42.76 & 1.69 & 13.78 & 41.76 & 0.00 & 5.06 & 94.94 & 0.07 & 9.64 & 9.94 & 0.03 & 0.03 & 10.63 & 0.05 & 26.68 & P \\ 
4C+30.19 & 7.23 & 6.55 & 54.44 & 31.79 & 0.04 & 2.33 & 97.63 & 0.51 & 8.74 & 9.97 & 0.01 & 0.03 & 11.59 & 3.83 & 24.81 & I/M \\ 
4C+39.29 & 2.81 & 3.92 & 7.43 & 85.84 & 0.01 & 2.77 & 97.22 & 0.13 & 9.70 & 9.89 & 0.01 & 0.01 & 10.70 & 0.11 & 26.05 & P \\ 
4C+39.32 & 46.40 & 7.33 & 18.14 & 28.13 & 0.04 & 6.11 & 93.85 & 0.25 & 9.18 & 9.91 & 0.03 & 0.03 & 11.29 & 1.57 & 26.13 & S \\ 
4C+40.20 & 59.35 & 0.31 & 40.22 & 0.13 & 0.01 & 97.37 & 2.62 & 0.60 & 8.85 & 8.90 & 0.03 & 0.03 & 10.44 & 0.06 & 26.60 & P \\ 
4C+40.34 & 20.20 & 9.84 & 20.10 & 49.86 & 0.04 & 1.89 & 98.08 & 0.54 & 9.31 & 9.98 & 0.02 & 0.02 & 11.53 & 2.97 & 25.38 & S \\ 
4C+46.23 & 12.08 & 7.35 & 1.00 & 79.56 & 0.01 & 0.13 & 99.85 & 0.21 & 9.67 & 10.00 & 0.02 & 0.03 & 11.99 & 2.75 & 25.08 & E \\ 
4C+49.25 & 0.64 & 27.23 & 59.09 & 13.04 & 1.51 & 52.27 & 46.21 & 0.56 & 8.36 & 9.33 & 0.02 & 0.02 & 10.97 & 10.30 & 25.96 & E \\ 
4C+52.37 & 5.68 & 3.32 & 46.60 & 44.40 & 0.02 & 6.89 & 93.10 & 0.48 & 9.26 & 9.91 & 0.01 & 0.02 & 11.69 & 1.69 & 25.11 & S \\ 
4C+62.22 & 19.72 & 12.02 & 12.66 & 55.59 & 0.04 & 0.43 & 99.54 & 0.73 & 9.30 & 10.04 & 0.02 & 0.02 & 11.44 & 2.30 & 27.06 & P \\ 
B2-0026+34 & 9.41 & 19.10 & 37.79 & 33.69 & 0.08 & 20.97 & 78.95 & 0.86 & 8.92 & 9.79 & 0.03 & 0.03 & 11.32 & 3.65 & 26.81 & P \\ 
B3-0754+401 & 21.79 & 5.75 & 19.35 & 53.11 & 0.01 & 2.89 & 97.09 & 0.24 & 9.50 & 9.99 & 0.02 & 0.02 & 11.39 & 0.72 & 23.95 & S \\ 
B3-0833+442 & 2.38 & 10.45 & 0.00 & 87.17 & 0.01 & 0.00 & 99.99 & 0.49 & 9.68 & 10.07 & 0.02 & 0.02 & 11.60 & 1.15 & 23.95 & S \\ 
B3-0902+468 & 2.39 & 13.07 & 0.36 & 84.19 & 0.03 & 0.06 & 99.92 & 0.40 & 9.58 & 10.04 & 0.02 & 0.02 & 11.75 & 3.22 & 24.66 & E \\ 
B3-1128+455 & 6.30 & 3.68 & 75.09 & 14.93 & 0.05 & 41.20 & 58.74 & 0.49 & 8.89 & 9.59 & 0.01 & 0.02 & 10.71 & 0.65 & 26.68 & P \\ 
B3-1206+415 & 0.16 & 9.72 & 0.42 & 89.70 & 0.01 & 0.02 & 99.97 & 0.07 & 9.66 & 10.04 & 0.03 & 0.03 & 11.69 & 1.24 & 24.42 & E \\ 
B3-1241+411 & 25.89 & 7.63 & 31.81 & 34.68 & 0.03 & 6.57 & 93.40 & 0.17 & 9.16 & 9.91 & 0.02 & 0.03 & 11.45 & 1.91 & 25.59 & S \\ 
B3-1308+451 & 10.61 & 1.86 & 0.51 & 87.03 & 0.00 & 0.01 & 99.99 & 0.06 & 9.97 & 10.05 & 0.02 & 0.03 & 11.71 & 0.33 & 25.35 & I/M \\ 
B3-1315+415 & 0.00 & 15.50 & 6.09 & 78.41 & 0.02 & 0.20 & 99.77 & 0.74 & 9.44 & 10.06 & 0.03 & 0.03 & 12.11 & 6.28 & 24.35 & S \\ 
B3-1412+461 & 2.41 & 12.94 & 29.82 & 54.83 & 0.04 & 1.95 & 98.01 & 0.36 & 9.20 & 10.03 & 0.02 & 0.03 & 10.59 & 0.36 & 25.41 & P \\ 
B3-1445+410 & 11.65 & 1.01 & 0.09 & 87.25 & 0.00 & 0.02 & 99.98 & 0.03 & 9.95 & 10.02 & 0.02 & 0.03 & 11.33 & 0.05 & 25.43 & E \\ 
FBQSJ094525 & 19.51 & 27.23 & 37.31 & 15.95 & 0.29 & 18.68 & 81.03 & -0.05 & 8.29 & 9.77 & 0.02 & 0.02 & 10.88 & 4.69 & 25.05 & I/M \\ 
J0945+1737 & 39.13 & 36.33 & 1.94 & 22.60 & 0.34 & 2.70 & 96.95 & 0.41 & 7.77 & 9.95 & 0.02 & 0.01 & 10.94 & 6.45 & 24.15 & S \\ 
J0958+1439 & 39.98 & 20.86 & 5.69 & 33.48 & 0.13 & 4.08 & 95.79 & 0.69 & 8.55 & 9.94 & 0.01 & 0.01 & 11.16 & 3.75 & 23.39 & S \\ 
J1000+1242 & 19.10 & 34.20 & 15.90 & 30.81 & 0.23 & 6.23 & 93.55 & 0.14 & 8.26 & 9.93 & 0.01 & 0.01 & 10.81 & 3.02 & 24.12 & I/M \\ 
J1010+0612 & 4.21 & 28.69 & 57.98 & 9.12 & 7.57 & 55.80 & 36.57 & 0.52 & 8.24 & 9.10 & 0.01 & 0.02 & 10.74 & 12.70 & 24.28 & I/M \\ 
J1010+1413 & 33.54 & 28.96 & 1.06 & 36.44 & 0.15 & 1.22 & 98.63 & 0.38 & 8.27 & 9.97 & 0.01 & 0.01 & 11.31 & 6.33 & 23.80 & I/M \\ 
J1100+0846 & 35.59 & 30.03 & 5.87 & 28.52 & 0.15 & 5.46 & 94.39 & 0.21 & 8.08 & 9.91 & 0.02 & 0.02 & 11.45 & 8.65 & 24.09 & S \\
J1316+1753 & 37.88 & 37.04 & 0.00 & 25.08 & 0.16 & 0.00 & 99.83 & 0.87 & 7.71 & 10.06 & 0.02 & 0.02 & 11.39 & 8.20 & 23.69 & I/M \\ 
J1338+1503 & 17.39 & 23.01 & 43.97 & 15.64 & 0.41 & 16.55 & 83.03 & 0.43 & 8.46 & 9.82 & 0.01 & 0.02 & 11.11 & 11.70 & 23.18 & S \\ 
J1356+1026 & 20.34 & 45.21 & 0.00 & 34.45 & 0.16 & 0.00 & 99.84 & 0.41 & 7.80 & 10.03 & 0.02 & 0.02 & 11.67 & 15.30 & 24.25 & I/M \\ 
J1430+1339 & 14.08 & 40.69 & 37.08 & 8.15 & 0.87 & 30.50 & 68.64 & 0.21 & 7.89 & 9.67 & 0.02 & 0.02 & 11.03 & 19.20 & 23.59 & S \\ 
PKS1543+005 & 20.96 & 1.34 & 9.59 & 68.12 & 0.00 & 3.03 & 96.97 & 0.31 & 9.78 & 9.94 & 0.03 & 0.03 & 11.39 & 0.11 & 26.85 & P \\ 
PKS1607+26 & 10.56 & 8.38 & 29.88 & 51.19 & 0.03 & 5.03 & 94.94 & 0.59 & 9.35 & 9.98 & 0.02 & 0.02 & 11.23 & 1.05 & 27.16 & P \\ 
SDSSJ103507 & 15.78 & 2.19 & 6.77 & 75.26 & 0.00 & 0.41 & 99.59 & 0.31 & 9.84 & 10.05 & 0.02 & 0.02 & 11.11 & 0.11 & 26.72 & P \\ 
SDSSJ114311 & 42.64 & 24.46 & 31.19 & 1.71 & 3.02 & 77.79 & 19.19 & -0.00 & 8.22 & 9.08 & 0.03 & 0.03 & 10.22 & 6.35 & 25.82 & P \\ 
SDSSJ141327 & 0.22 & 2.43 & 84.50 & 12.86 & 0.06 & 24.12 & 75.82 & 0.76 & 8.85 & 9.73 & 0.02 & 0.02 & 11.33 & 2.58 & 25.22 & E \\ 
SDSSJ142104 & 1.71 & 1.49 & 43.47 & 53.34 & 0.01 & 11.83 & 88.17 & 0.23 & 9.45 & 9.86 & 0.02 & 0.03 & 11.77 & 0.81 & 25.91 & P \\ 
UGC05771 & 1.11 & 11.72 & 0.94 & 86.23 & 0.02 & 0.05 & 99.93 & 0.28 & 9.53 & 10.00 & 0.03 & 0.03 & 11.63 & 1.69 & 23.23 & E \\
    \hline
  \end{tabular}
  \begin{minipage}[t]{\textwidth}{\raggedright $^{1}$Galaxy name;  $^{2-8}$ Percentage contribution of a Featureless Continuum, Light- and Mass-weighted young SSPs, intermediate-age SSPs and old SSPs respectively; $^{9}$ Dust extinction; $^{10-13}$ Light- and Mass-weighted average age and metallicity of the stellar population; $^{14}$ Present stellar mass of the galaxy; $^{15}$ Star Formation Rate over the last 50\,Myr; $^{16}$ Radio luminosity estimated assuming H$_{\rm 0}$=69.8 (Freedman et al., 2019); $^{17}$ Morphology determined from SDSS images.\par}
\end{minipage}
\end{center}
\end{table*}

\begin{table*}
\centering
\caption{Stellar Population Synthesis Results for MPS Sources}
\label{tab:starlight_ps}
\setlength{\tabcolsep}{3.5pt}
\begin{center}
\begin{tabular}{lrrrrrrcccccccccl}
  \hline\hline
  (1) & (2) & (3) & (4) & (5) & (6) & (7) & (8) & (9) & (10) & (11) & (12) & (13) & (14) & (15) & (16) & (17) \\
Object & FC  & {\it x}$_y$ & {\it x}$_i$ & {\it x}$_o$ & {\it m}$_y$ & {\it m}$_i$ & {\it m}$_o$ & A$_{\rm V}$ & $\langle \rm{log}\, t_\star \rangle_L$ & $\langle \rm{log}\,t_\star \rangle_M$ & $\langle\rm{Z_\star}\rangle_L$ & $\langle\rm{Z_\star}\rangle_M$ & M$_\star$ & SFR & Log L$_R$  & Morph \\

 & (\%) & (\%)  &(\%)   &(\%)  &(\%) & (\%) & (\%) & (mag) &   (yr) &  (yr)      & (*) & (*) & (M$_\odot$) & (M$_\odot$/yr)  & (W/Hz) & \\ 
  \hline \hline
WISEAJ023516.81 & 6.21 & 5.31 & 22.69 & 65.79 & 0.03 & 0.97 & 99.00 & 0.34 & 9.41 & 10.00 & 0.02 & 0.02 & 11.41 & 1.62 & 26.28 & P \\ 
4C+17.44 & 8.39 & 0.00 & 9.10 & 82.51 & 0.00 & 2.80 & 97.20 & 0.18 & 9.84 & 9.93 & 0.02 & 0.02 & 11.53 & 0.00 & 27.44 & I/M \\ 
WISEAJ082926.57 & 12.68 & 3.54 & 0.00 & 83.78 & 0.00 & 0.00 & 100.00 & 0.55 & 9.94 & 10.08 & 0.01 & 0.01 & 10.76 & 0.06 & 26.47 & P \\ 
WISEAJ083121.42 & 11.07 & 0.06 & 70.14 & 18.73 & 0.00 & 9.56 & 90.44 & 0.49 & 8.83 & 9.98 & 0.02 & 0.03 & 11.27 & 0.04 & 27.04 & P \\ 
PKS0829+18 & 5.28 & 6.86 & 0.78 & 87.08 & 0.01 & 0.01 & 99.98 & 0.11 & 9.80 & 10.06 & 0.02 & 0.02 & 11.55 & 0.60 & 26.59 & P \\ 
WISEAJ085010.43 & 0.00 & 0.00 & 75.30 & 24.70 & 0.00 & 15.16 & 84.84 & 0.21 & 8.88 & 9.80 & 0.02 & 0.02 & 11.22 & 0.00 & 26.12 & P \\ 
WISEAJ085323.42 & 18.13 & 1.79 & 71.70 & 8.38 & 0.08 & 45.00 & 54.92 & 0.52 & 8.82 & 9.60 & 0.01 & 0.01 & 10.92 & 1.47 & 25.51 & P \\ 
PKS0937+033 & 0.00 & 21.13 & 44.40 & 34.47 & 0.07 & 0.95 & 98.98 & 0.92 & 8.57 & 10.06 & 0.02 & 0.03 & 11.41 & 4.06 & 27.24 & I/M \\ 
WISEAJ115654.48 & 0.00 & 7.10 & 0.00 & 92.90 & 0.01 & 0.00 & 99.99 & 0.41 & 9.84 & 10.06 & 0.02 & 0.02 & 11.73 & 1.11 & 26.59 & I/M \\ 
WISEAJ122933.48 & 24.89 & 0.85 & 15.58 & 58.69 & 0.00 & 1.07 & 98.93 & 0.45 & 9.58 & 9.93 & 0.02 & 0.02 & 11.42 & 0.07 & 26.99 & P \\ 
NGC5506 & 61.98 & 17.91 & 7.79 & 12.32 & 0.22 & 1.38 & 98.40 & 1.75 & 8.07 & 10.07 & 0.01 & 0.02 & 10.54 & 1.51 & 23.45 & S \\ 
$[$HB89$]$1413+135 & 0.00 & 21.16 & 25.54 & 53.30 & 0.05 & 1.63 & 98.32 & 0.92 & 8.95 & 10.05 & 0.02 & 0.02 & 11.27 & 1.82 & 27.09 & S \\ 
3C315 & 0.00 & 0.00 & 5.59 & 94.41 & 0.00 & 1.66 & 98.34 & 0.09 & 9.85 & 9.91 & 0.02 & 0.02 & 11.26 & 0.00 & 25.90 & I/M \\ 
3C316 & 0.00 & 61.67 & 0.00 & 38.33 & 0.41 & 0.00 & 99.59 & 0.62 & 7.53 & 9.95 & 0.01 & 0.01 & 11.01 & 8.51 & 27.74 & P \\ 
\hline
\end{tabular}
\end{center}
\end{table*}

\section{Results and Discussion}

\begin{figure}
\includegraphics[width=\columnwidth]{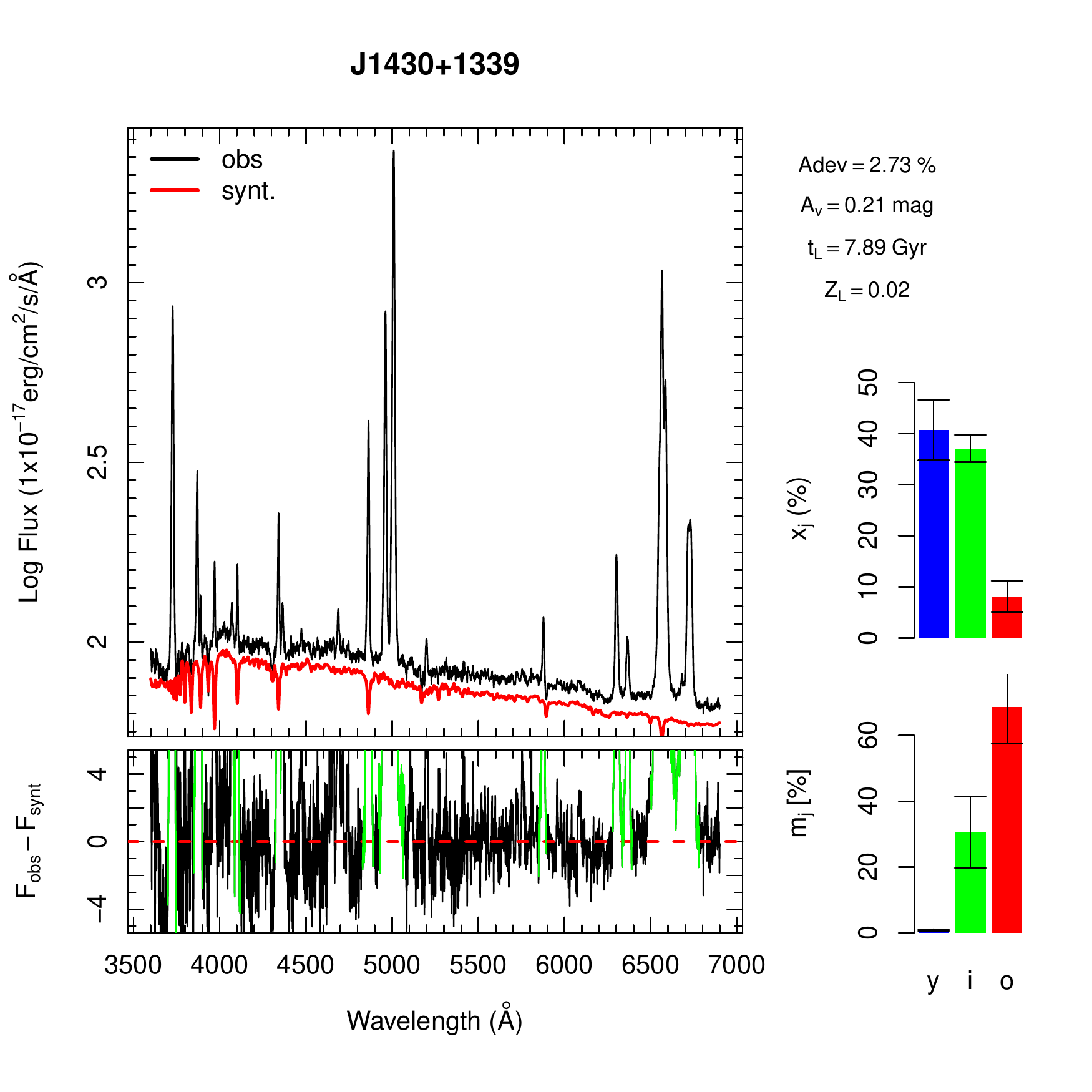}
\caption{Example of the stellar continuum fit made with {\sc starlight} for the galaxy J1430+1339.The upper panel shows the observed spectrum (black) and the best fit stellar population model (red) shifted down for clarity. The bottom panel shows the galaxy spectrum after subtraction of the stellar contribution. Emission lines masked during the fit are shown in green. The horizontal dashed-line (red) shows the residual line equal to zero. The percentage distribution of young (y), intermediate (i) and old (o) stellar populations weighted by luminosity and mass in each galaxy is plotted in the histogram at right panel of each plot.}
\label{fig:fit1}
\end{figure}

\begin{figure}
\includegraphics[width=\columnwidth]{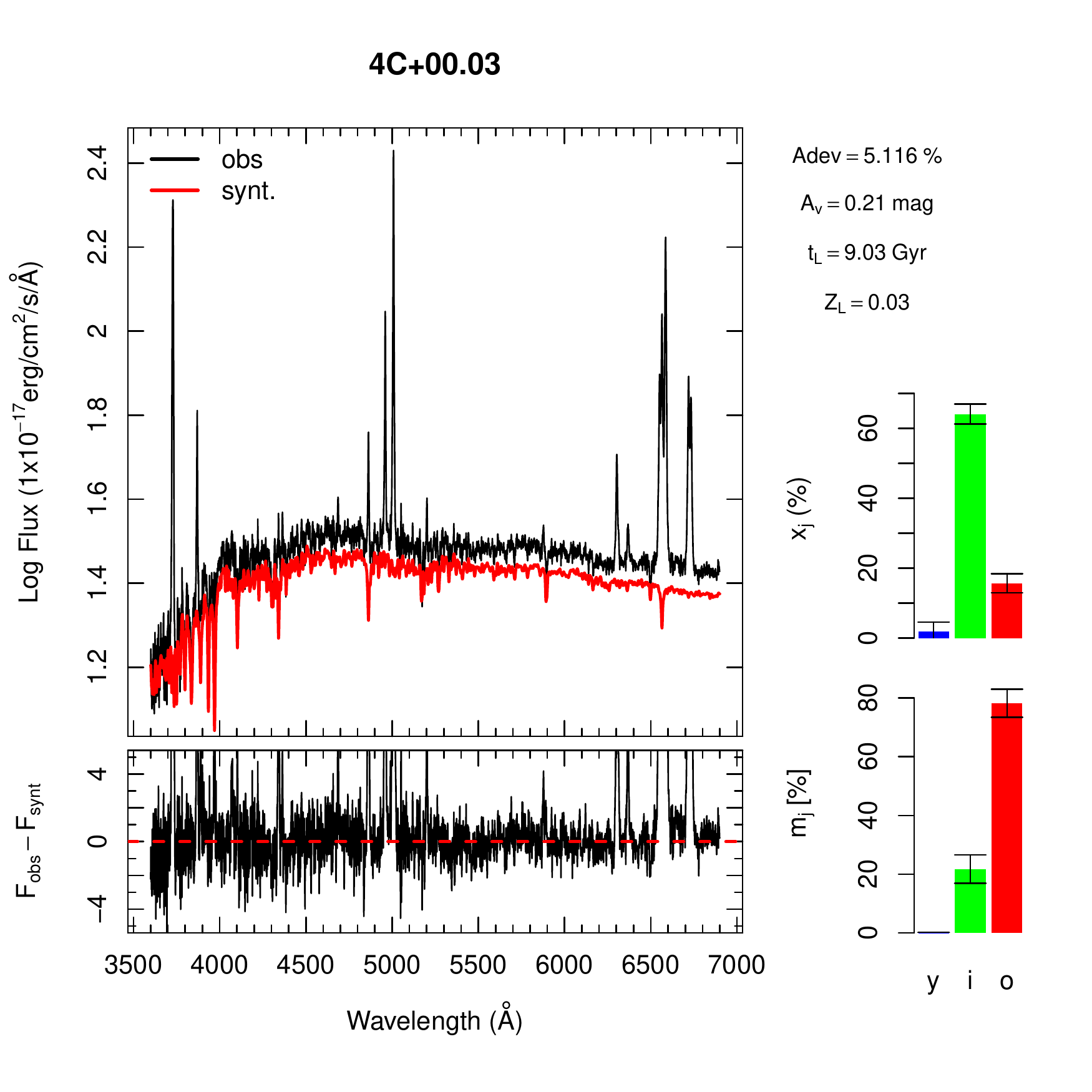}
\includegraphics[width=\columnwidth]{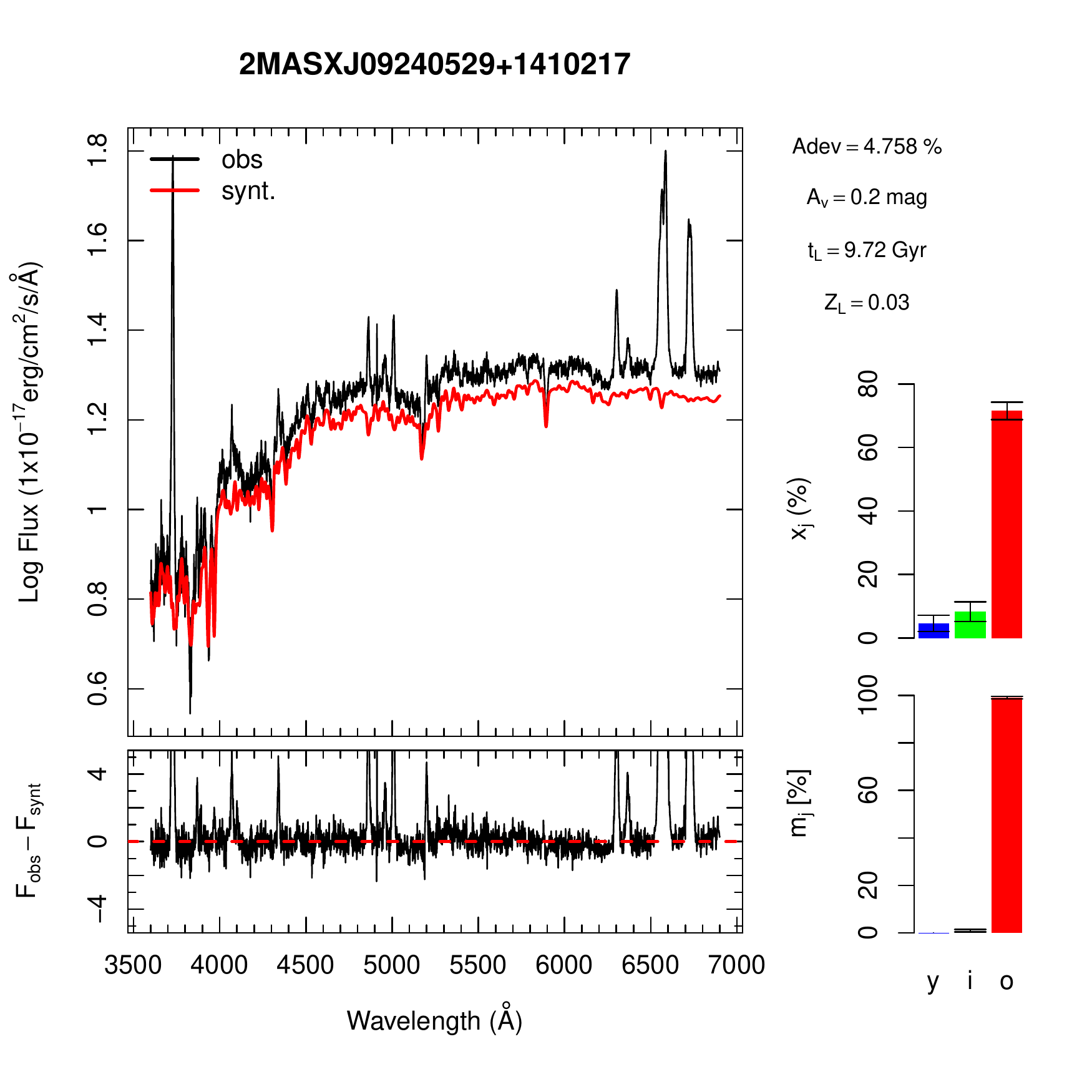}
\caption{The same as figure~\ref{fig:fit1} but for the galaxy 4C+00.03 (upper panel) and 2MASXJ09240529+1410217 (bottom panel).}
\label{fig:fit2}
\end{figure}

The mean deviation from the best fit of the model to the data, indicated by the ADEV parameter, suggests that, in general, the {\sc starlight} code showed satisfactory results (ADEV = 12\%) for the whole sample of Peaked-spectrum sources. The spectral fitting, as well as the percentage of young, intermediate and old stellar populations for three different types of galaxies representative of our sample, are shown in Fig.~\ref{fig:fit1} and \ref{fig:fit2}. The remaining results can be found in the online supplementary material. 

In Fig.~\ref{fig:fit1} we show the synthesis results for the source J1430+1339. The spectrum of this object displays a steep blue continuum and strong [\ion{O}{ii}]~$\lambda$3727 emission, both features frequently employed as indicators of recent star formation activity \citep{K98,R02,We07}. Prominent nebular emission lines of \ion{H}{i}, [\ion{O}{iii}]~$\lambda\lambda$4959,5007, [\ion{N}{ii}]~$\lambda\lambda$6548,6584, and [\ion{O}{i}]~$\lambda\lambda$6300,6363 are also evident in the spectrum. As we can see from the right-side panel of Fig.~\ref{fig:fit1}, young/intermediate-age stellar populations contribute with $\sim$75\% of the integrated flux of the source, while old stars contribute with less than 10\%. On the other hand, the contribution of the old stellar population is enhanced in the mass fraction histogram. This occurs because the older the stellar population is, less light it will radiate, which decreases the flux contribution and increases the mass contribution of this age in the stellar population. The source have mean age $\langle \rm{Log}\, t \rangle_L$ = 7.9\,Gyr, solar metallicity ($\langle Z \rangle$ = 0.02) and low internal reddening (A$_v$=0.21\,mag). Such results along with the high star formation rate (19.2M$_\odot/yr$) estimated in Section 4.2, suggest that J1430$+$1339 experienced recent episodes of star formation. According to \cite{J19}, all ten sources contained in their catalogue -- and included in our sample -- would be considered as star forming galaxies by the method of \citet[][]{BH12} but they claimed that only one object (J1338+1503, see the spectral fitting in the online supplementary material) follows the FIR-radio correlation established for star-forming galaxies (see Fig.~3 in their paper). Their sample, however, contains objects classified as type 2 obscured quasars with moderate radio luminosities (Log$[L_{\rm{1.4GHz}}/\rm{W\,Hz}^{-1}]=23.3-24.4$) and ionized outflows based on broad [\ion{O}{iii}] emission-line components \citep{Mu13,H14}. Recently, \citet[][]{J20}, who used observations of the  CO(2-1) and CO(6-5) transitions found that the targets represent a gas-rich phase of galaxy evolution with simultaneously high levels of star formation and nuclear activity. They also pointed out that the jets and outflows do not have an immediate impact on the global molecular gas reservoirs. All of these results are consistent with ours.

The top panel of Fig.~\ref{fig:fit2} illustrates the results for 4C+00.03. The optical spectrum of this object is  dominated by a stellar population of intermediate age. It has a red continuum and strong emission lines. We find the source being older and of higher metallicity than J1430+1339. The internal reddening is similar in both sources. It is noteworthy that 4C+00.03 displays spectral features typical of transitioning post-starburst or E+A galaxies. These objects have high-order, deep Balmer absorption lines and measurable nebular emission lines. Such features are indicative of either an ongoing starburst with the youngest stellar population obscured by dust or that the star formation has not been fully quenched in these galaxies. \citep{D99,P99,H06,W09}. From the morphological point of view, it is believed that post-starburst galaxies are dominated by bulges and, in some cases with an underlying disk component \citep{T04,Q04}. These features put these galaxies in the transition between the star-forming blue cloud and the quiescent red sequence galaxies \citep{Y14,R18}. We estimate that 21\% of our young AGN sample has a typical post-starburst spectrum, which is not surprising since several studies in the literature revealed a connection between the AGN activity and the post-starburst phase \citep[][]{K03,W07,W10,Y14}. 
Based on line ratios in the SDSS spectra, \citet[][]{K10} classified this source as high excitation galaxy (HEG), which means it has $\rm{log}[\ion{O}{II}]\lambda\lambda3727,3729/[\ion{O}{iii}]\lambda5007 > 0.2$ and $\rm{log}[\ion{O}{iii}]\lambda5007/H\beta \geq 0.75$. Another important aspect of this classification is that HEGs are powered by the accretion of cold gas that could be provided by a merger with a gas-rich galaxy and show a stronger jet contribution to the ionization of the ISM than, for example, low excitation galaxies. In addition, HEGs seems to have high Eddington ratios, which suggests a high mass accretion rate or radiative efficiency \citep[][]{Son}. Although these findings seem to disagree with those obtained from stellar population synthesis, where we detected a high fraction of intermediate-age stellar population, they reinforce how complex the interaction between the ISM and the newly launched jet of the AGN can be. It is widely believed that AGN outflows can remove large amounts of gas from their host galaxy, preventing it from forming stars. But, under specific physical conditions, the outflows can compress the dense gas cloud, potentially enhancing or triggering star formation. \citet[][]{Zub} showed from a set of idealized simulations that there is a critical AGN luminosity that results in a high fragmentation rate of gas clouds. The compression occurs while the AGN is active and the fragmentation takes place after the AGN switches off. The duration of the AGN episode in the simulations is $\sim$1\,Myr, much larger than the lifetime of the radio jet of the CSS/GPS ($\sim 10^3 - 10^4$ yr). This result suggests that the young- and intermediate-aged stellar populations fraction in our sample of peaked-spectrum sources may have been generated in response to gas compression that occurred as the jet advanced through the interstellar medium.

We present the results for 2MASXJ09240529+1410217 in the bottom panel of Fig.~\ref{fig:fit2}. This object is dominated by old stars, has a red continuum and a prominent 4000~\AA\ break. Notice the presence of absorption lines produced in the atmospheres of giant and cold stars such as Ca\,{\sc k} $\lambda$3933\AA,~Mg\,{\sc i} $\lambda$5175\AA~and NaD $\lambda$5892\AA. Although the spectrum of this source displays characteristics of early-type galaxies, there is a small fraction of young and intermediate age stars (less than $\sim$ 10\%) indicating that there is a residual star formation in the galaxy. A similar result was found by \citet{Diniz} for an early-type galaxy hosting an AGN. Their findings indicated that while the old-aged (13\,Gyr) stellar population is confined to the central region, the intermediate-aged (900\,Myr) one surrounds the galaxy nucleus. Unfortunately, we do not have enough spectral and spatial resolution to state whether this is the case of the 2MASXJ09240529+1410217 source. The origin of residual star formation in early-type galaxies is still under debate. Two mechanisms have been suggested as driver of this phenomenon: Galaxy merger \citep{K09,Y05} and gas cooling in elliptical galaxies \citep{MB03,VB15}. As we will discuss in the next section, the visual inspection of the optical images revealed an elliptical structure for this source with no evidence of galaxy merger (see Table \ref{tab:starlight} for more details). 

In summary, our results indicate that most CSS/GPS sources in our sample ($\sim$ 60\%) are dominated by an old-age ($t > 2 \times 10^9$\,yr) stellar population. About 23\% of the objects are composed of intermediate age stellar populations ($50 \times 10^6 < t \leq 2\times 10^9$\,Gyr). The contribution of young stellar population is small, with 5\% of the sources being dominated by young stars ($t \leq 50\times 10^6$\,yr). Although the fraction of young stellar populations in our sample is slightly larger than those found by \citet{R16} for interacting systems, overall both results seem to be in agreement. Our findings also confirm recent reports by \citet[][]{K20}, who found in a sample of 29 GPS by means of MIR colours, a wide range of pronounced  star formation activity ($\geq 0.5$\,M$\odot$ yr$^{-1}$) within their hosts. It implies that the radio-jet activity seems to be independent of the stellar age of the populations. Note that \citet{johnston08+} and \citet{rees16+} had already found no significant differences in the stellar populations of galaxies with and without radio AGN (once the samples are carefully matched in stellar mass).

\subsection{Reddening and Metallicity}

We divided our compact radio sources sample into four categories: $S_y$, $S_i$, $S_o$ and $S_f$. The first three sub-samples contain objects of young-, intermediate- and old-aged SPs. The $S_f$ sub-sample, however, contains objects dominated by a featureless continuum ($F_\nu\propto \nu^{-1.5}$). These objects differs from the ones we removed in Sect. 2 as they have a significant contribution of the stellar populations. It is worthwhile to mention that the FC is commonly used as a signature of the active galactic nuclei, but as discussed in \citet[][]{R09} in optical wavelengths this component cannot be easily distinguished from a reddened young starburst ($t \leq$ 5\,Myr). That is why we choose to keep them in subsequent analyses.

\begin{figure*}
	\includegraphics[width=0.7\textwidth]{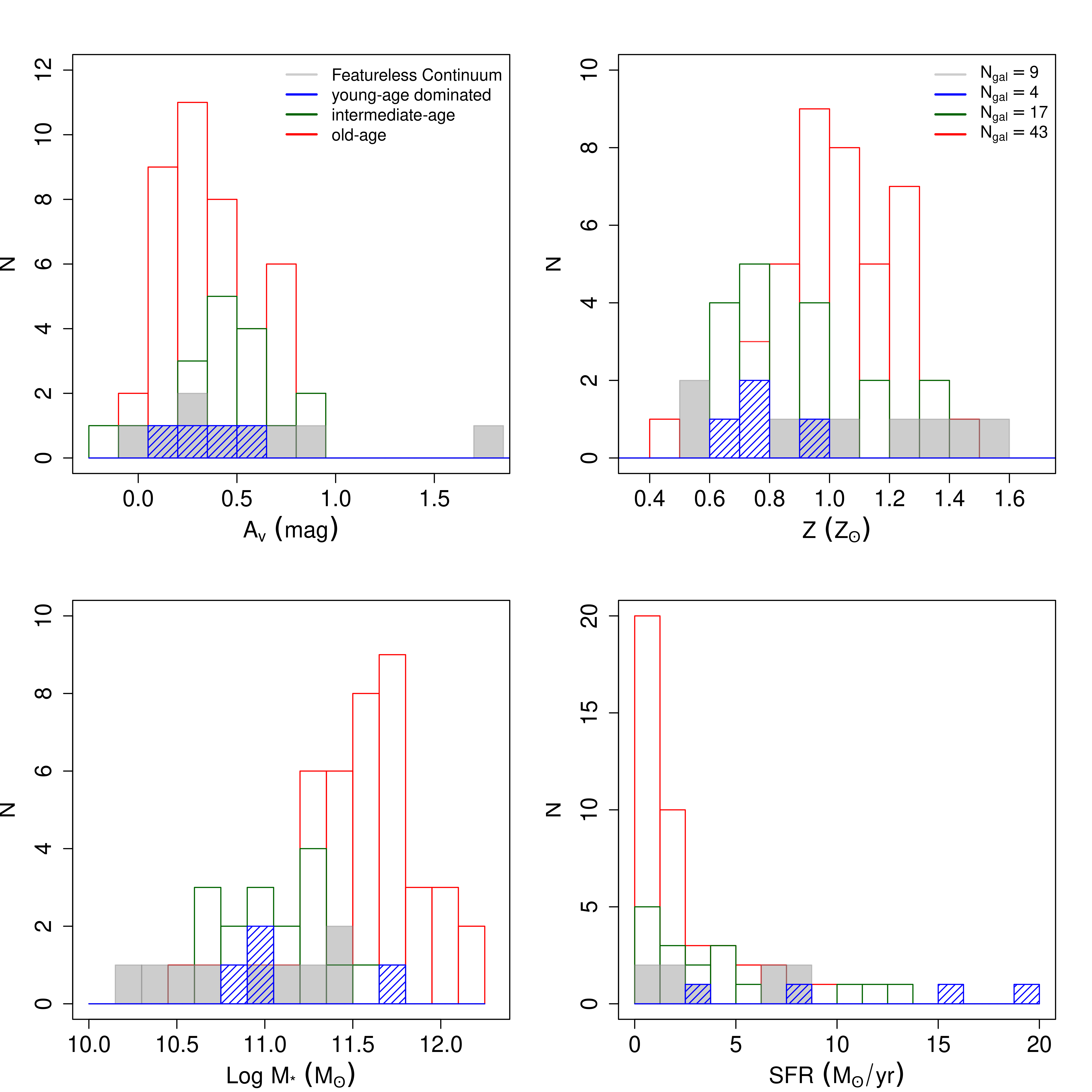}
	\caption{In the top row are the distribution of reddening (left) and light-weighted average metallicity (right) while in bottom panel are the stellar mass (M$_\star$, left) and the Star formation rate (SFR, right). In all four panels the results are  distributed between objects dominated by young- (blue), intermediate- (green), old-age (red) stellar population and by a featureless continuum.}
	\label{fig:hist}
\end{figure*}

We illustrate in the top left panel of Fig \ref{fig:hist} the average distribution of dust extinction (A$_v$) for each sub-sample. As we can see, $S_y$ cover a smaller range of $A_v$ while the remaining sample has a widespread dust extinction distribution. $S_y$  has also less average dust reddening ($A_v$ = 0.34\,mag) in comparison to $S_i$ ($A_{v}$ = 0.44\,mag), $S_o$ ($A_{v}$ = 0.37\,mag) and $S_f$ ($A_{v}$ = 0.54\,mag), which is somewhat surprising since we would expect star-forming galaxies to have more dust than quiescent objects. It is worth mentioning that the young SP dominated sub-sample is composed of four objects. Thus, we cannot confirm that it is a universal result.

In the top right panel of figure~\ref{fig:hist} we show the light-weighted metallicity (Z$_L$) of the SP for our sub-samples. As expected, $S_y$ followed by $S_i$ has the lowest metallicity ($Z_L$ = 0.78 Z$_\odot$) values. This is consistent with a galaxy chemical enrichment scenario where a young population is enriched by the evolution of the early massive stars. 

\subsection{Star formation rate and stellar mass}

One of the data products that can be derived from the stellar population fitting process  is the star formation rate ($SFR_\star$) over an user-defined age interval ($\Delta t = t_{j_f}-t_{j_i}$). This can be derived since the SSPs model spectra used in the fits are in units of L$_{\odot}$ \AA$^{-1} M_{\odot}^{-1}$, and the observed spectra ($O_{\lambda}$) are in units of $\rm erg/s/cm^2/$\AA. The SFR$_\star$ over the chosen $\Delta t$ interval can then be computed assuming that the mass of each base component ($j$) that has been processed into stars can be obtained as:
\begin{equation}
    M_{\star,j}^{\rm ini} = \mu_{j}^{\rm ini} \times \frac{4\pi d^2}{3.826\times 10^{33}},
\end{equation}
where $M_{\star,j}^{\rm ini}$ is given in $M_\odot$, $\mu_{j}^{\rm ini}$ represents the mass that has been converted into stars for the $j$-th element and its flux. This parameter is given in $M_\odot\,{\rm erg s^{-1} cm^{-2}}$; $d$ is the distance to the galaxy in cm and 3.826$\times10^{33}$ is L$\odot$ in erg\,s$^{-1}$. Thus, the SFR over the $\Delta t$ as defined above
can be obtained from the equation \citep{Riffel+20,Riffel+21}:
\begin{equation}
    {\rm SFR_\star} = \frac{\sum_{j_i}^{j_f} M_{\star,j}^{\rm ini}}{\Delta t}.
\end{equation}
In the present work we have computed the $SFR_\star$ over the last 50\,Myr. In addition to the above, {\sc starlight} also computes the current stellar mass of the galaxy inside the studied aperture (M$_\star$). 
\par
To be consistent in comparing galaxies in our sample, we performed an aperture correction on SFR and M$_\star$. For this purpose, fiber and model magnitudes were obtained from the SDSS database. The fiber magnitude reflects the flux contained in the SDSS spectrograph aperture (3'' in diameter), while the model magnitude represents the integrated flux of the galaxy. The corrected M$_\star$ and SFR are thus given by:
\begin{equation}
\textrm{M}_{\star} = \frac{\textrm{M}_{\star,fit}}{f_{corr}} \quad  \textrm{and} \quad SFR = \frac{SFR_{fit}}{f_{corr}}
\end{equation}
where M$_{\star,fit}$ and $SFR_{fit}$ are the values returned by {\sc starlight} code; The correction factor is defined as $f_{corr} = 10^{[-0.4*(m_1 - m_2)]}$ with $m_1$ and $m_2$ representing, respectively, the fiber and model magnitudes. 
\par
Since the SP of the bulge is older than the SP of the disc, the SFR/flux ratio varies with the radius. However, the SDSS apperture is fixed, and thus for local galaxies the bulge plays a major role in the spectra if compared to high-$z$ objects. This effect has been estimated by \citet{Kewley+05}, who analyzed nuclear and integrated spectra for 101 galaxies from the Nearby Field Galaxy Survey. The authors reported that $z>0.05$ is sufficient to avoid a strong scatter in SFR, but ideally an object should have at least $z=0.1$ for an optimal correction. In our final sample, only one CSS/GPS and one MPS have $z<0.05$, and only 10 of our targets have $z\leq0.1$, implying the SFR correction is optimal for 86\% of our sample. For the $M_{\star}$, our correction should be more accurate, since the red light is dominated by older stars, which dominate the mass of the galaxy. The corrected SFR and M$_{\star}$ are listed in Columns 14 and 15 of tables \ref{tab:starlight} and \ref{tab:starlight_ps}.
\par
Another source of uncertainty in our derived values is due to statistical uncertainties in the code and the spectra. \citet{Burtscher+21} employed a special version of {\sc starlight}, that saves the full Markov Chain of the Monte Carlo sampling. From these, they were able to compute the actual statistical uncertainties involved in the spectral fitting process. Their stellar populations were divided in four bins instead of three, and derived average uncertainties of 0.4\%, 0.3\%,  2.4\% and 2.2\% for the young, young-intermediate, intermediate-old and old populations, respectively. This puts the SFR uncertainty close to 4\%, since the average contribution from young SSPs is 10\% in our sample.
\par
Lastly, our estimates are based on more external SPs, since optical spectra is very sensible to dust reddening. The region close to the SMBH is embedded in dust \citep[e.g. higher than 20 magnitudes for all Seyfert\,2 galaxies in][]{Burtscher+16}, and thus inaccessible to optical spectroscopy.
\par
The stellar mass of the sample varies between Log$\,\rm{M}_\star = 10.22\,\rm{M}_\odot$ and Log$\,\rm{M}_\star = 12.13\,\rm{M}_\odot$. We show in the bottom left panel of figure~\ref{fig:hist} the stellar-mass distribution for our sub-samples. Clearly, the objects dominated by old-age stars have high stellar masses ($\rm{Log}\,M_\star$ = 11.52M$_\odot$, on average) while the remaining sub-samples dominate the low-mass regime. The featureless continuum dominated systems, however, possess the lowest stellar mass values.
\par
The SFR varies between 0 and 19.20$~\rm{M}_\odot$~yr$^{-1}$ with most sources with star formation rates lower than $< 5 \rm{M}_\odot$~yr$^{-1}$, as shown in the bottom-right panel of Fig.~\ref{fig:hist}. As previously mentioned, the objects dominated by old-age stars have high stellar masses and low SFR ($\sim2.16\,\rm{M}_\odot$~yr$^{-1}$), which is consistent with the findings of \citet[][]{K20} and \citet[][]{W10} using MIR data. Interestingly, three of these sources (2MASXJ14005162, J1010+1413 and 2MASXJ1530) have mo\-de-ra\-te star formation rate (3.84, 6.33 and 8.15 M$_\odot$~yr$^{-1}$, respectively). This occurs because such objects have a significant fraction of young and intermediate-age stars. As we will discuss in the following section, they are classified, respectively, as elliptical, interacting systems and spiral galaxies. In general, our results indicate that $\sim$ 24\% of the compact radio sources show moderate levels of star formation ($\rm{SFR} > 5\,\rm{M}_\odot/\rm{yr}$), which seems to indicate that compact symmetric objects avoid environments with extreme SFR. 

Some authors have found different results based on Infrared (IR) data. For example, \citet[][]{D12} using detection of polycyclic aromatic hydrocarbon (PAH) features found evidence for recent star formation activity at optical and/or Mid-to-Far IR wavelengths for 75\% of the CSS/GPS sources. \citet[][]{OdeaSaikia} detected 7.7$\mu$m PAH emission in 4 compact sources and estimated an upper limit to the star formation rates of 10 - 60\,M$_\odot$~yr$^{-1}$. They also used SFR based on FIR luminosity and found a high fraction of sources with substantial star formation rate (SFR > 10\,M\,$_\odot$/yr). Their sample includes very compact radio sources (lobe separation $\sim$ 10 pc) called High-Frequency Peaked sources and objects at z > 1, which could explain the high values of star formation rate found by them. Moreover, this SFR may be associated with the presence of cold gas embedded in a very obscure dust circumnuclear region, which is undetectable in the optical wavelengths.

\section{Host Galaxies Morphologies}

Because the stellar population of a galaxy is very dependent on the morphology \citep[e.g.][]{CD19}, we obtained SDSS images and performed a visual classification of these objects. This classification was made independently by three distinct classifiers. When at least two of them agreed, the morphological class was established. For objects where the three classifiers disagreed, each classifier explained his choice and tried to convince one of the other two into changing their classification. We split our objects into four categories: (1) elliptical, (2) spiral, (3) irregular/merger and (4) point source. Readers interested in taking a closer look at the SDSS images used to perform the visual classification of the sample are referred to Figures 2 and 3 of the supplementary material. Out of the 58 CSS/GPS in our sample, we classified 15 as ellipticals, 18 as spirals, 12 as irregular/mergers and 13 as point sources. We found that the MPS sample is composed of 2 spirals, 4 irregular/mergers and 8 point-like sources. We do not identify elliptical galaxies in this sample. To validate our morphological classification, we used data from Galaxy Zoo 2 \footnote{Data publicly available at http://data.galaxyzoo.org} \citep{Wi13}, which provides detailed morphological classifications of nearly 250,000 galaxies drawn from 7th SDSS data release. Of the 30 objects with Galaxy Zoo counterpart, 15 are elliptical and 6 are spiral. The remaining 9 objects are classified as merger or irregular.  It is worth mentioning that such morphological classification was defined taking into account 43\% of the votes of Galaxy Zoo users. So, although the results agreed in 75\% of the cases, we chose to keep our visual classification in order to be complete and homogeneous.

\begin{table*}
\centering
\caption{Median properties of the galaxies classified as spiral, elliptical, irregular/merger and point-like source.}
\label{tab:morp}
\begin{center}
\begin{tabular}{ccccccc}
  \hline
 Morph & z & $\langle Z\rangle_L$ & $\langle t\rangle_L$ & Log$L_R$ & Log M$_\star$ & SFR \\
       &   &  & (yr)   &  (W/Hz) &   ($M_\odot$)&($M_\odot$/yr)\\
  \hline
  S   & 0.11$\pm$0.06 & 0.02$\pm$0.01 & 9.14$\pm$0.49 & 24.40$\pm$1.01 & 11.34$\pm$0.31 & 2.17$\pm$2.25 \\ 
  E   & 0.14$\pm$0.06 & 0.02$\pm$0.01 & 9.60$\pm$0.17 & 24.68$\pm$0.59 & 11.69$\pm$0.18 & 2.16$\pm$1.57 \\ 
  I/M & 0.17$\pm$0.07 & 0.02$\pm$0.01 & 8.93$\pm$1.00 & 25.20$\pm$1.41 & 11.47$\pm$0.33 & 3.94$\pm$4.76 \\ 
  P   & 0.41$\pm$0.15 & 0.02$\pm$0.01 & 9.30$\pm$0.62 & 26.60$\pm$0.65 & 11.11$\pm$0.49 & 0.60$\pm$0.80 \\ 
   \hline
\end{tabular}
\end{center}
\end{table*}

We highlight in Table \ref{tab:morp} the main properties of the objects classified as elliptical, spiral, irregular/merger and point-like sources, according to our classification criteria. The uncertainties associated to each value correspond to the median absolute deviation calculated using the {\sc stats} library in R software package. Despite the small difference observed in the median redshift of E and S, the z distribution, as a whole, is similar according to the KS test. However, as expected, the point-like sources have higher-z and radio luminosities compared to other galaxy distributions. In terms of stellar mass and metallicities the values are comparable. The I/M morphological class contain, in general, the youngest ($\langle t \rangle_L$ = 8.93 yr) objects and those with higher SFR ($>$3.94$M_\odot$/yr). This results is particularly important since it is well established that merger and interaction between galaxies can deposit gas in the circumnuclear region triggering star formation and possibly AGN activity \citep{H09,W08}. 

Our results indicate that there is no preferred morphological type of host galaxies for Peaked-spectrum sources, confirming previous findings. For example, \citet[][]{DV98,DV00,S1,S2} and \citet[][]{L07} pointed that CSS/GPS host galaxies tend to be large, bright elliptical galaxies dominated by old stellar population, while others studies indicate a significant disk component host galaxy \citep[][]{H82,J10,A13,M11}. Also, there are few examples of host galaxies showing evidence of merger or tidal interactions with close companions \citep{S07,Jo10,E16}.

\begin{figure*}
	\includegraphics[width=0.98\textwidth]{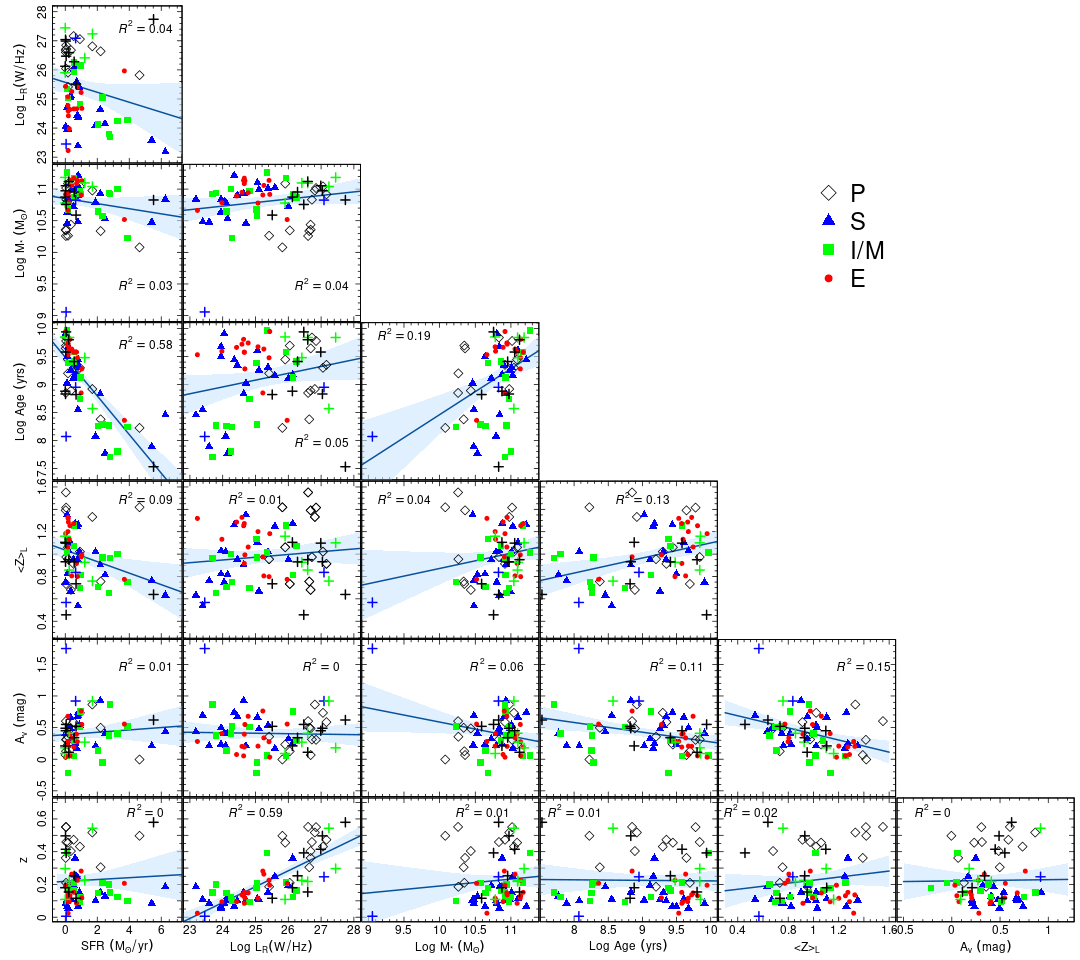}
	\caption{Comparison between redshift, radio luminosity and stellar properties. Elliptical galaxies are denoted red, spirals are blue, irregular/merger purple and point sources green. The black cross symbol depicts the MPS sources. For each graph, we show in blue the linear regression model, assuming the x axis variable as the independent one. The shaded area represents the 95\% confidence interval for the whole sample.}
	\label{fig:2dhist}
\end{figure*}

In order to compare the stellar population properties, and how they correlate with redshift, radio luminosity and morphology, we present in Figure~\ref{fig:2dhist} the properties derived with {\sc starlight}, as well as redshift, radio luminosity and morphology. The values for individual galaxies are presented in Table~\ref{tab:starlight}. As we can see in Fig.~\ref{fig:2dhist}, there are no significant differences between MPS and CSS/GPS sources properties, which reinforce how optically similar they are. Moreover, there are small trends among the properties. The radio luminosity increases weakly with z. It reflects the incompleteness of our sample and the fact that all sources here were first detected in radio. As the distance to the source increases, only the brightest sources are detected. Also, the SFR showed a remarkable correlation with the mean age of the system for all galaxy populations. As expected, the younger the system, the higher the star formation rate. Interestingly, we see a large number of spiral and irregular/merger galaxies with SFR typical of elliptical galaxies. This may be related to the fact that the SDSS uses a fixed aperture (3'') to obtain the galaxy spectra, which would imply we are sampling different regions of a galaxy along the redshift distribution. Another possibility is that these galaxies are going through a strong phase of AGN activity, which prevents the galaxy to form stars. In the next section, we will discuss the role of the AGN and its implications in the host galaxy.

\section{Mid-Infrared Photometry}

\begin{figure}
	\includegraphics[width=\columnwidth]{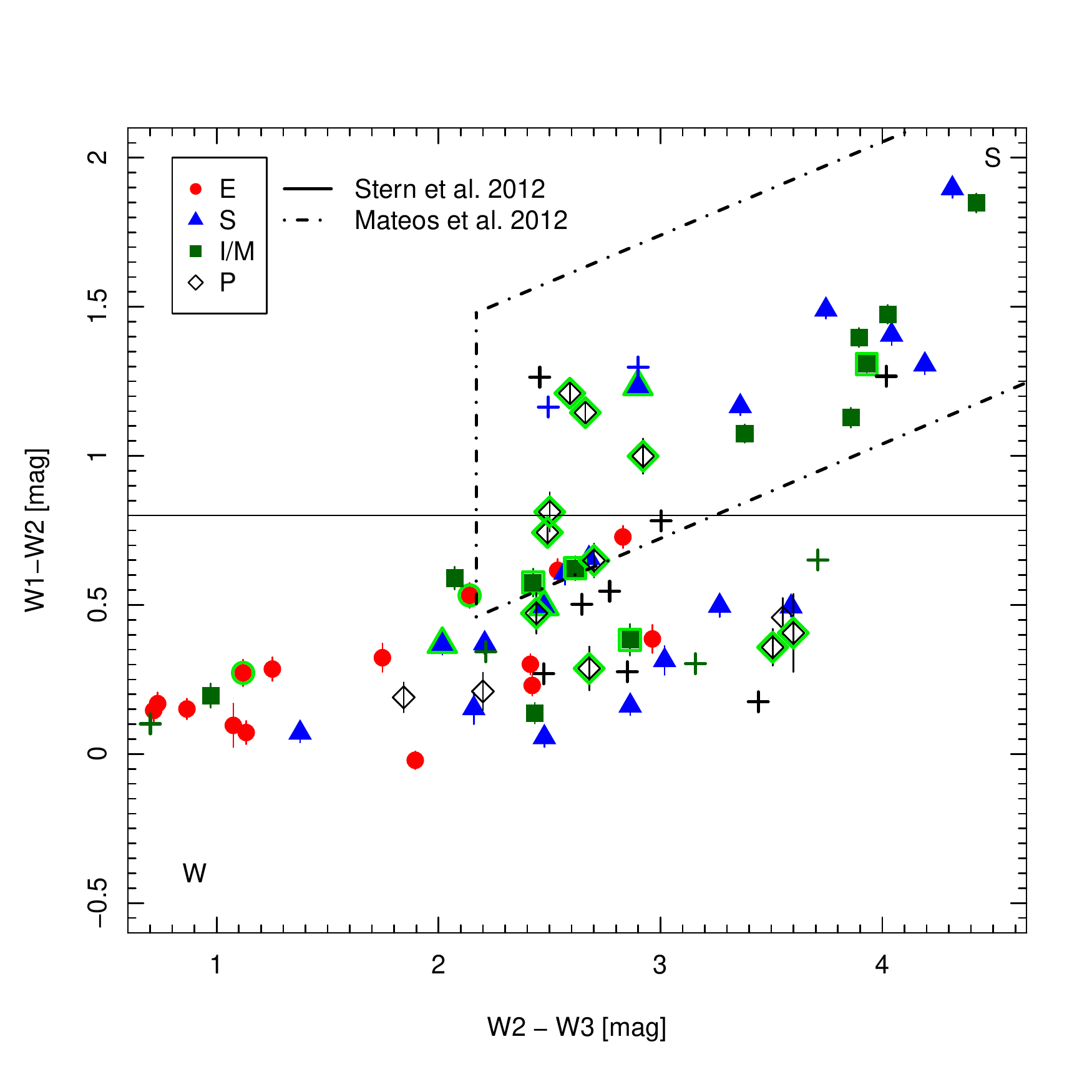}
	\caption{Wise colour diagram for our sample of compact radio sources. Red circles and blue triangles represent elliptical and spiral galaxies, respectively. Green squares denote objects classified as irregular or mergers while open diamonds represent point sources. The cross symbols corresponds to MPS sources. Green contours refer to X-ray emitting objects. For comparison, we illustrate the \citet[][]{S12} (horizontal solid line) and \citet[][]{M12} (dotted, dashed lines) cut. The MIR emission of sources dominated by AGN component is located above the \citet[][]{S12} cut and within wedge defined by \citet[][]{M12}.}
	\label{fig:MIR}
\end{figure}

\begin{figure}
	\includegraphics[width=\columnwidth]{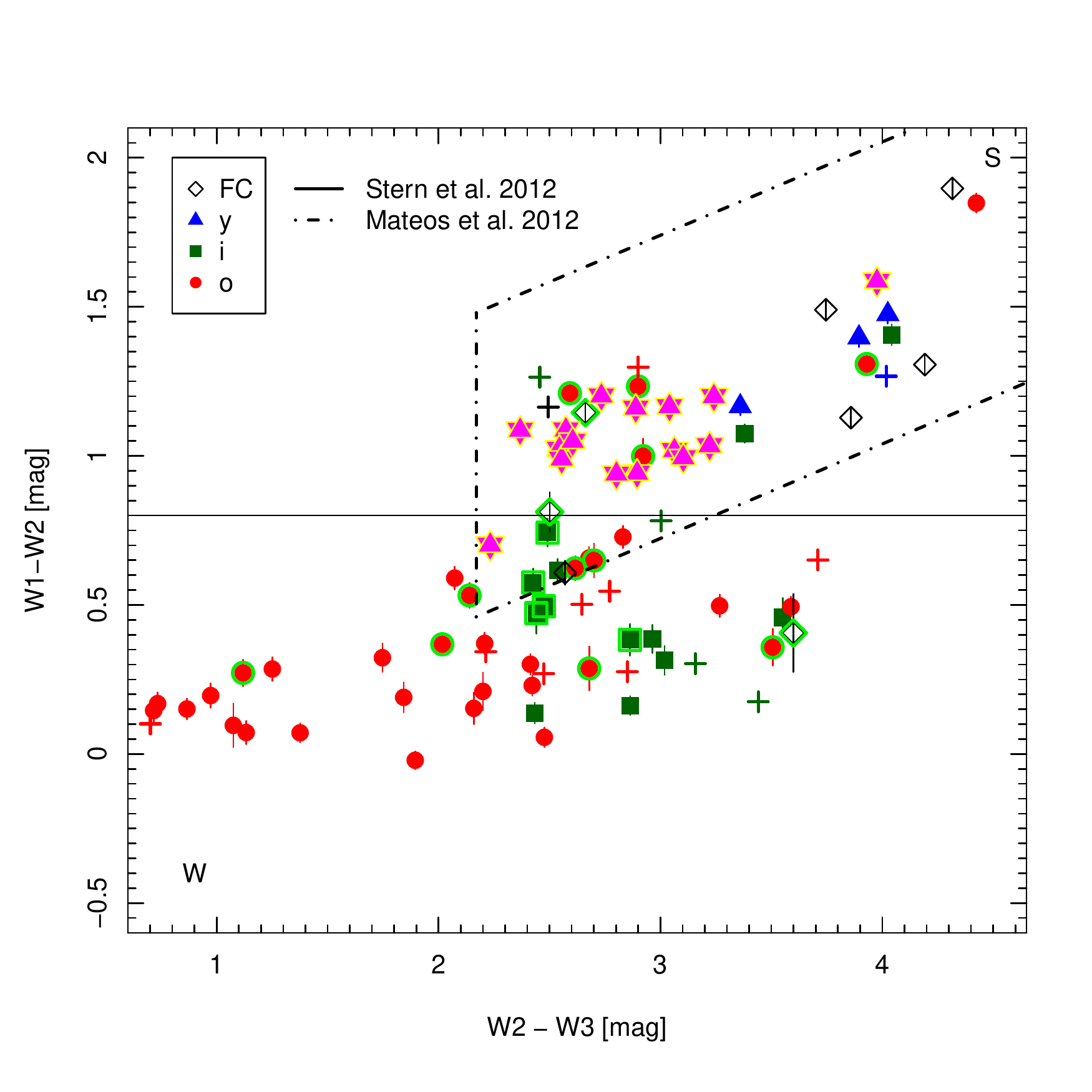}
	\caption{The same as figure~\ref{fig:MIR} but with diamond, triangle, circle and square symbols denoting featureless continuum dominated objects and those dominated by young-, intermediate- and old-age stellar populations. The cross symbols corresponds to MPS sources. The magenta stars represent classical AGNs, i.e., young radio sources that were not included in the population synthesis because the stellar contribution is completely diluted in the AGN continuum.}
	\label{fig:MIR2}
\end{figure}

We extended our analysis to the MIR regime to check the consistency of the results on the stellar content of the galaxies. In addition, we want to investigate the main excitation mechanism that drives the observed spectrum, i.e., AGN, starburst or composite. For this purpose, we obtained WISE data in three bands (W1, W2 and W3) centered at 3.4, 4.6 and 12$\mu$m, respectively. The W1-W2 and W2-W3 colours were derived for all sources and the values are listed in Table \ref{tab:radio} and \ref{tab_mps:radio}. Figures~\ref{fig:MIR} and~\ref{fig:MIR2} show the results. The symbols and colours in the former refer to the morphological class while in the latter to the dominant stellar population in the sample. For comparison, we depicted the megahertz-peaked spectrum sources with the cross symbol and its colours following the above description. We also included in Fig.~\ref{fig:MIR2} the objects set as as classical AGNs (magenta symbols). In both figures, the light green contour depicts the X-ray emitting objects. The horizontal line represents the classification proposed by \citet{S12}. It divides AGN-driven sources (W1$-$W2~$>$~0.8~mag) from those powered by star formation (W1$-$W2~$<$~0.8~ mag). In addition, we added the region delimited by the dotted-dashed lines, first depicted by \citet{M12}. According to these authors, the sources located within these wedges are considered classical, $bona-fide$ AGNs. 

We found that almost all sources hosted by elliptical galaxies and/or dominated by old stellar population fall into the region where the MIR emission is dominated by the stellar population. On the other hand, inside the wedge dominated by AGNs, there is a prevalence of CSS/GPS + MPS sources whose host galaxy is classified as spiral, merger/interacting, and point-like. These results are in agreement the ones obtained by \citet{K20} and reinforce the idea that MIR colours may successfully distinguish galaxies with different levels of star formation, but they are slightly related to morphological type. Moreover, according to \citet{M12}, the W2-W3 colour defines a sequence of increasing level of star formation from the left to the right axis of the plot. It is possible to see that most elliptical CSS/GPS display W2$-$W3 values $<$~2 while most spirals and I/M sources are located in the region of increasing levels of star-formation, consistent with the morphological classification done by our group. This result is also consistent with an evolution scenario where compact radio sources evolve into extended radio sources (FRI and FRII). In the WISE colour diagram, the evolved FRIs are clustered in the region occupied by elliptical galaxies with little star-formation activity (W1-W2 < 2) and FRII, the region of  high star-formation activity (W2-W3 > 2). 

Interestingly, we see a handful of spiral and irregular/merger CSS/GPS sources with a high fraction of old-age stellar population. There are two possible explanations for the discrepancies found between the stellar population synthesis results and morphological classification. The first one could be related to the bias introduced by the fixed aperture in the SDSS data, which samples different portions of the galaxy according to the redshift. It affects mostly the spirals and I/M host galaxies. At low $z-$values, the innermost kiloparsec - where old age stars dominate - is mapped. For the most distant ones, the light from the galaxy disc dominates. The second explanation may reflect uncertainties in the morphological classification of the sources. Although we have performed a rigorous visual inspection, we are not able to overcome this barrier because we are limited by the SDSS image resolution. The reader may keep in mind that both issues may affect our results to some extent. However, note that the overall, very good agreement between galaxy morphology and stellar content points out to the consistency of our approach.

In order to better explore our data, we divided the sample into two data sets based on the position in the WISE colour diagram. Objects whose colours meet the Mateos et al. and Stern et al. criteria are defined as strong AGN (S) and those that do not as weak AGN (W). Figure~\ref{fig:MIR} illustrates the results. While strong AGNs are preferentially found in gas-rich galaxies, the weak ones can be hosted by both quenched and gas-rich galaxies. Looking at Fig.~\ref{fig:MIR2} we can see a more clear trend for weak AGNs. The compact radio sources with colours W1-W2 $\leq 0.5$ and dominated by intermediate-age stellar populations are located in the region associated to starburst galaxies, i.e. W2-W3 $\geq 2.2$\,mag. On the other hand, the sources dominated by old-age stars are restricted to the region occupied by elliptical galaxies with little star formation activity. The region dominated by strong AGN, however, is filled by sources with all kinds of dominant stellar population. Note that if the FC component is associated to very young stars, we may interpret that strong AGNs are dominated by the young- and intermediate-age stellar populations, confirming the tendency observed in the figure~\ref{fig:MIR}. These results are important once they may help to establish a possible mechanism to drive material towards the galaxy center and trigger the AGN activity in the young radio sources. 

Another interesting point to be addressed is how different is the WISE colour distribution of our young radio sources and those known as classical AGNs. The spectra of the latter are dominated by a non-stellar continuum and strong emission lines, with some of them displaying broad components emitted by a broad line region. For comparison, we include in Fig.~\ref{fig:MIR2} fifteen compact radio sources (magenta star symbols) with spectra similar to those of {\it bona-fide} AGN. In the literature, these sources are classified as QSO and almost all of them are X-ray emitters. As expected, classical and strong AGNs are confined to the same region in the WISE colour diagram, where powerful AGNs are located. However, only strong AGNs have the most extreme colour distribution. At first, these results suggest that there is no big difference, from the standpoint of MIR colours, between compact radio sources whether they are classic AGNs with a power-law continuum or AGNs with a dominant stellar continuum. This result, along with the visual inspection of weak and strong AGNs spectra, leads us to state that our sample is mostly composed of compact radio sources with nuclear activity similar to LINERs/Seyfert galaxies. 

Several studies in the literature have investigated the connection between young AGNs and Narrow-line Seyfert 1 galaxies \citep[NLS1,][]{Wu09,Gu16,Berton16,Singh18}. In these objects, the broad component of the Balmer lines has full-width at half maximum values $<$ 2000~km\,s$^{-1}$ \citep[][]{OP85}, that is, narrower than in classical Seyfert 1 \citep[see][for a review]{K08}. Moreover, they have smaller SMBH masses and higher accretion rates than classical Seyfert~1s, which suggests they are young AGNs similar to GPS and CSS sources \citep{Kom06,Gu10,Cacci14,Gu15}. \citet{Wu09} found that most of the 65 young radio AGNs in their sample have SMBH mass and Eddington ratios similar to those of NLS1s. On the other hand, \citet{L20} found that only one of the 126 young radio sources could be defined as NLS1. They claimed that there is a mixed of accretion modes in their sample which implies the early stage of jet evolution are not necessarily associated with high accretion systems.

We also search for X-ray counterpart from our young radio sources in the NASA/IPAC Extragalactic Database (NED)\footnote{https://ned.ipac.caltech.edu/}.\, In total, we find 21 X-ray emitters (19 CSS/GPS and 2 MPS) sources in our sample. These objects are identified as green contours in Fig.~\ref{fig:MIR} and \ref{fig:MIR2}. As we can see, the majority of our X-ray emitting sources are located either very close to the S region bottom edge or much above it, i.e. are considered luminous AGNs. On the other hand, the ones located in the W region have WISE colours consistent with star-forming galaxies. Two main hypothesis are considered to explain the origin of the X-ray emission in compact radio sources. In the high excitation radio galaxies (HERGs) it can be produced by the accretion disk and corona \citep[][]{T09} while in low excitation radio galaxies (LERGs) the X-ray emission comes from the radio source via synchrotron or inverse Compton scattering \citep[][]{S08,O10}. We will further explore this topic in a future work.

 
\section{Final Remarks}

In this work we have explored the optical and MIR properties of the most compact radio galaxies in order to determine the dominant stellar population and the impact of the radio jet on the circumnuclear environment of these objects. Our sample consists of 58 CSS/GPS and 14 MPS sources located at low redshift ($z \leq 1$) and with a SDSS-DR12 counterpart. 

From the stellar population synthesis, performed with {\sc starlight} code, we find that our sample is dominated by old- and intermediate-age stars, with just a marginal contribution of young stellar populations. Regarding the stellar masses and star formation rates our results indicate that old-age dominated objects have high stellar masses and low SFR ($\sim$2.16\,M$_\odot$/yr) which is consistent with findings in the literature. In addition, we noted that $\sim$24 percent of the compact radio sources show moderate levels of SFR indicating that these objects avoid extreme environments. 

The inspection of the source spectra along with synthesis results indicates that our sample is composed of a mix of star-forming, post-starburst, and quiescent galaxies. These results are also consistent with those obtained through the morphological analysis of our sample. Of the 72 compact radio sources, 15 are elliptical, 20 spirals, 16 irregular or merger, and 21 point-like sources. Also, we found that CSS/GPS have similar optical properties MPS sources. 

The MIR colours analysis of these sources revealed that elliptical galaxies and some spirals lie in the region of the WISE colour diagram where the sources are powered by stellar activity, while the powerful AGNs region is, in general, composed of spiral and interacting systems. Such a result is particularly interesting as it may shed light on the possible mechanism responsible for triggering AGN activity in young radio sources.

\section*{Acknowledgements}
A.R.A acknowledges partial support from Conselho Nacional de Desenvolvimento Cient\'{\i}fico e Tecnol\'ogico (CNPq) Fellowship (311935/2015-0). RR thanks to Conselho Nacional de Desenvolvimento Cient\'{i}fico e Tecnol\'ogico  ( CNPq, Proj. 311223/2020-6,  304927/2017-1 and  400352/2016-8), Funda\c{c}\~ao de amparo 'a pesquisa do Rio Grande do Sul (FAPERGS, Proj. 16/2551-0000251-7 and 19/1750-2), Coordena\c{c}\~ao de Aperfei\c{c}oamento de Pessoal de N\'{i}vel Superior (CAPES, Proj. 0001). RSN thanks the financial support from CNPq, grant 301132/2020-8. 

Funding for SDSS-III has been provided by the Alfred P. Sloan Foundation, the Participating Institutions, the National Science Foundation, and the U.S. Department of Energy Office of Science. The SDSS-III web site is http://www.sdss3.org/. SDSS-III is managed by the Astrophysical Research Consortium for the Participating Institutions of the SDSS-III Collaboration including the University of Arizona, the Brazilian Participation Group, Brookhaven National Laboratory, Carnegie Mellon University, University of Florida, the French Participation Group, the German Participation Group, Harvard University, the Instituto de Astrofisica de Canarias, the Michigan State/Notre Dame/JINA Participation Group, Johns Hopkins University, Lawrence Berkeley National Laboratory, Max Planck Institute for Astrophysics, Max Planck Institute for Extraterrestrial Physics, New Mexico State University, New York University, Ohio State University, Pennsylvania State University, University of Portsmouth, Princeton University, the Spanish Participation Group, University of Tokyo, University of Utah, Vanderbilt University, University of Virginia, University of Washington, and Yale University. 

\section*{data availability}

The data underlying this article are available in NASA/IPAC Extragalactic Database, at https://ned.ipac.caltech.edu/. The datasets were derived from sources in the public domain of  SDSS at https://www.sdss.org.

\bibliographystyle{mnras}

\onecolumn

\begin{figure*}
    \centering
    \includegraphics[width=\textwidth]{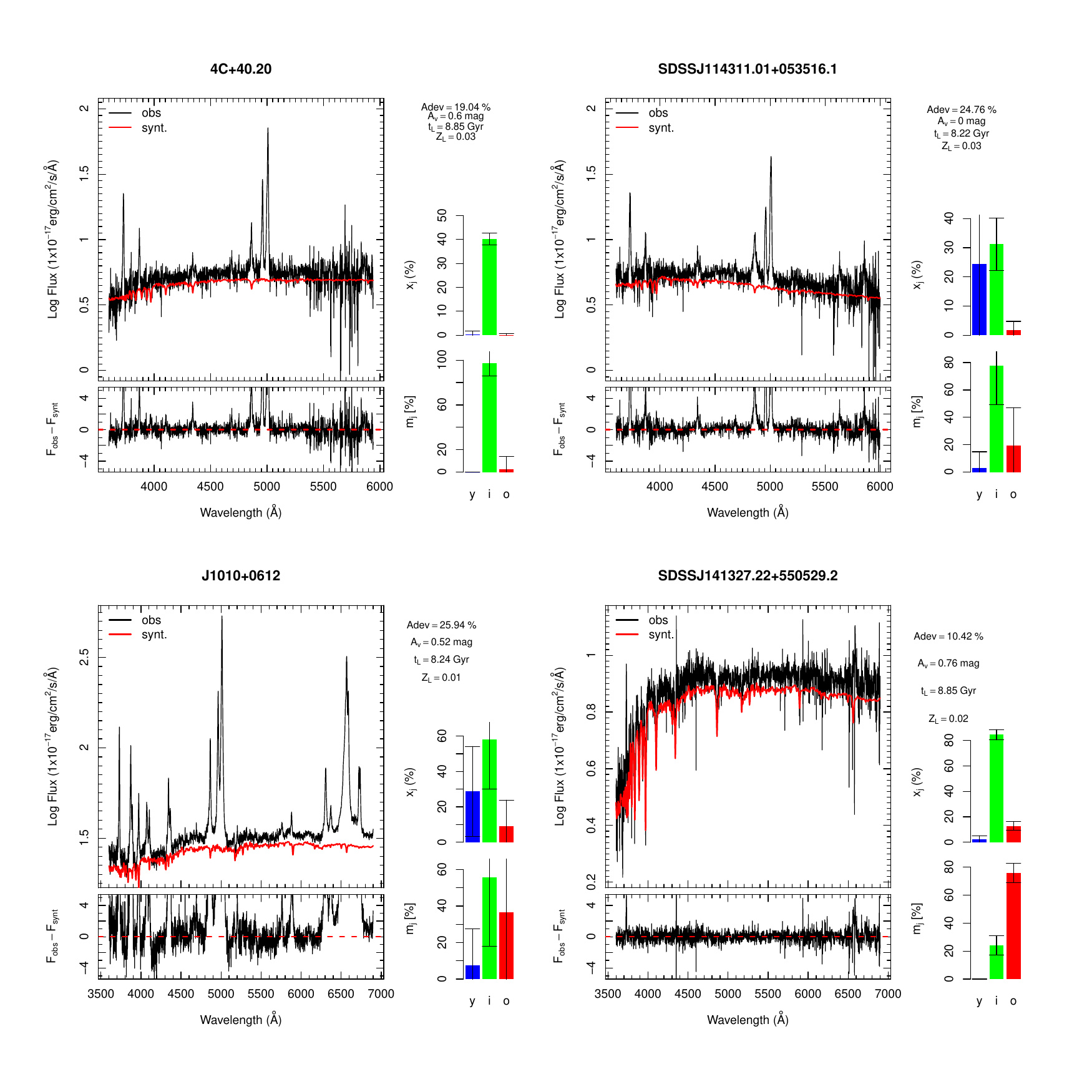}
    \caption{Stellar continuum fit made with starlight. The upper panel shows the observed spectrum (black) and the best fit stellar population model (red) shifted down for clarity. The bottom panel shows the galaxy spectrum after subtraction of the stellar contribution. The percentage distribution of young (y), intermediate (i) and old (o) stellar populations weighted by luminosity and mass in each galaxy is plotted in the histogram at right panel of each plot.}
    \label{A1}
\end{figure*}

\begin{figure*}\ContinuedFloat
    \centering
    \includegraphics{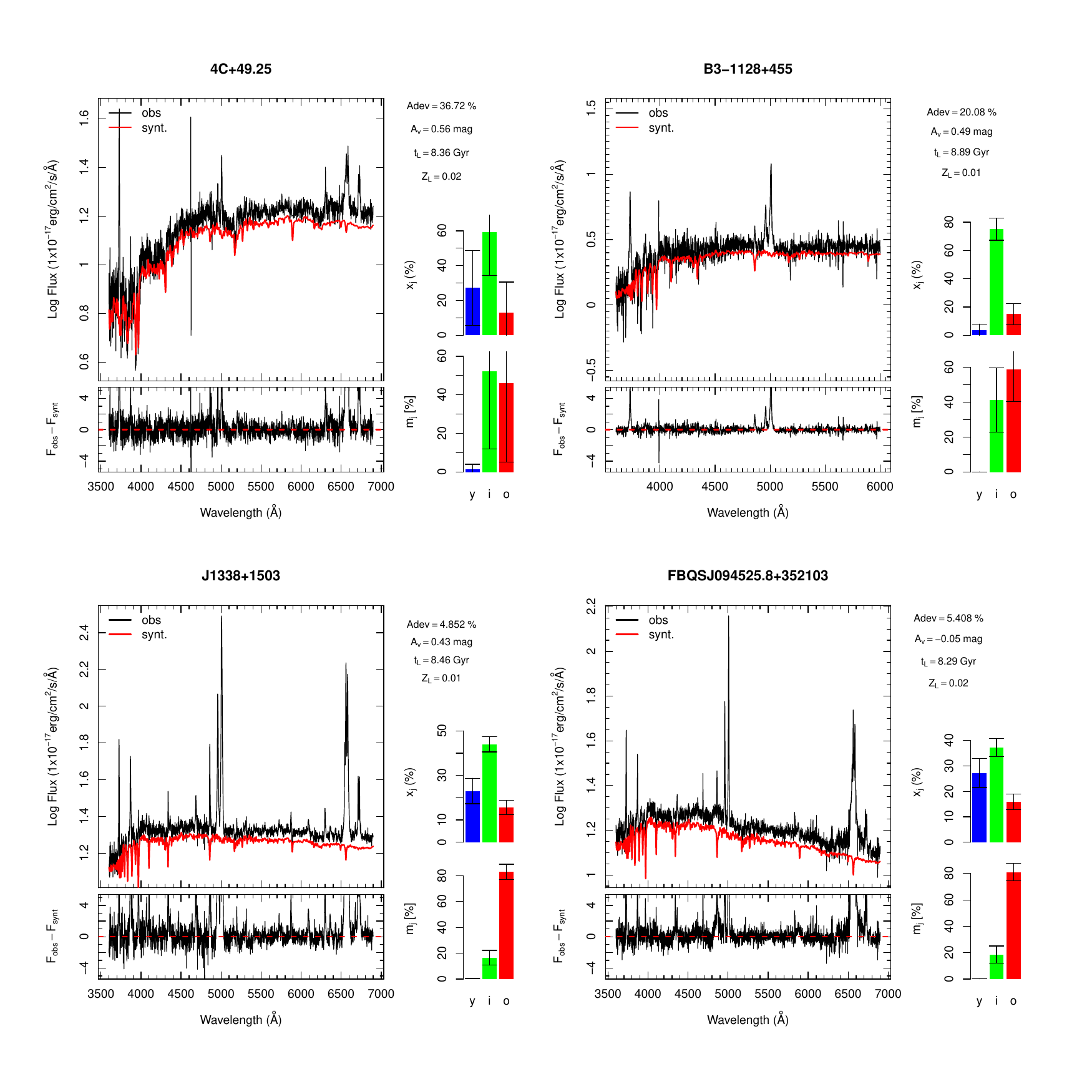}
    \caption{Continued}
\end{figure*}

\begin{figure*}\ContinuedFloat
    \centering
    \includegraphics{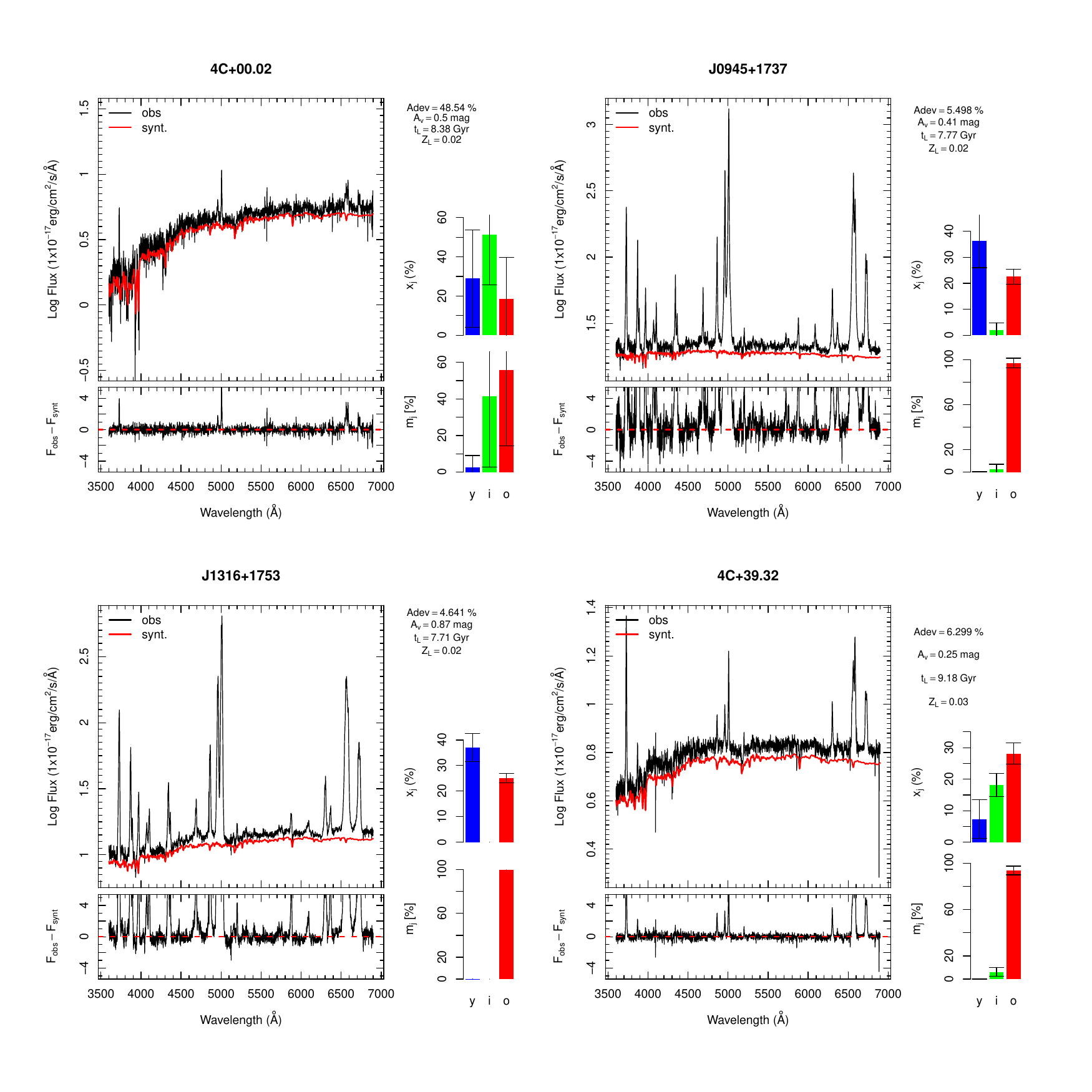}
    \caption{Continued}
\end{figure*}

\begin{figure*}\ContinuedFloat
    \centering
    \includegraphics{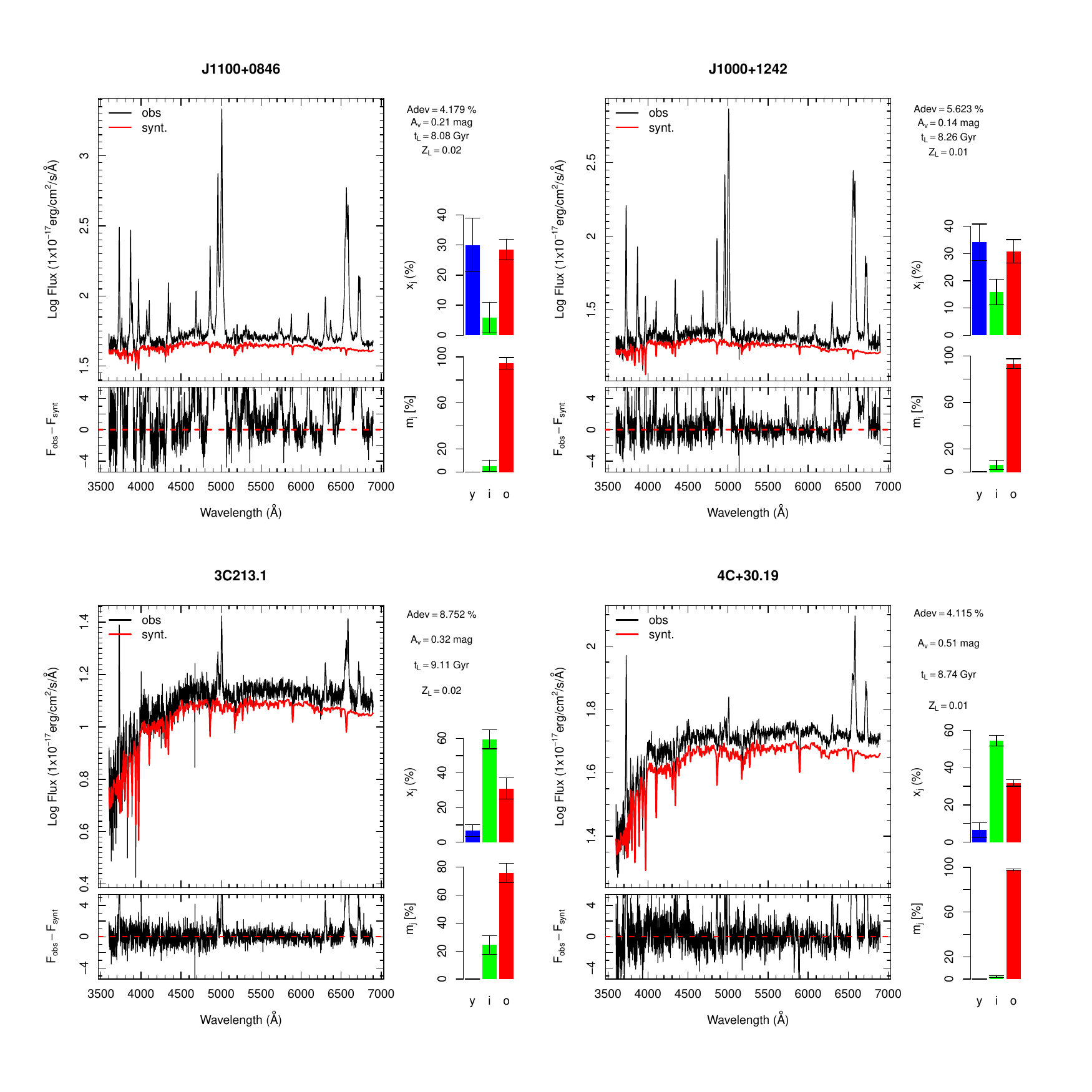}
    \caption{Continued}
\end{figure*}

\begin{figure*}\ContinuedFloat
    \centering
    \includegraphics{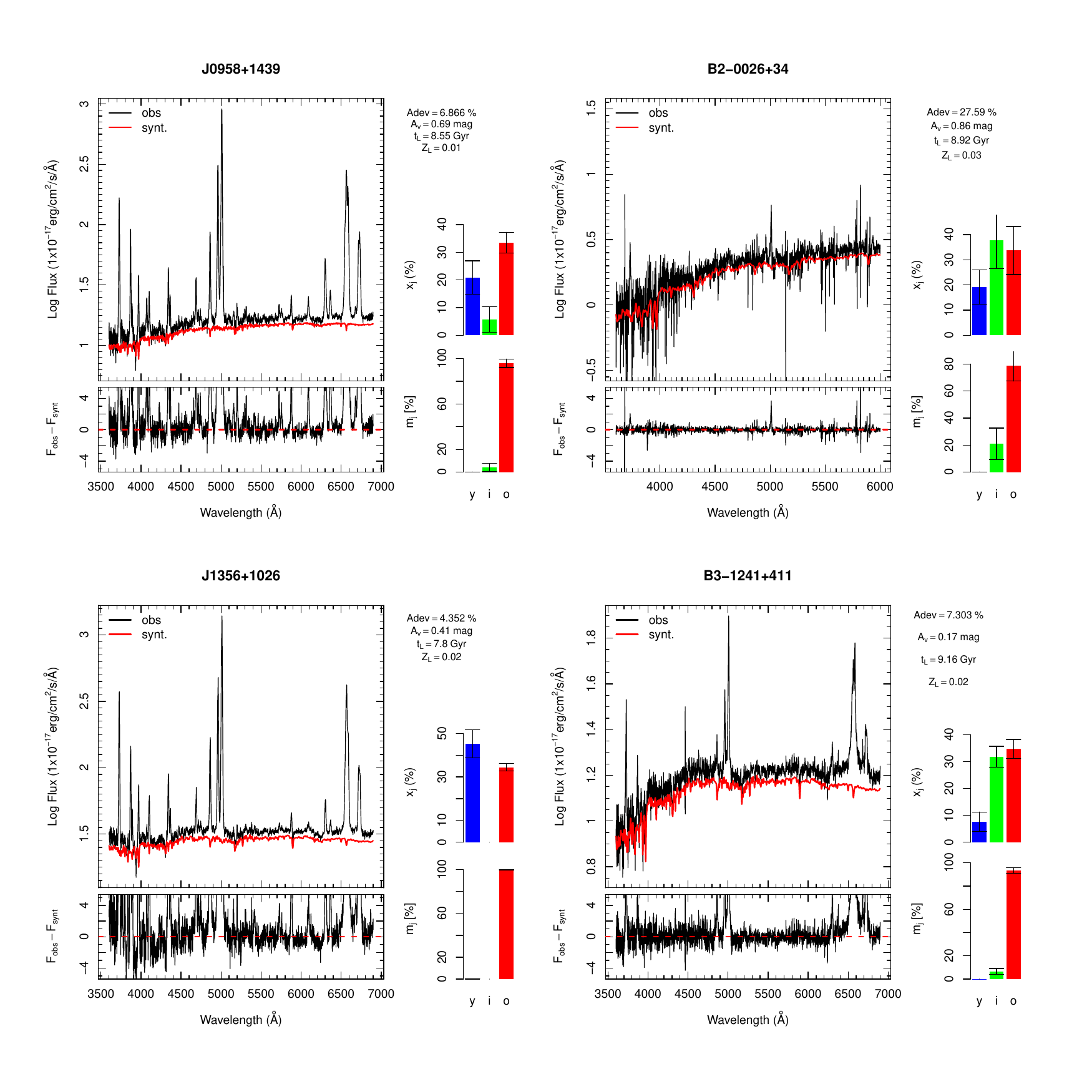}
    \caption{Caption}
\end{figure*}

\begin{figure*}\ContinuedFloat
    \centering
    \includegraphics{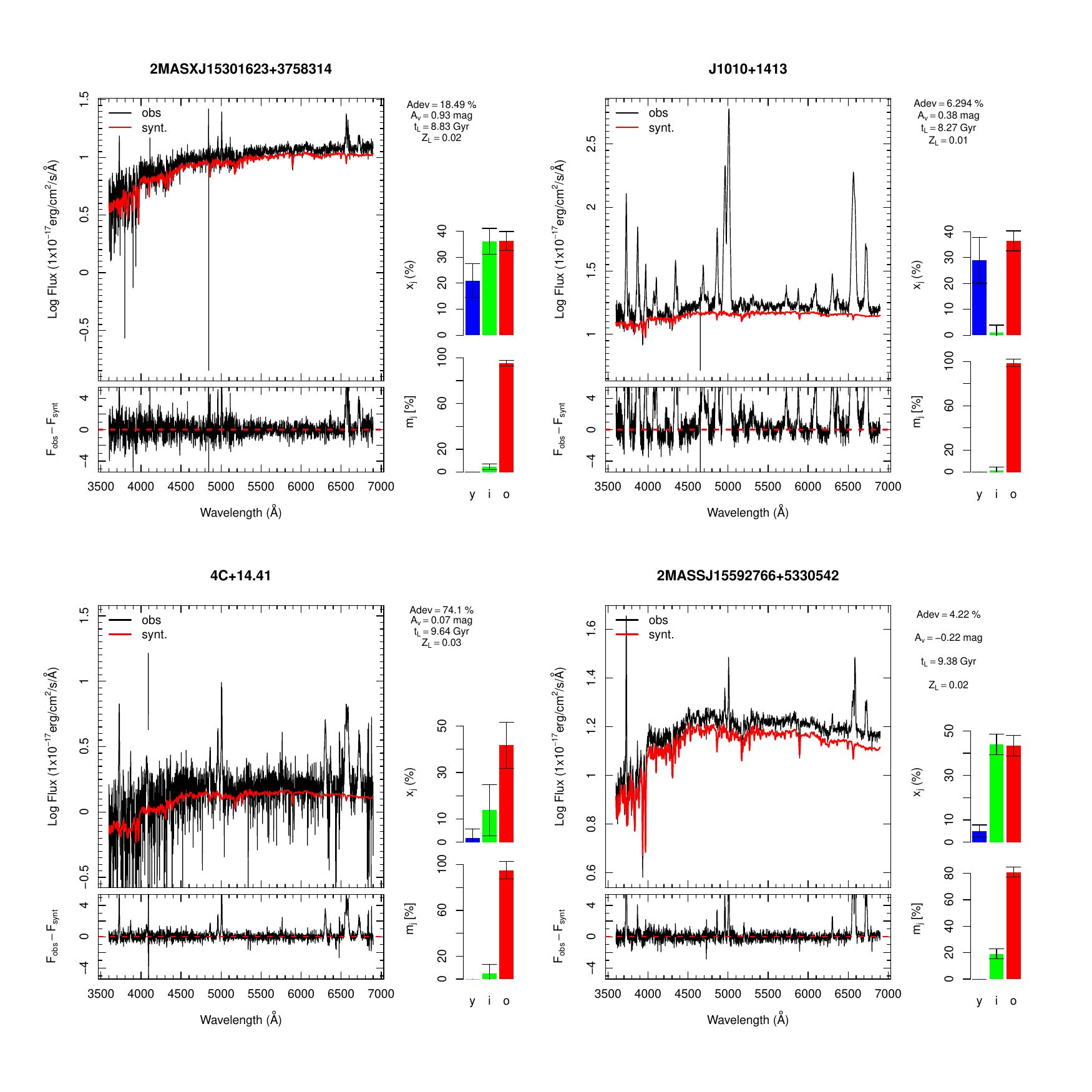}
    \caption{Continued}
\end{figure*}

\begin{figure*}\ContinuedFloat
    \centering
    \includegraphics{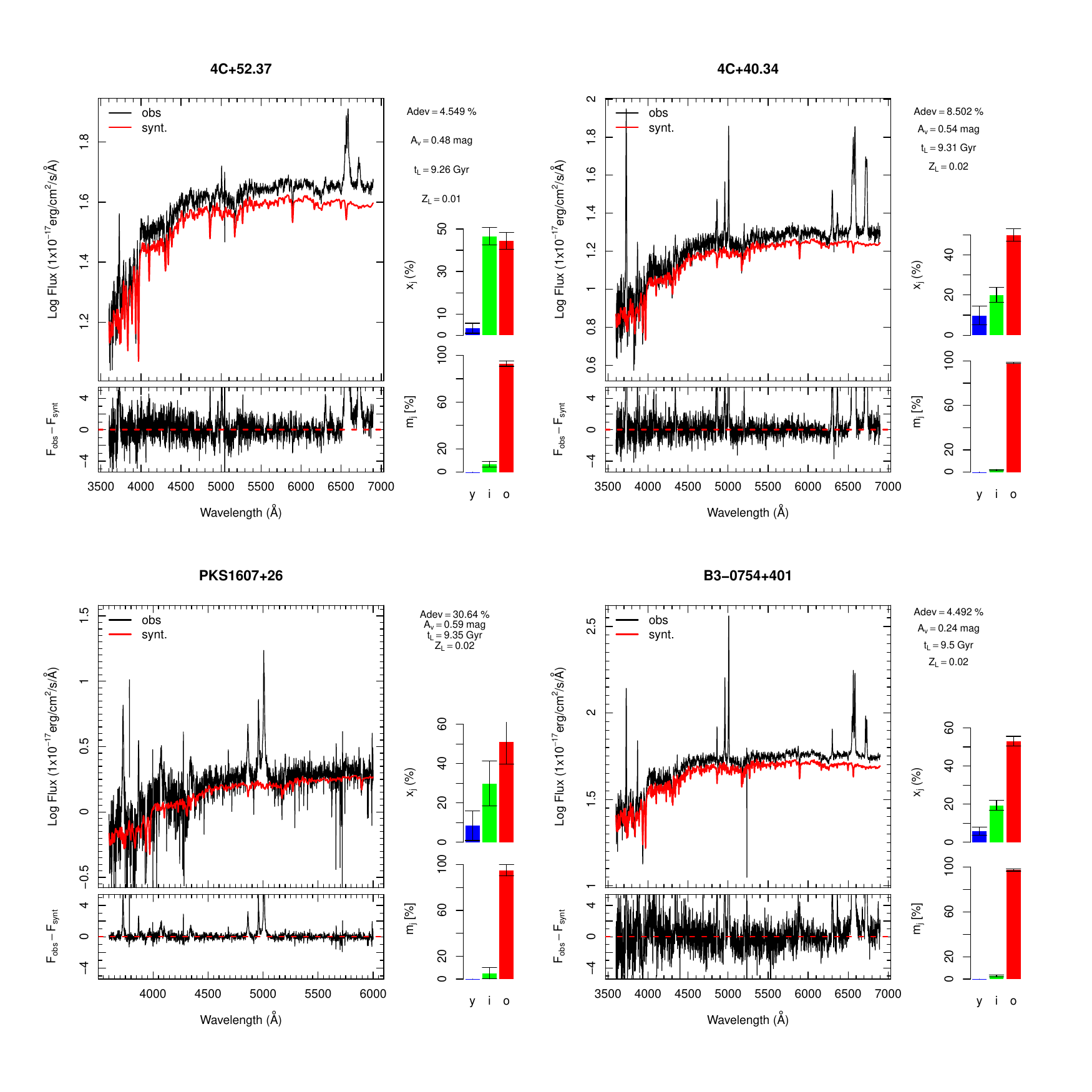}
    \caption{Continued}
\end{figure*}

\begin{figure*}\ContinuedFloat
    \centering
    \includegraphics{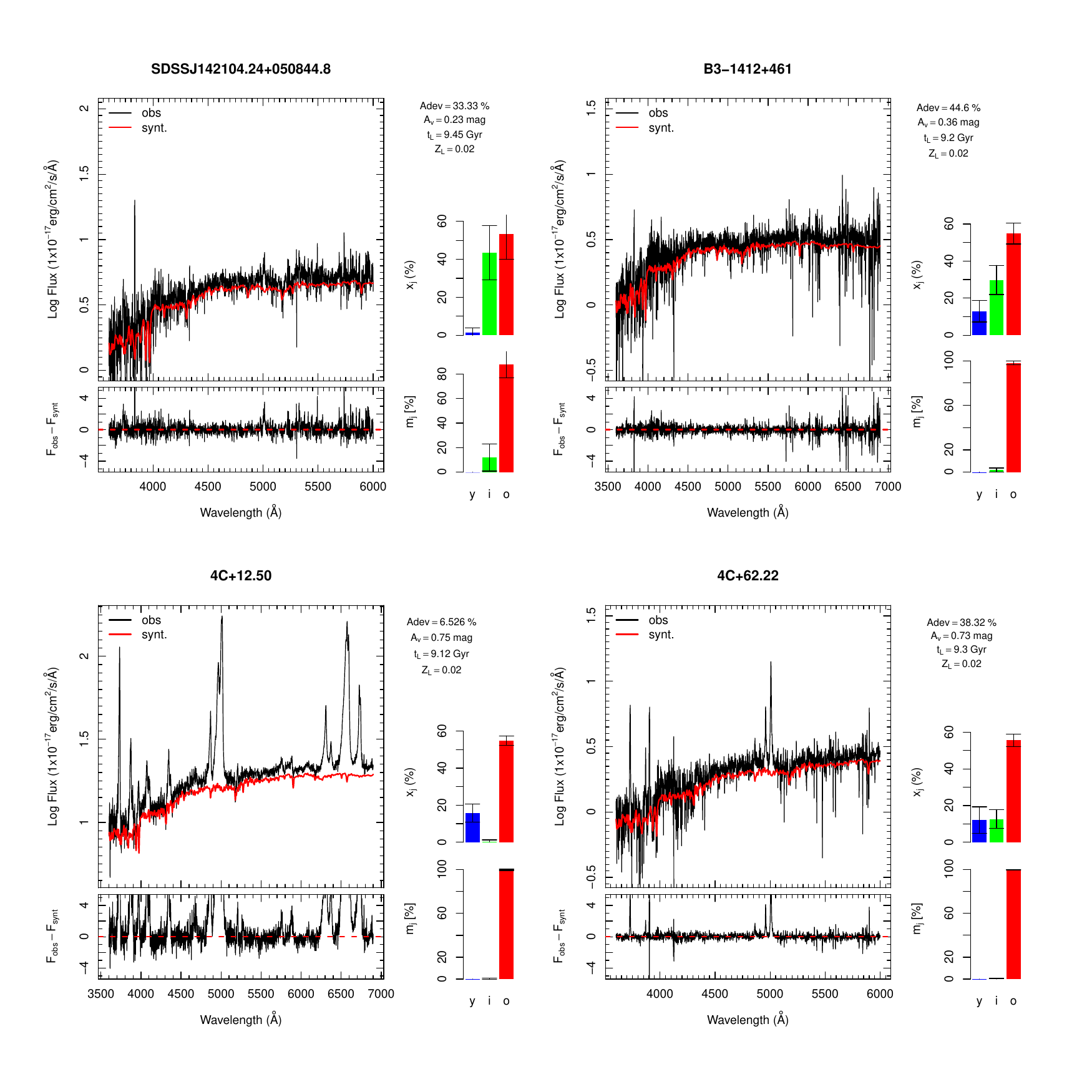}
    \caption{Continued.}
\end{figure*}

\begin{figure*}\ContinuedFloat
    \centering
    \includegraphics{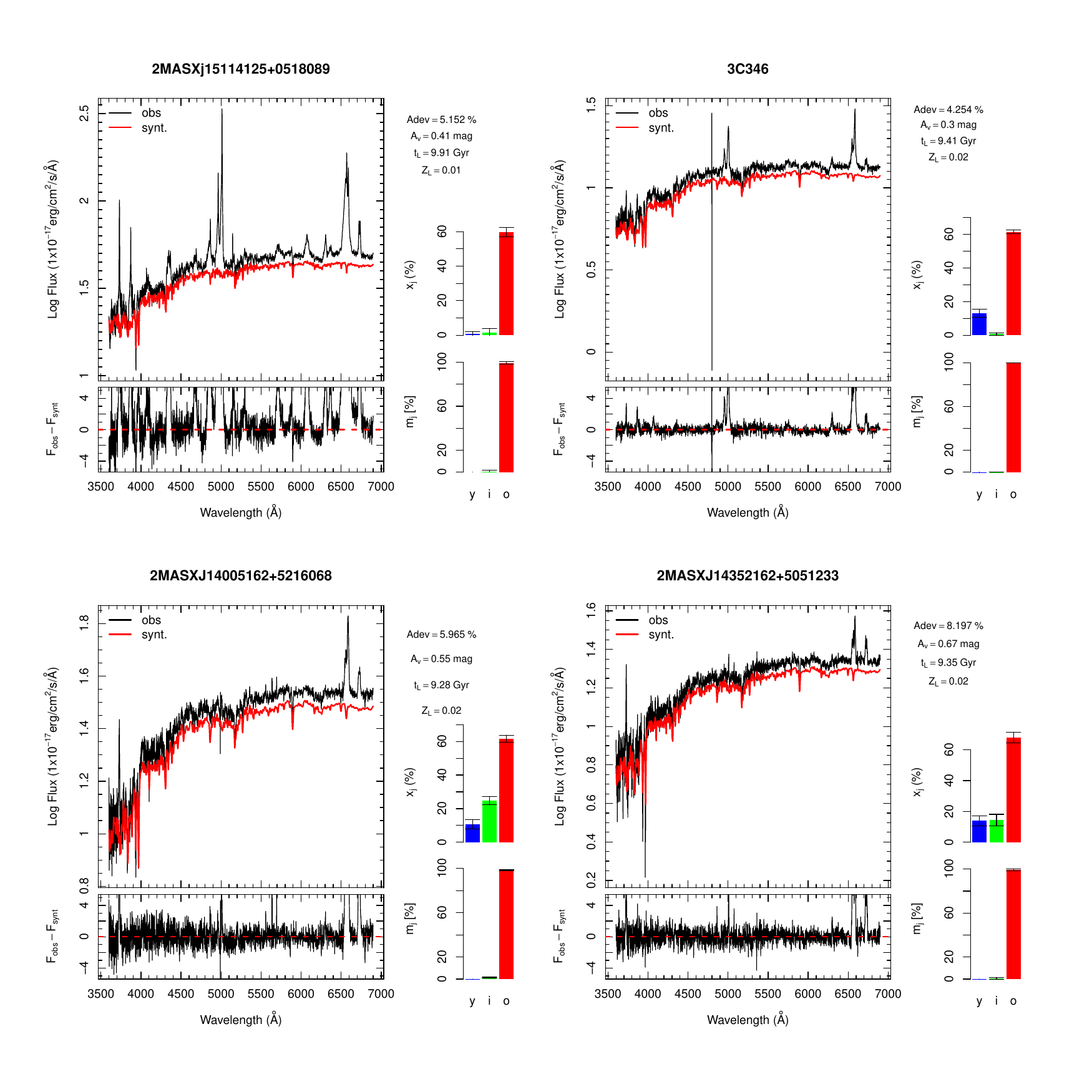}
    \caption{Continued}
\end{figure*}

\begin{figure*}\ContinuedFloat
    \centering
    \includegraphics{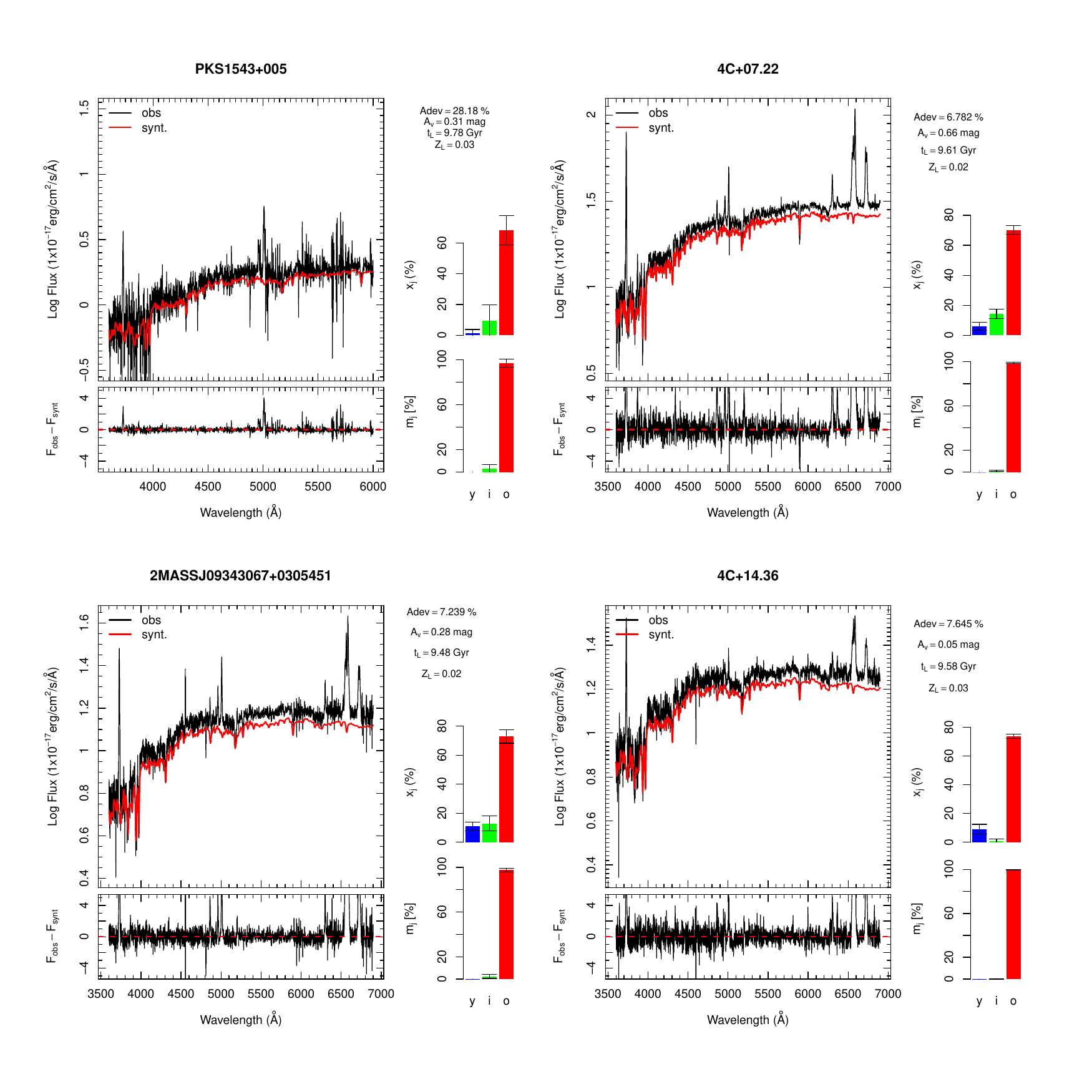}
    \caption{Continued}
\end{figure*}

\begin{figure*}\ContinuedFloat
    \centering
    \includegraphics{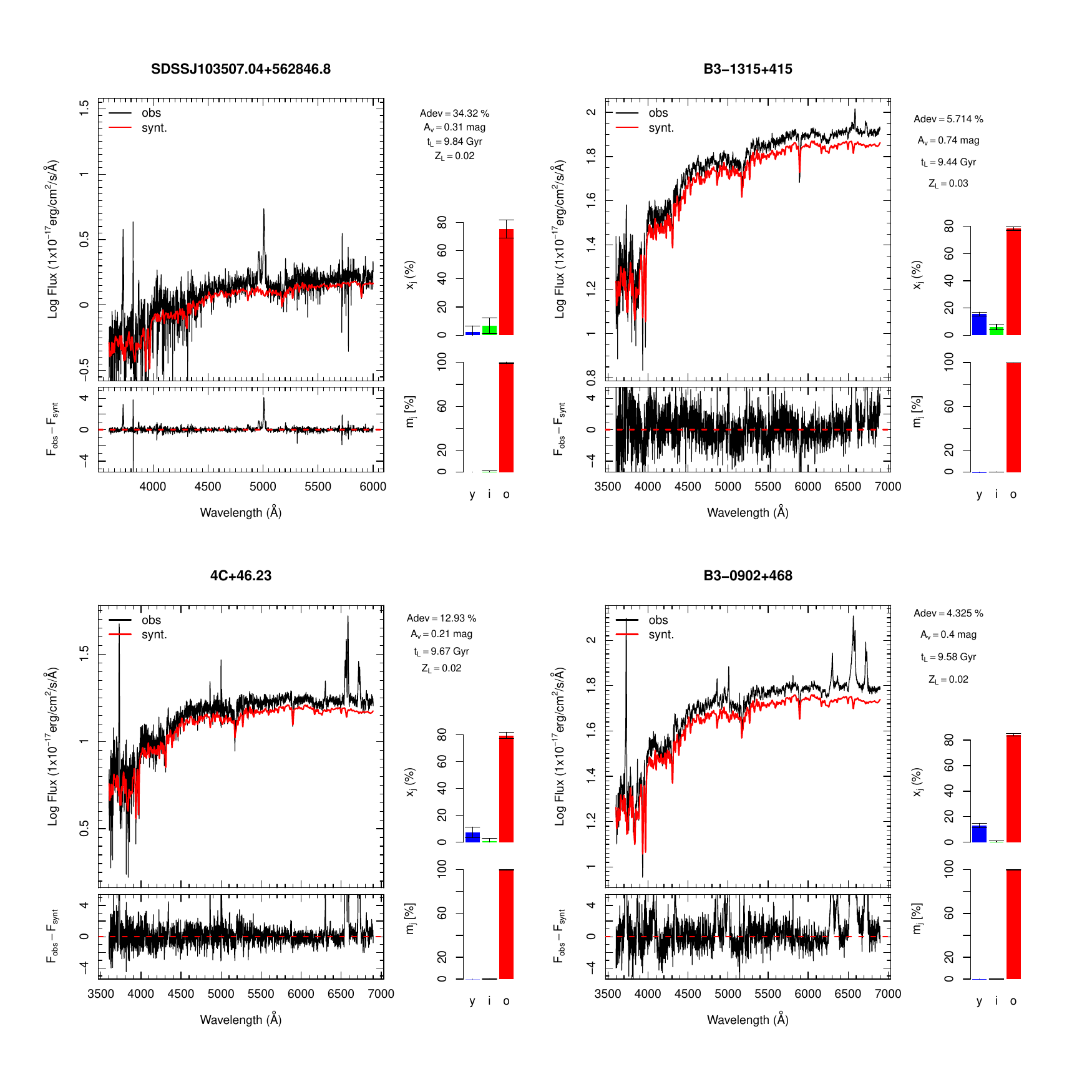}
    \caption{Continued}
\end{figure*}

\begin{figure*}\ContinuedFloat
    \centering
    \includegraphics{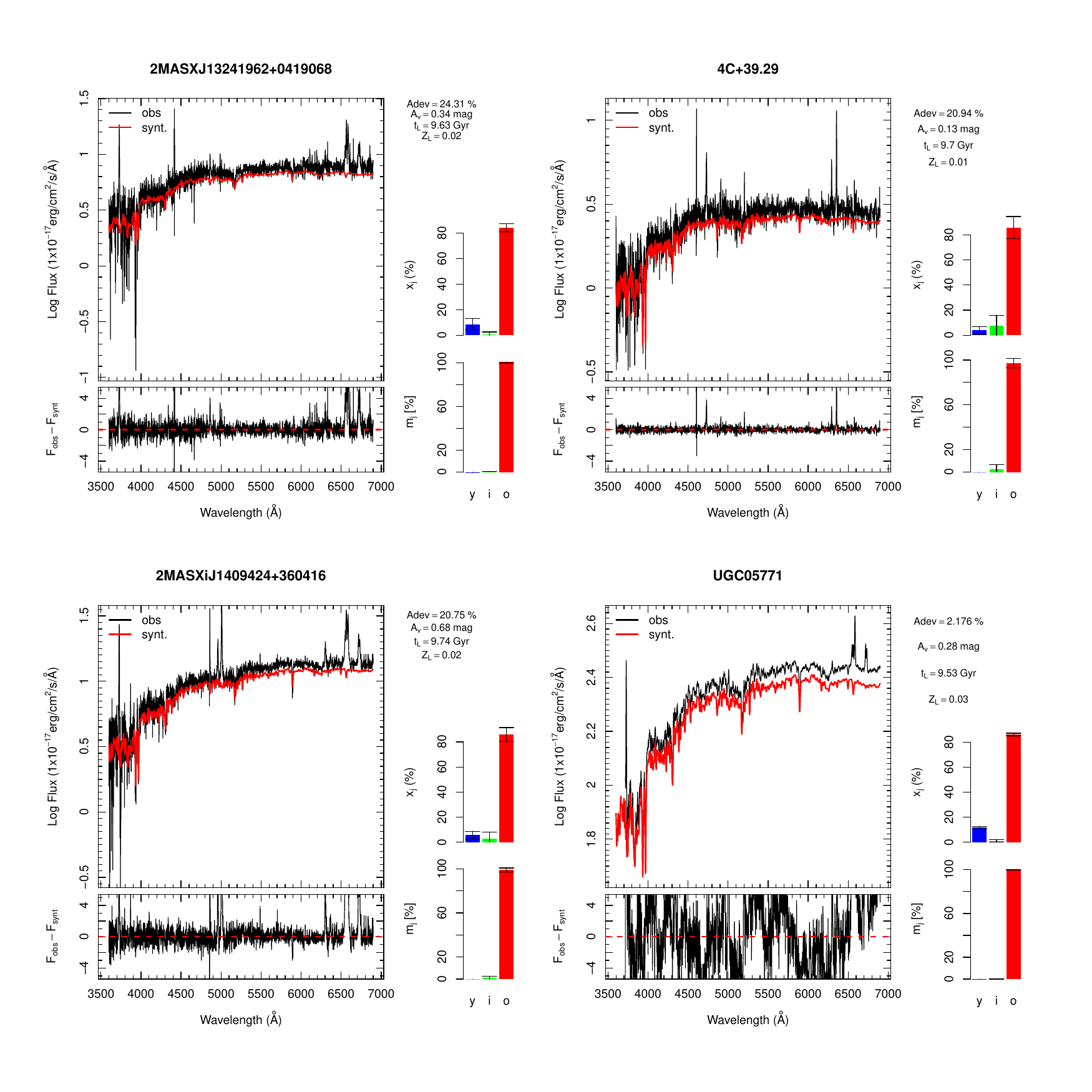}
    \caption{Continued}
\end{figure*}

\begin{figure*}\ContinuedFloat
    \centering
    \includegraphics{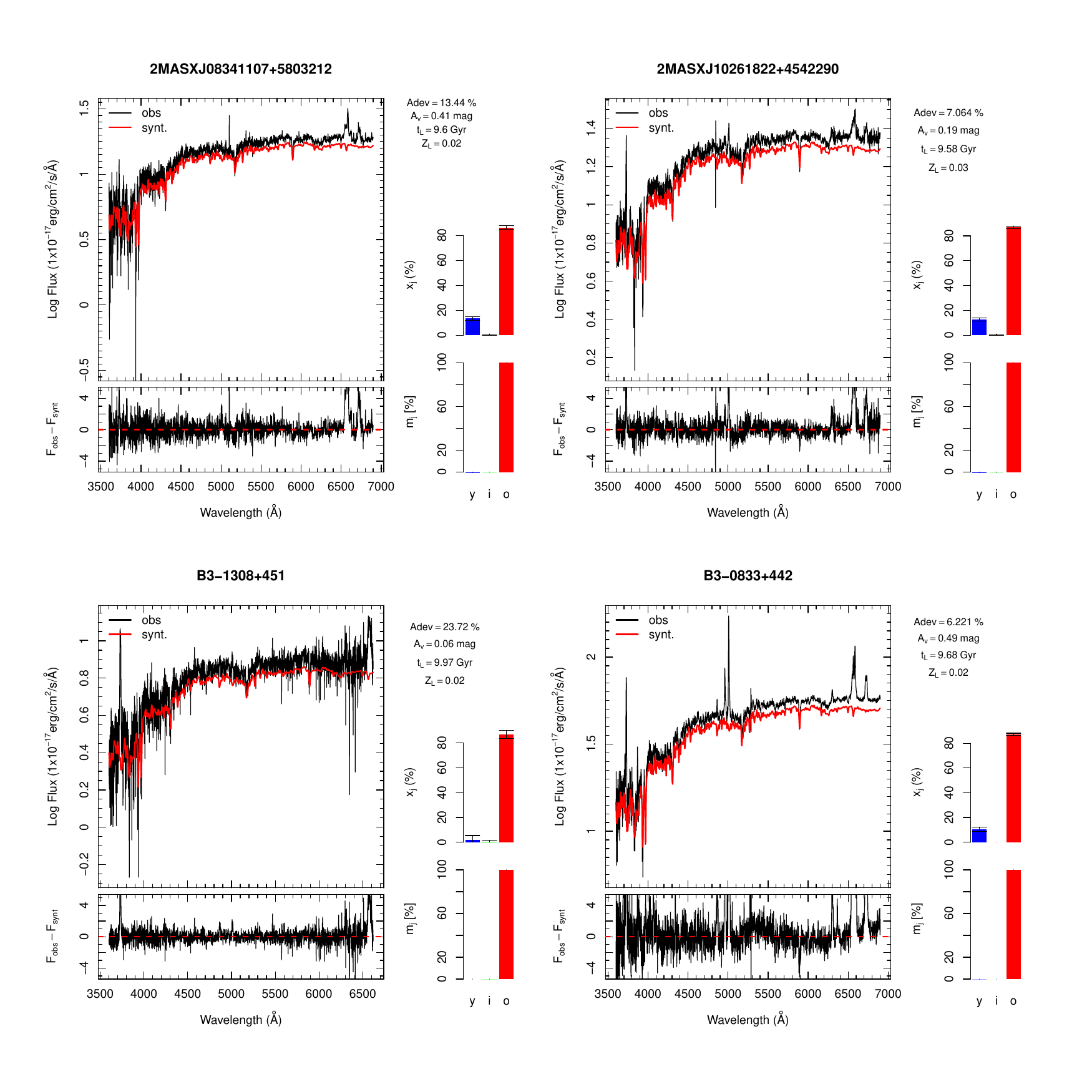}
    \caption{Continued}
\end{figure*}

\begin{figure*}\ContinuedFloat
    \centering
    \includegraphics{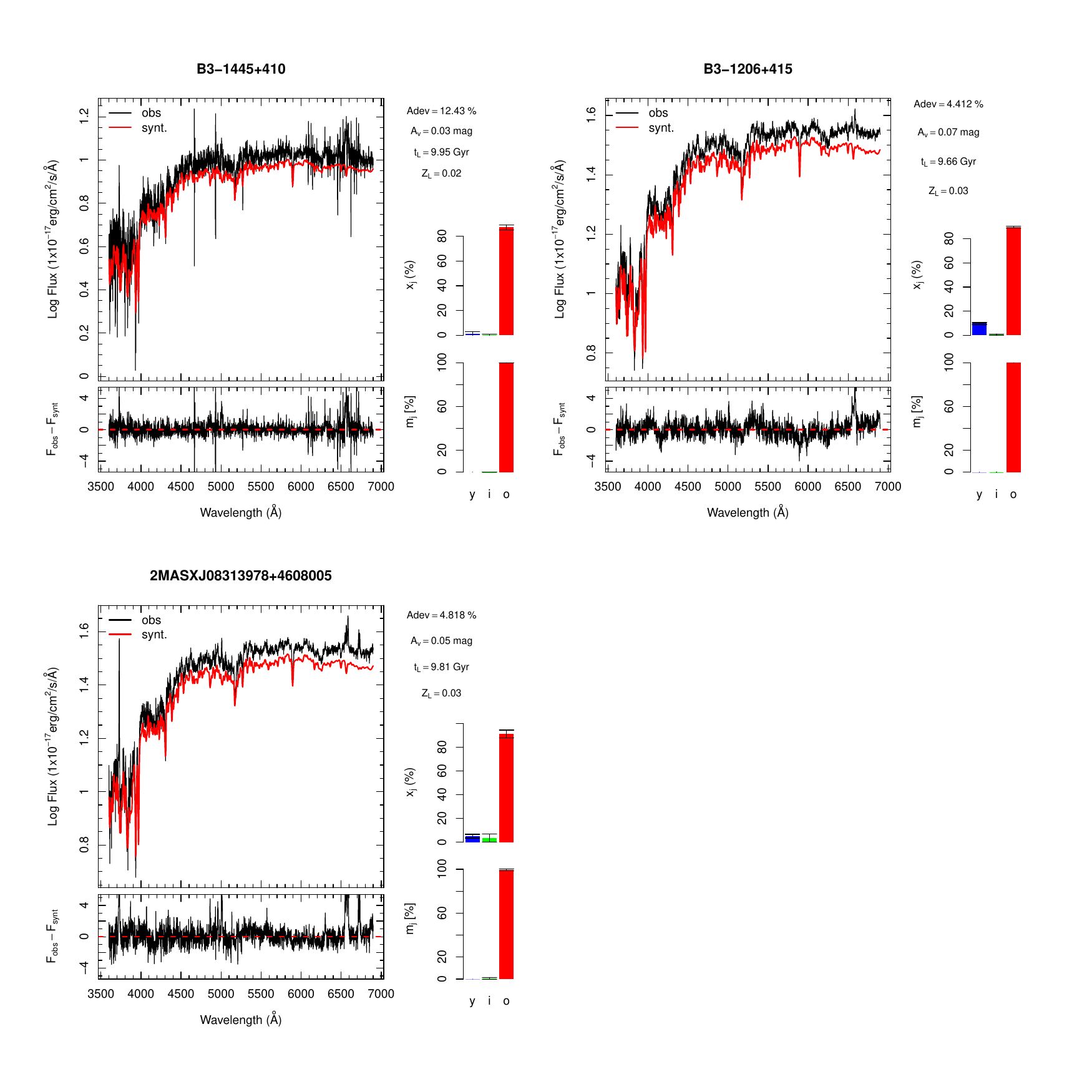}
    \caption{Continued}
\end{figure*}

\begin{figure*}\ContinuedFloat
    \centering
    \includegraphics{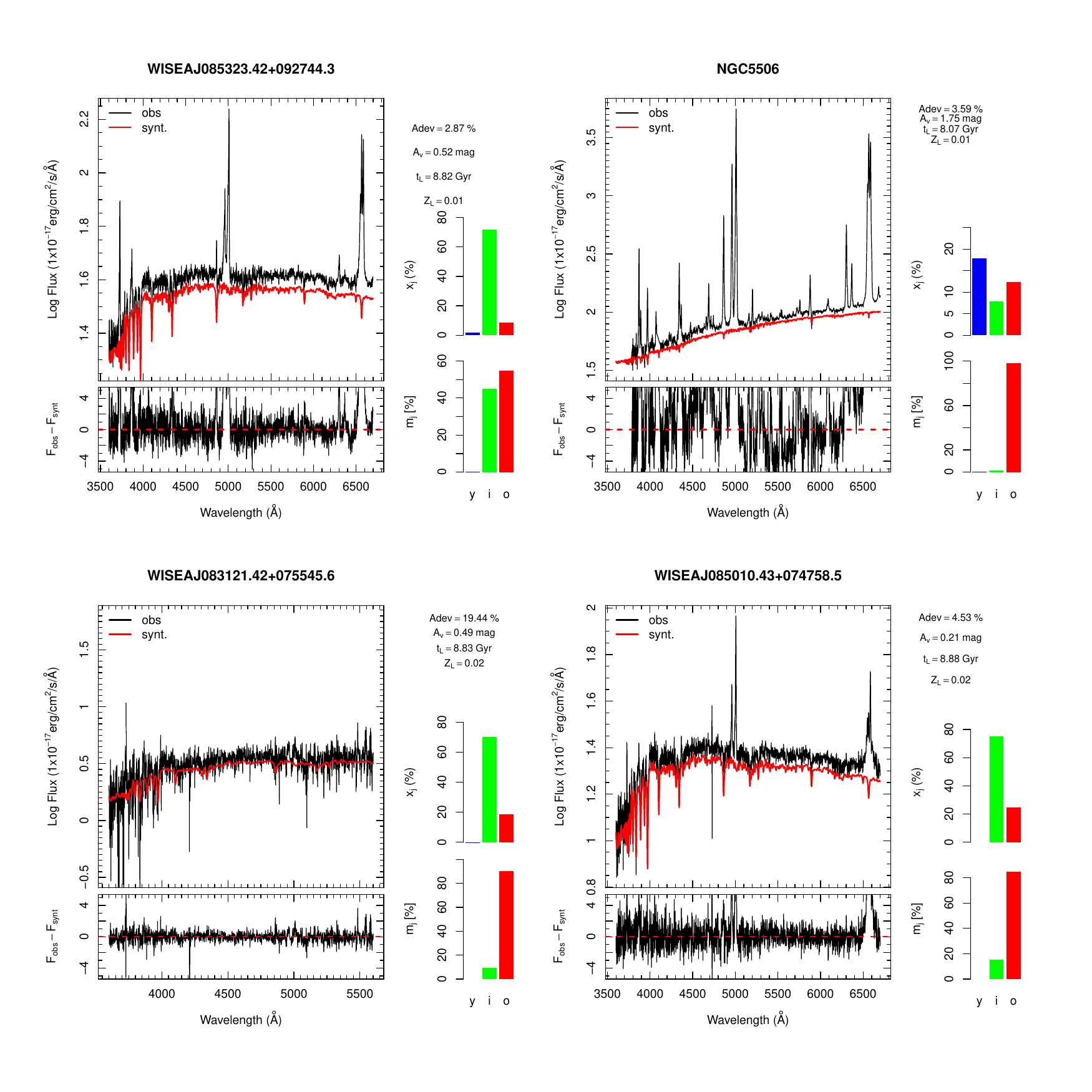}
    \caption{Continued}
\end{figure*}

\begin{figure*}\ContinuedFloat
    \centering
    \includegraphics{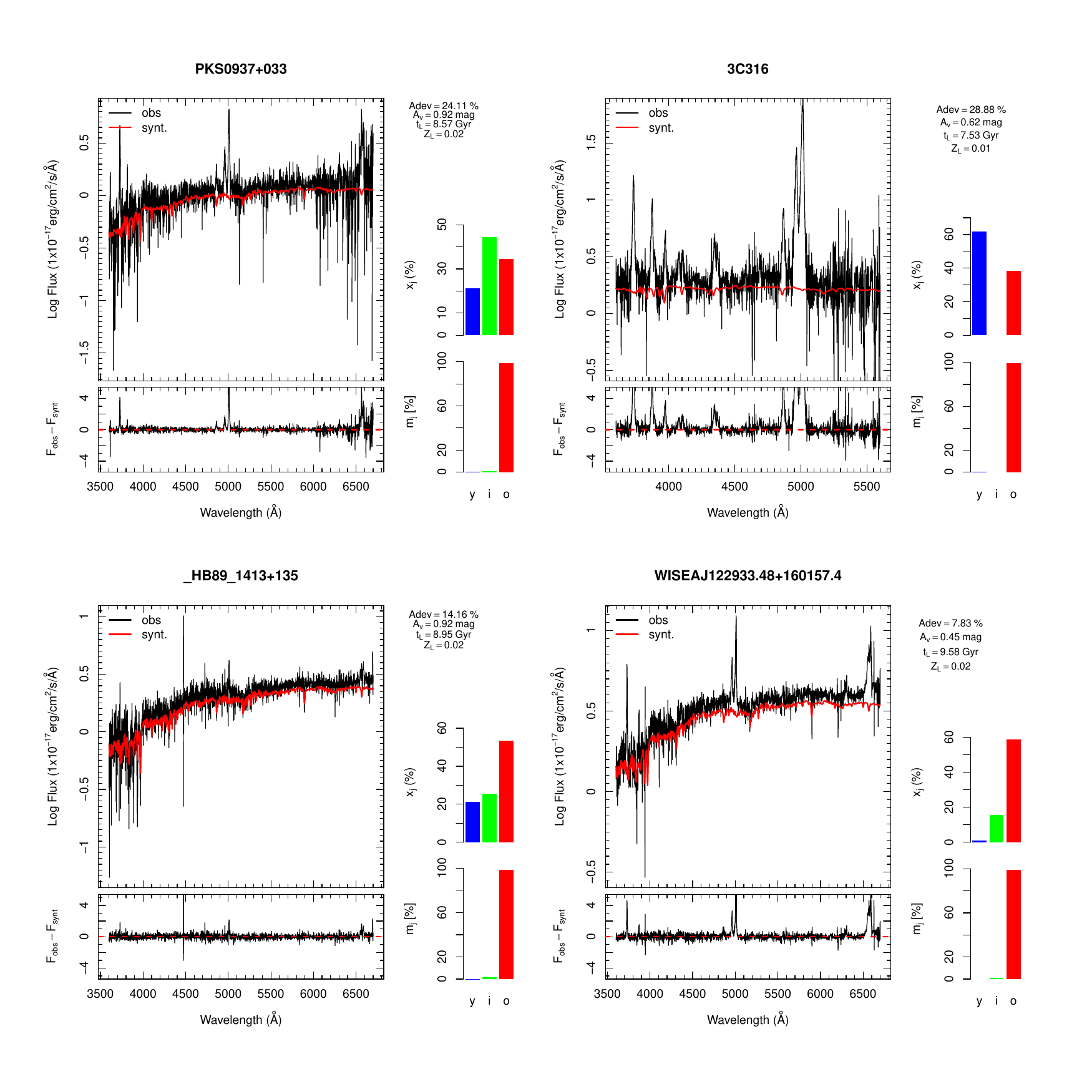}
    \caption{Continued}
\end{figure*}

\begin{figure*}\ContinuedFloat
    \centering
    \includegraphics{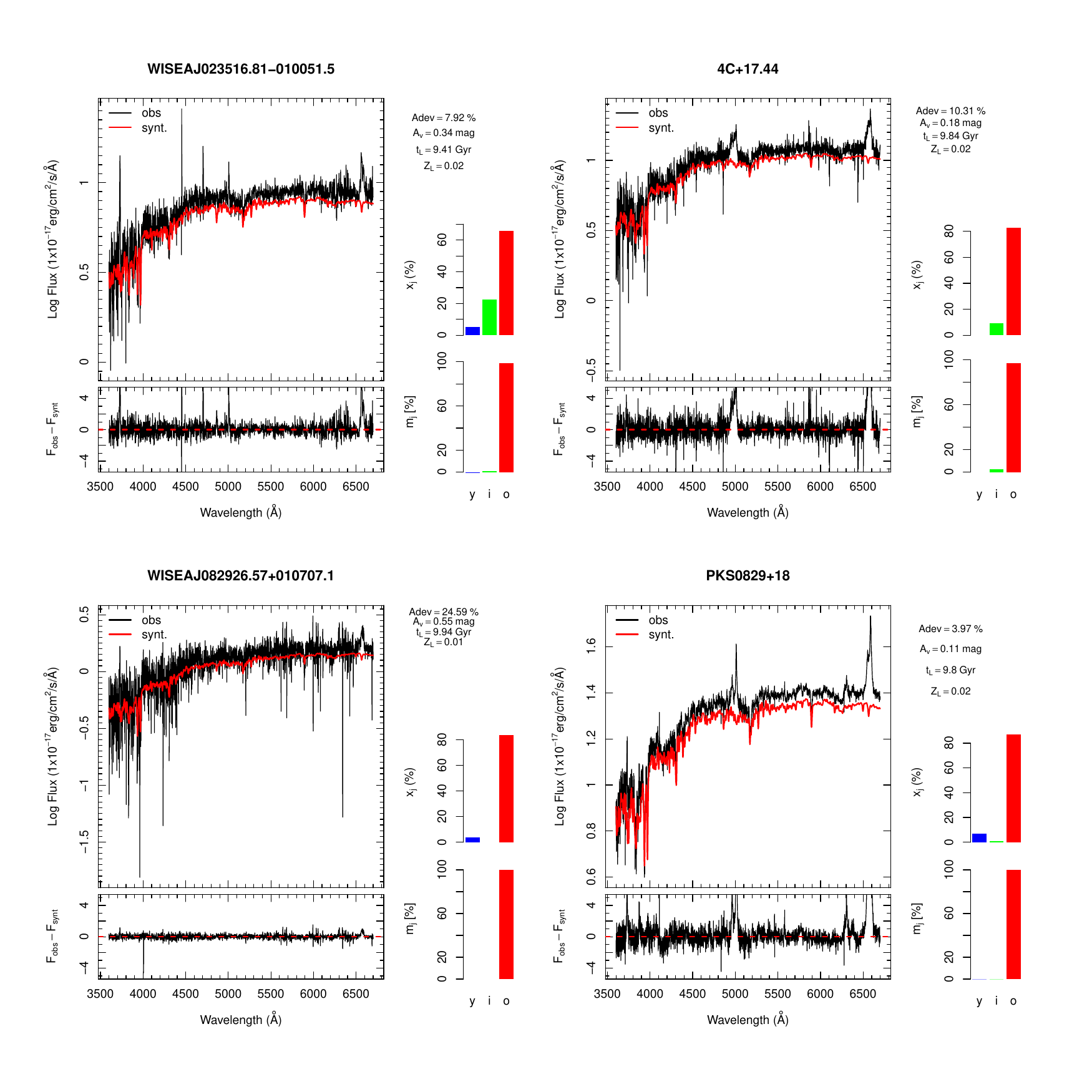}
    \caption{Continued}
\end{figure*}

\begin{figure*}\ContinuedFloat
    \centering
    \includegraphics{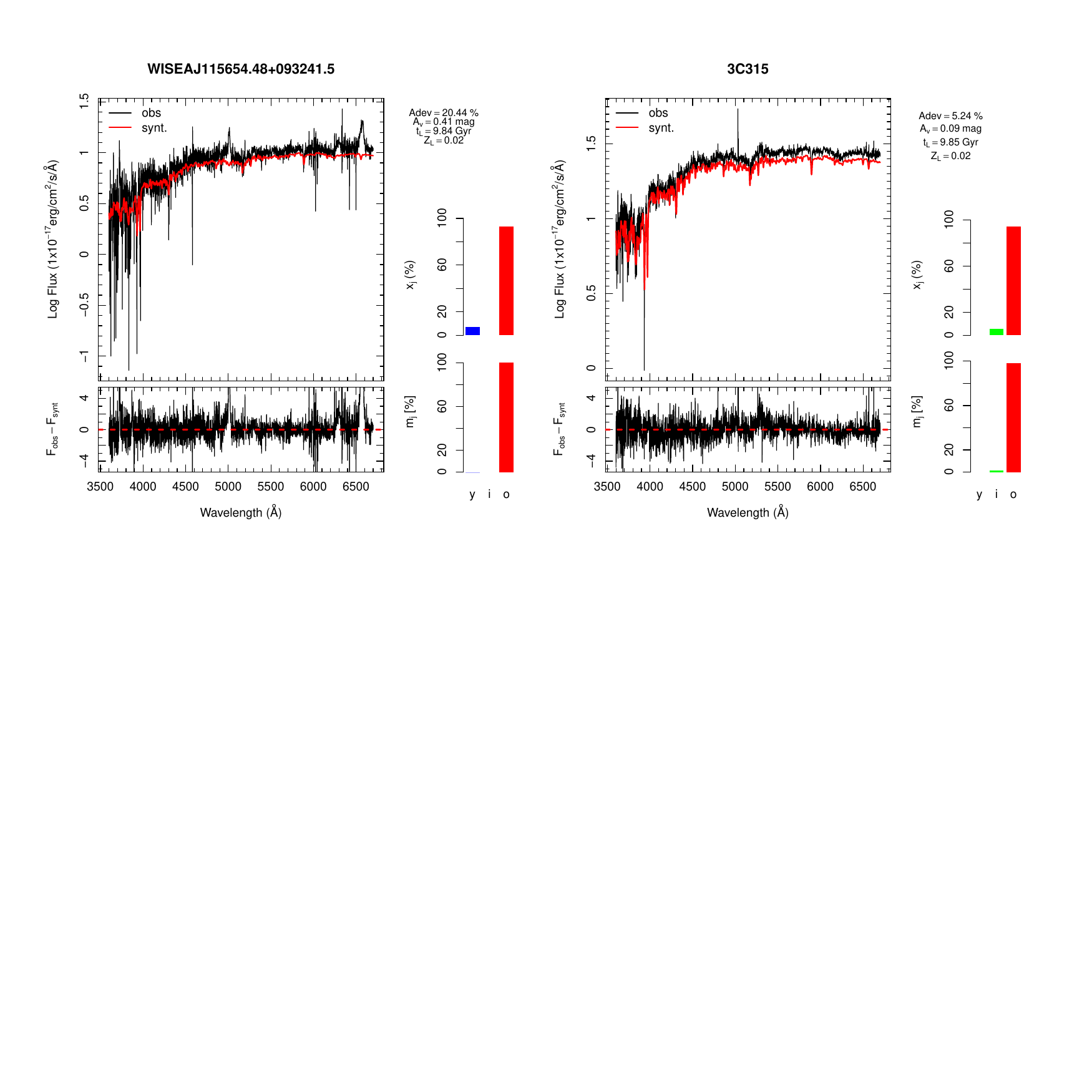}
    \caption{Continued}
\end{figure*}

\begin{figure*}
	\includegraphics[width=0.98\textwidth]{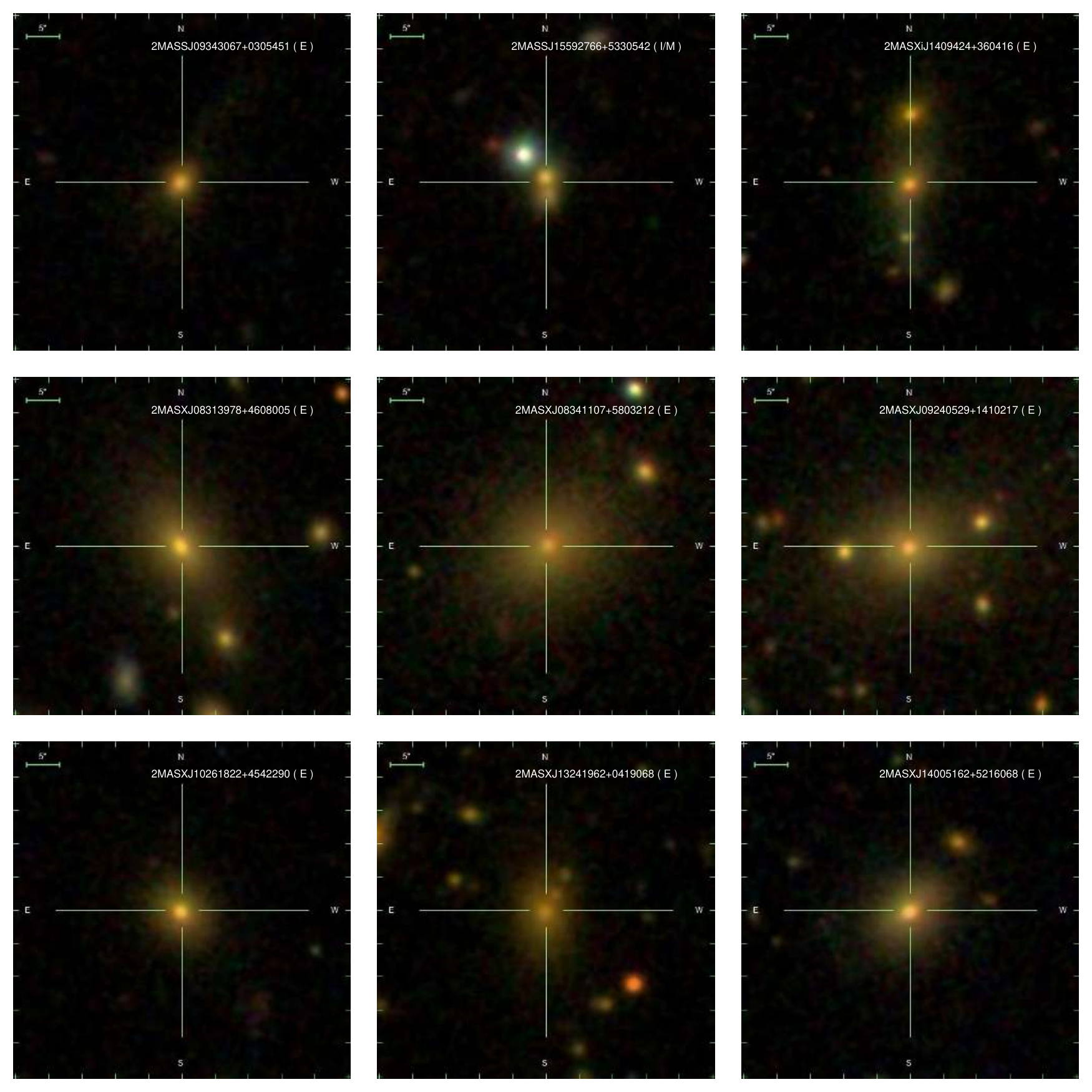}
	\caption{SDSS gri colour composite images for our CSS/GPS sample. In each panel, the position of the source is marked by a cross, while the image scale is shown in the upper left corner, the name and morphological classification of the source are indicated in the upper right corner. (E), (S), (I/M) and (P) represent, respectively, elliptical, spiral, irregular/merger and Point-like objects.  The images are equal in size, being a square 50''.}
	\label{fig:morph}
\end{figure*}

\begin{figure*}\ContinuedFloat
    \centering
    \includegraphics{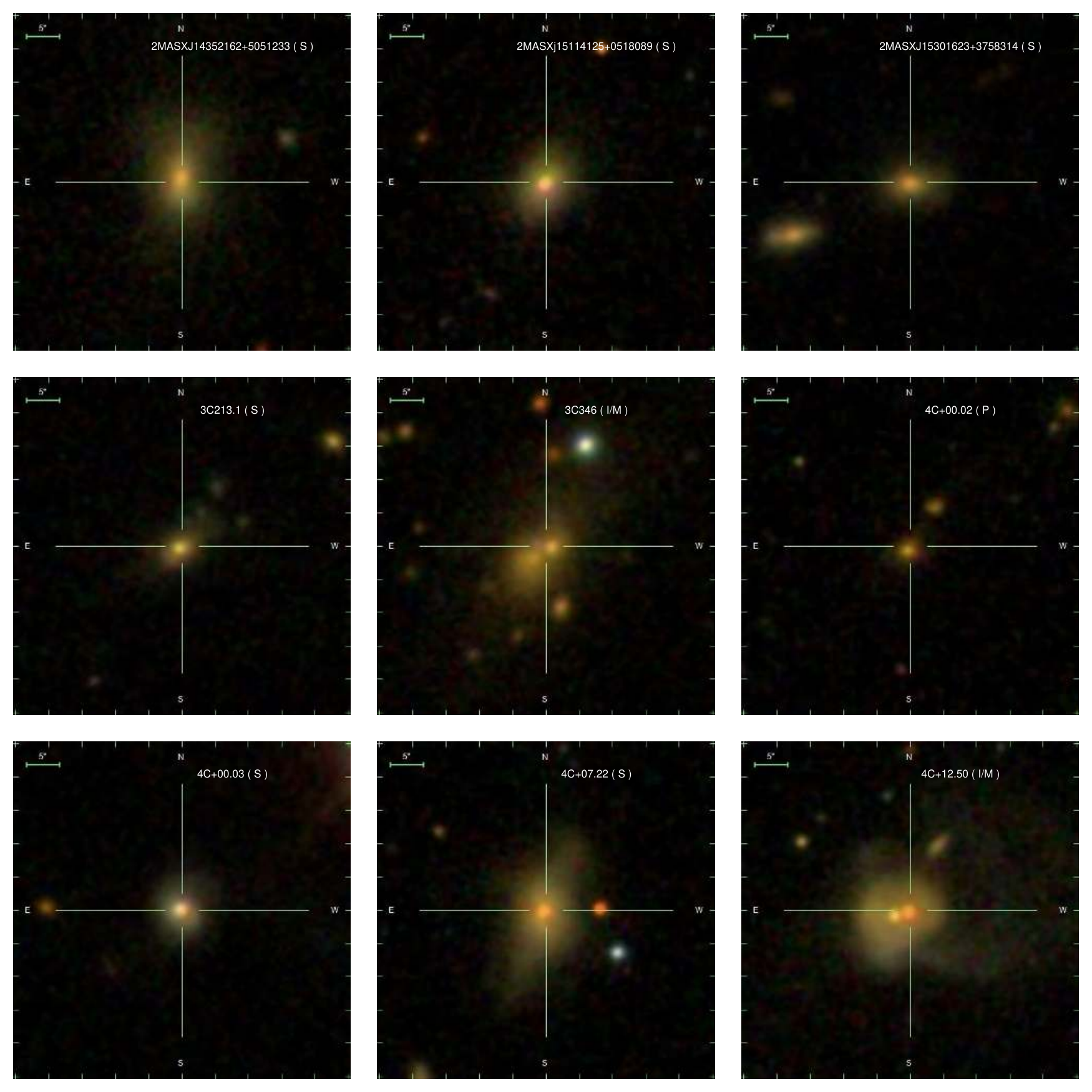}
    \caption{Continued}
\end{figure*}

\begin{figure*}\ContinuedFloat
    \centering
    \includegraphics{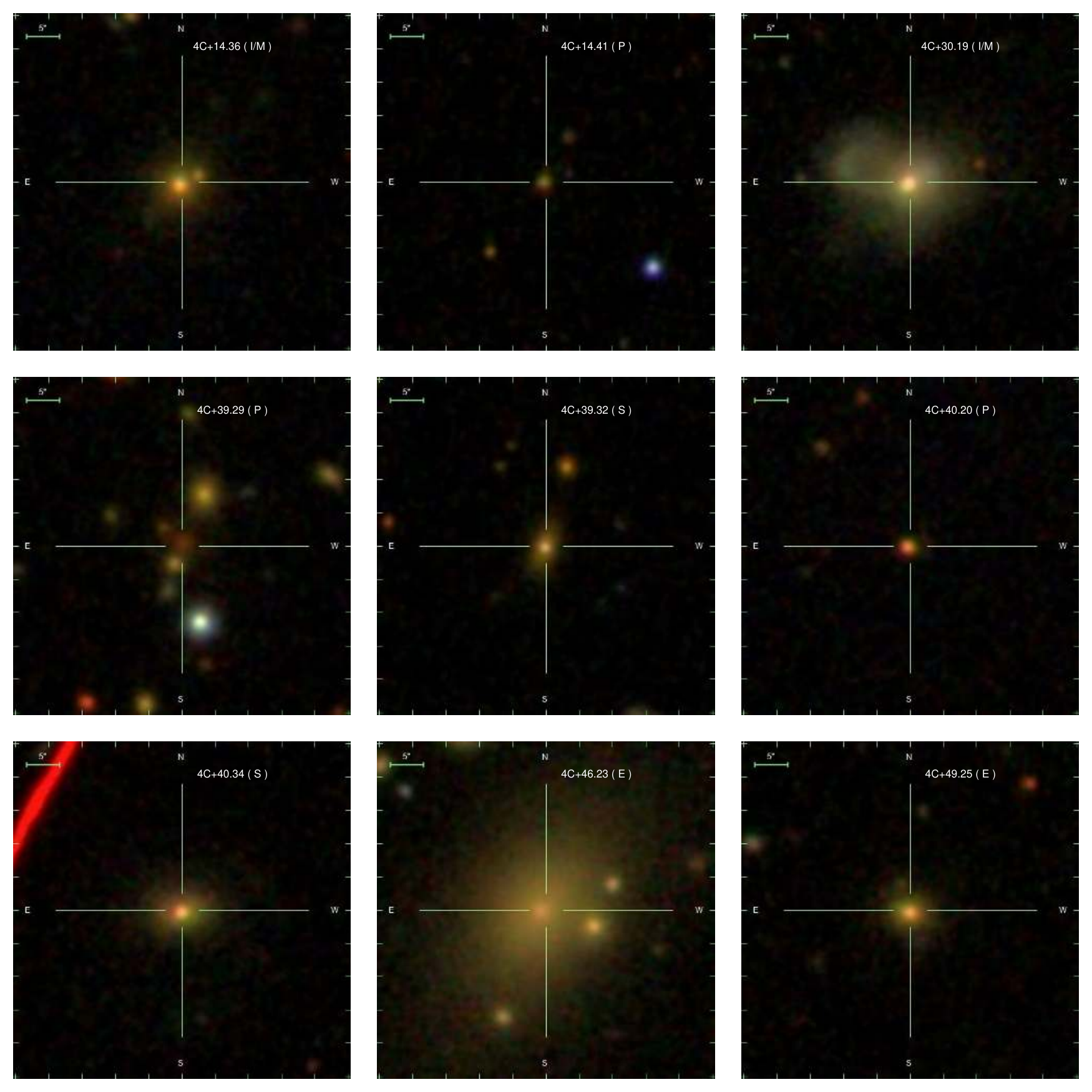}
    \caption{Continued}
\end{figure*}

\begin{figure*}\ContinuedFloat
    \centering
    \includegraphics{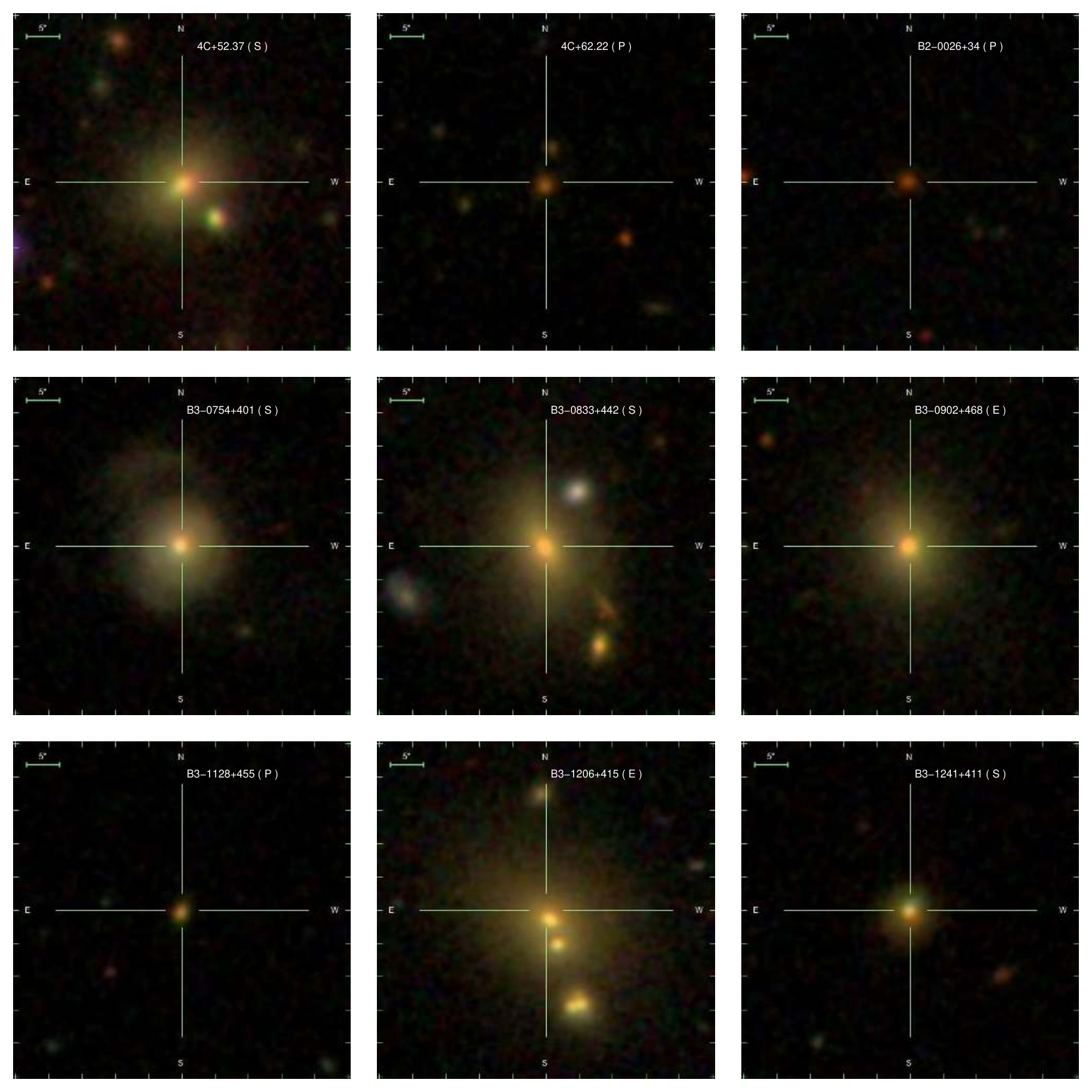}
    \caption{Continued}
\end{figure*}

\begin{figure*}\ContinuedFloat
    \centering
    \includegraphics{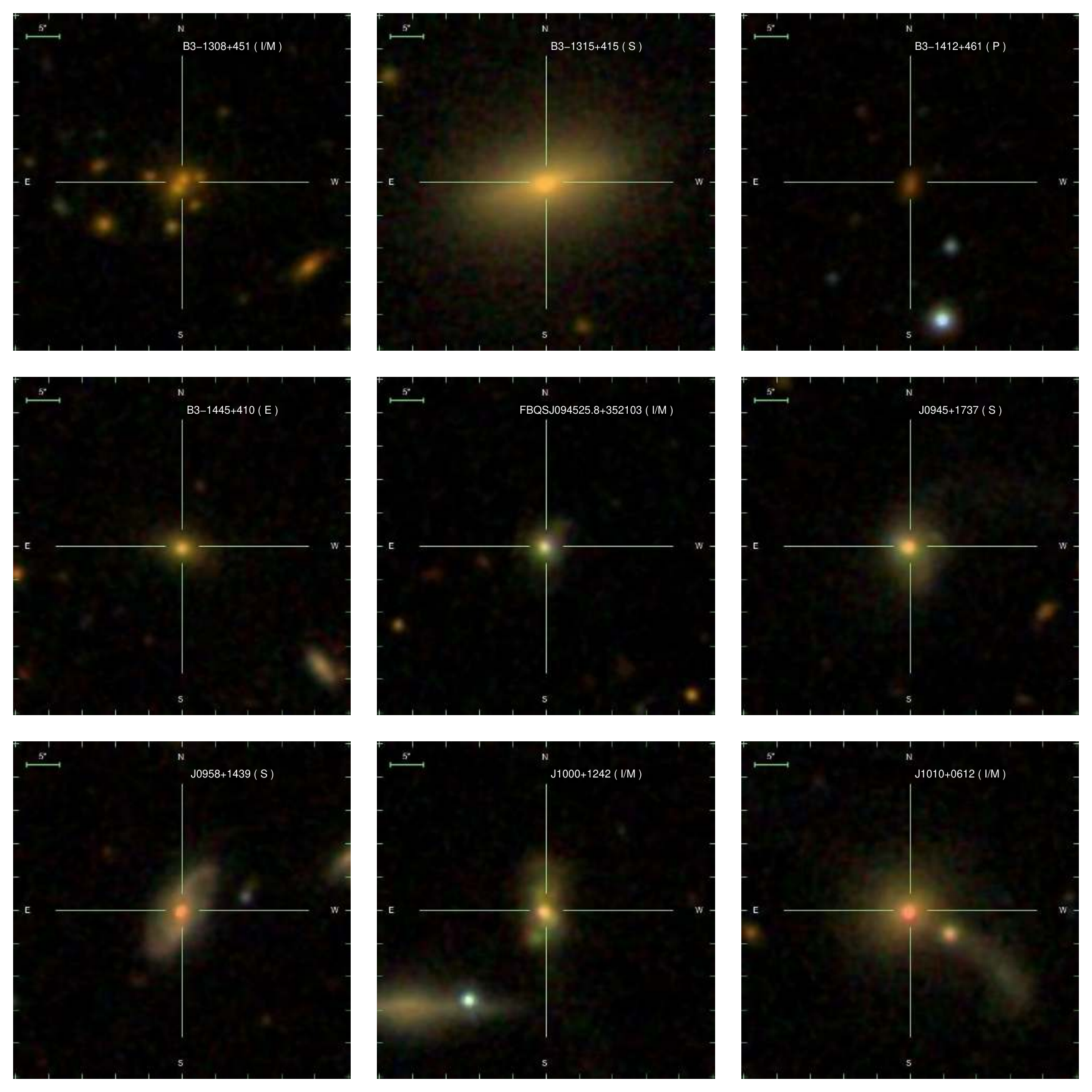}
    \caption{Continued}
\end{figure*}

\begin{figure*}\ContinuedFloat
    \centering
    \includegraphics{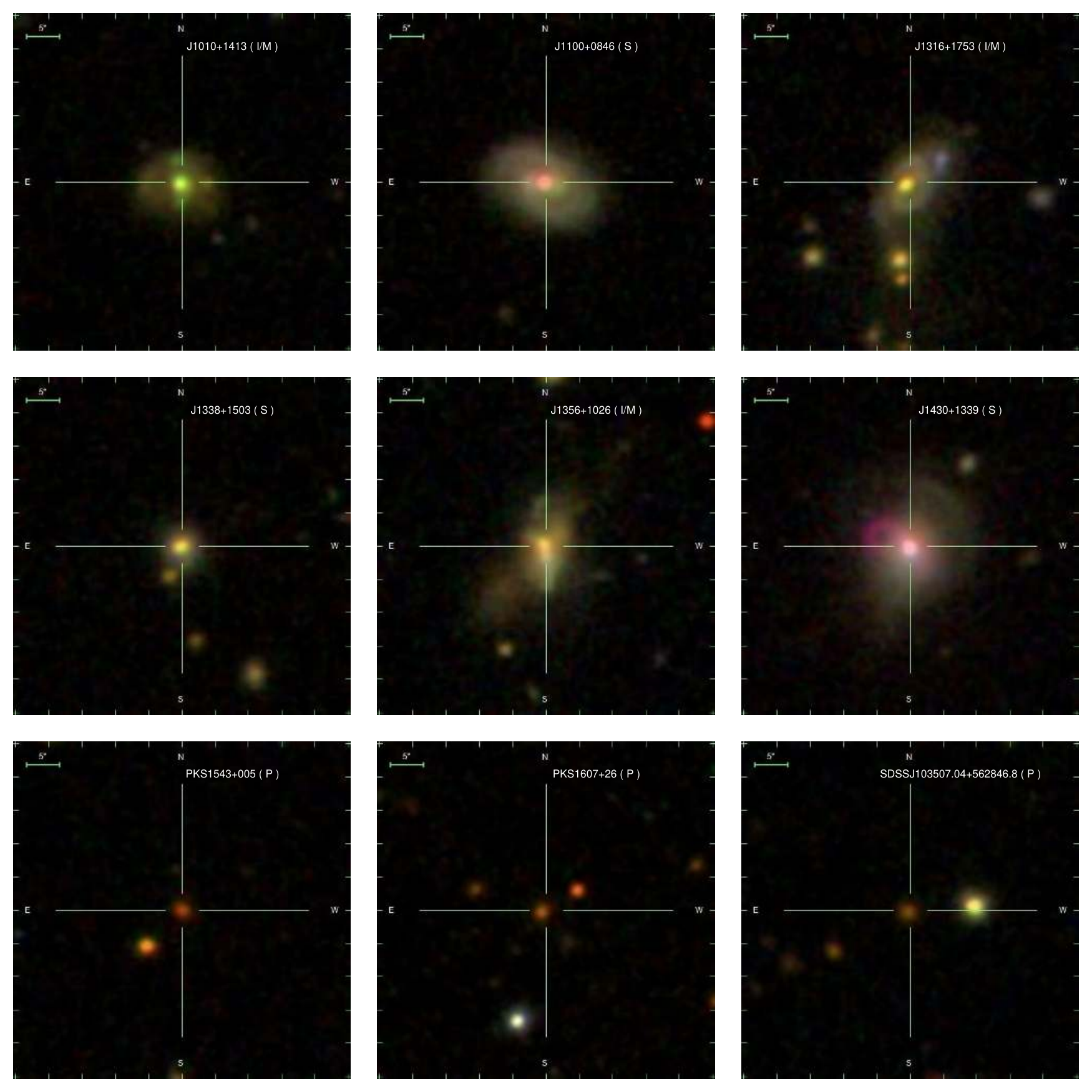}
    \caption{Continued}
\end{figure*}

\begin{figure*}\ContinuedFloat
    \centering
    \includegraphics{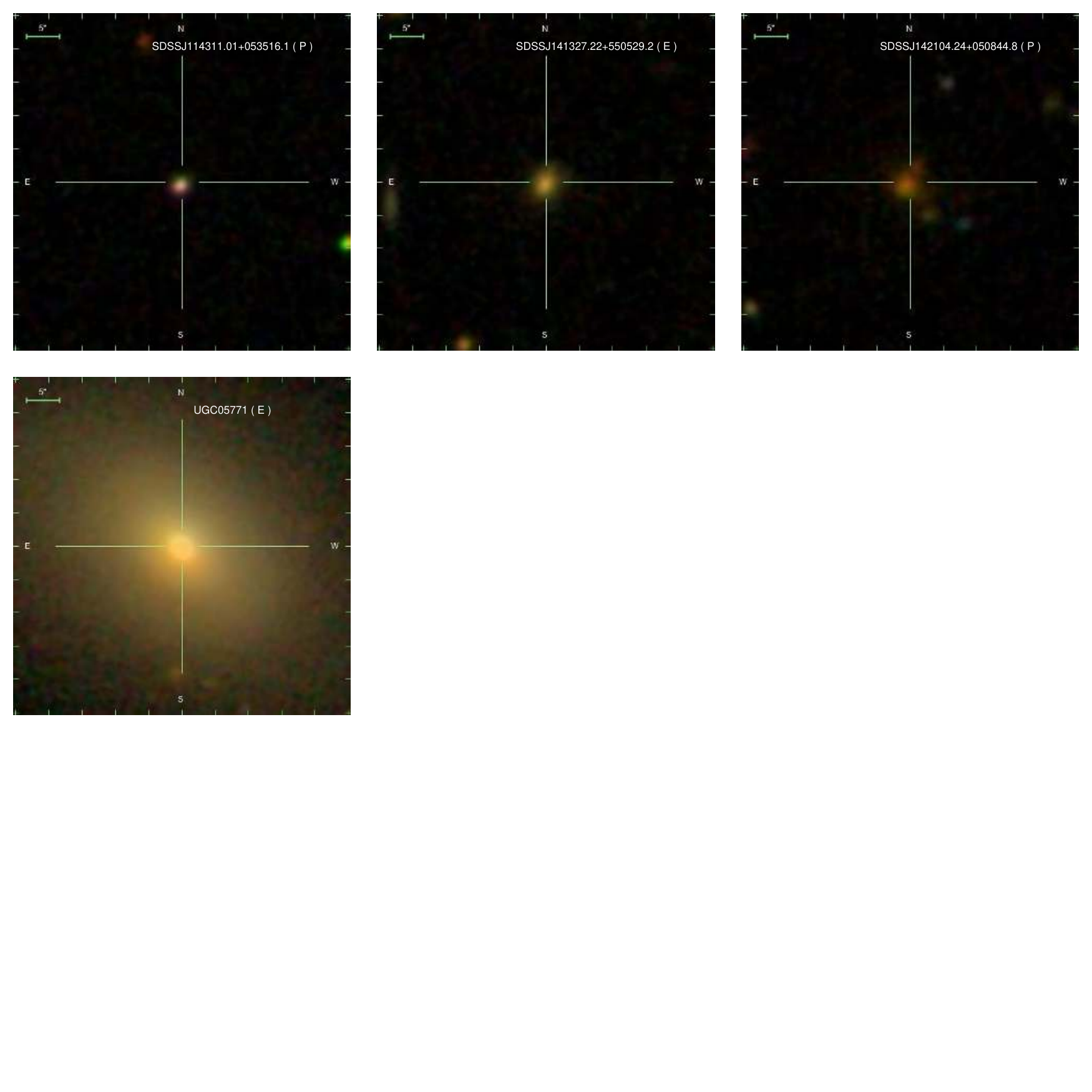}
    \caption{Continued}
\end{figure*}

\begin{figure*}
	\includegraphics[width=0.98\textwidth]{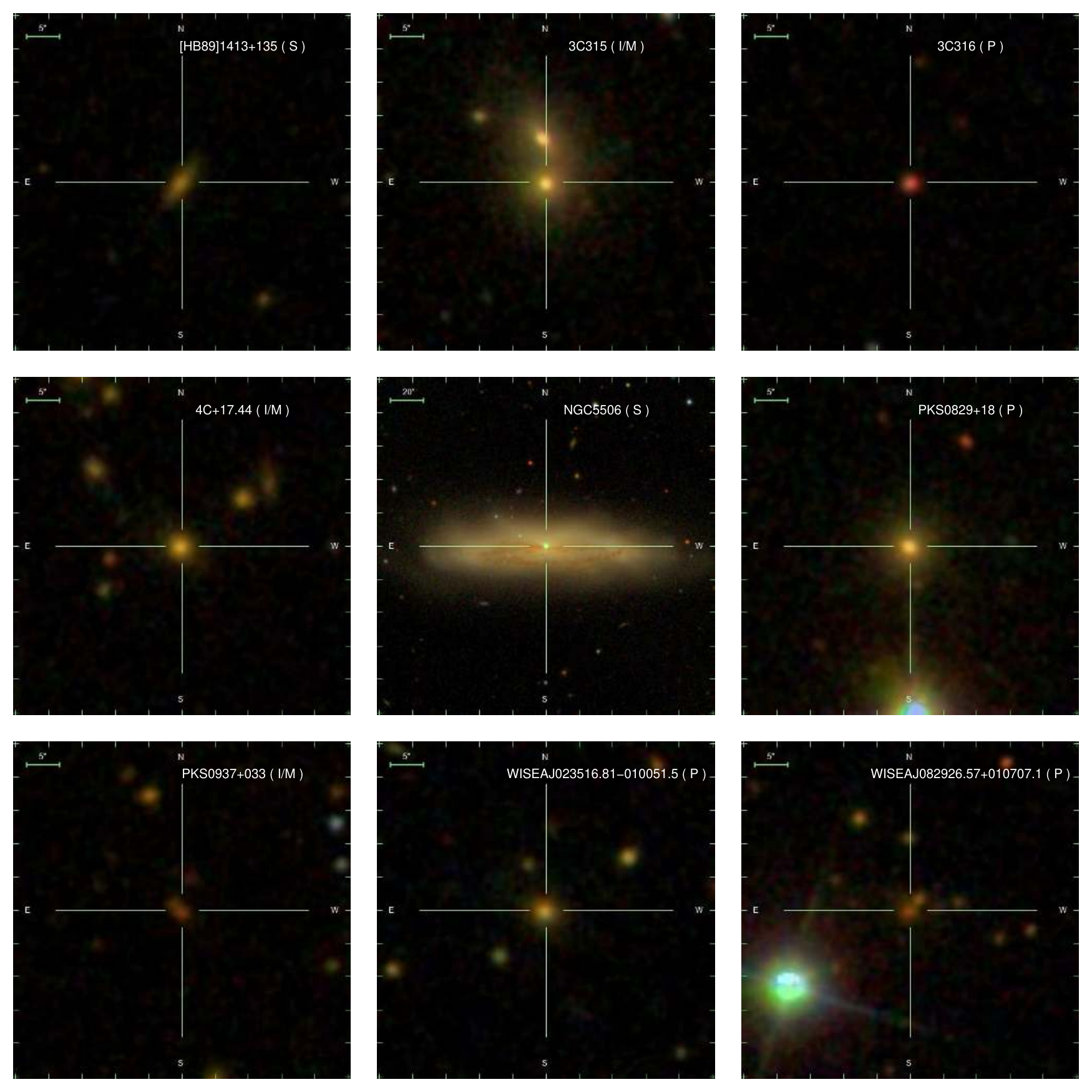}
	\caption{SDSS gri colour composite images for the MPS sources. In each panel, the position of the source is marked by a cross, while the image scale is shown in the upper left corner, the name and morphological classification of the source are indicated in the upper right corner. (E), (S), (I/M) and (P) represent, respectively, elliptical, spiral, irregular/merger and Point-like objects.  The images, except NGC5506, are equal in size, being a square 50''.}
	\label{fig:morph_Call}
\end{figure*}

\begin{figure*}\ContinuedFloat
    \centering
    \includegraphics{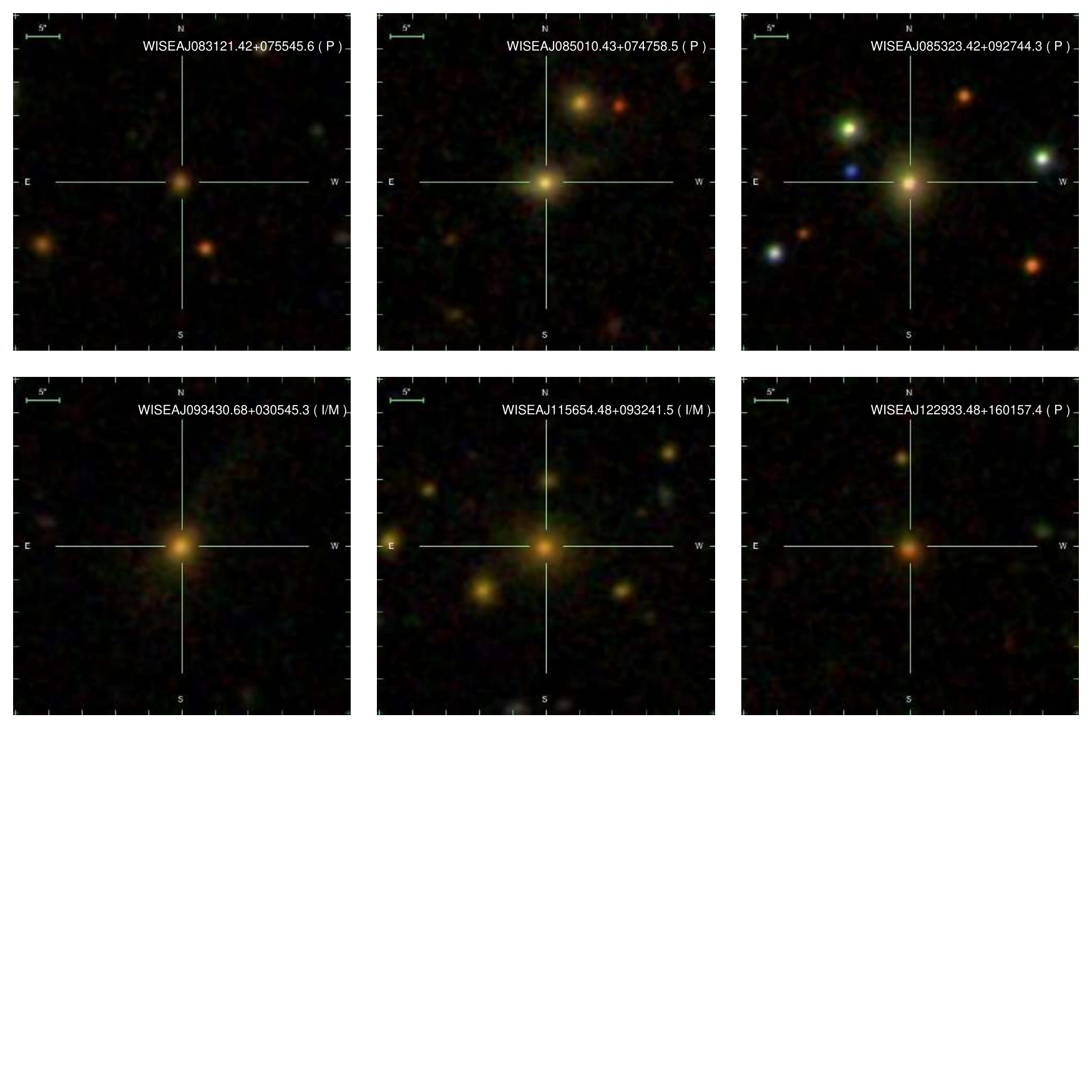}
    \caption{Continued}
\end{figure*}

\label{lastpage}
\end{document}